\documentclass[a4paper,11pt]{article}
\pdfoutput=1 
\usepackage{grffile}
\usepackage{aas_macros,braket}
\usepackage{jcappub,natbib} 
\DeclareUnicodeCharacter{2212}{-}
\usepackage[T1]{fontenc} 
\usepackage{color}
\usepackage{comment}

\usepackage{placeins}
\usepackage{mathtools}
\usepackage{amsmath}
\allowdisplaybreaks
\usepackage{float}

\usepackage{subcaption}

\usepackage{booktabs,siunitx}
\def\be{\begin{equation}}
\def\ee{\end{equation}}
\def\bea{\begin{eqnarray}}
\def\eea{\end{eqnarray}}

\def\ba#1\ea{\begin{align}#1\end{align}}

\def\up{\;\raise1.0pt\hbox{$'$}\hskip-6pt\partial\;}
\def\down{\;\overline{\raise1.0pt\hbox{$'$}\hskip-6pt
		\partial}\;}

\newcommand{\f}{\frac}

\def\apjs{Astrophys.\ J. Supp.}

\renewcommand{\v}[1]{\mathbf{#1}}

\newcommand{\vx}{\v{x}}
\newcommand{\vk}{\v{k}}
\newcommand{\vq}{\v{q}}
\newcommand{\bq}{\mathbf{q}}
\newcommand{\bk}{\mathbf{k}}
\newcommand{\bx}{\mathbf{x}}

\newcommand{\Omegagw}{{\Omega^I_{{\rm{{GW}}}}}}
\newcommand{\OmegagwV}{{\Omega^V_{{\rm{{GW}}}}}}
\newcommand{\deltagw}{\ensuremath{\delta^{{\rm{GW}}, \, I}}}
\newcommand{\deltagwV}{\ensuremath{\delta^{{\rm{GW}}, \, V}}}

\newcommand{\fnls}{\ensuremath{F^{\text{tts}}_{\rm NL}}}
\newcommand{\fnlt}{\ensuremath{F^{\text{ttt}}_{\rm NL}}}

\DeclareMathOperator{\sinc}{sinc}
\def\arrvline{\hfil\kern\arraycolsep\vline\kern-\arraycolsep\hfilneg}

\title{Probing parity-odd bispectra with anisotropies of GW $V$ modes}

\author[a]{Giorgio Orlando}

\affiliation[a]{Van Swinderen Institute for Particle Physics and Gravity, University of
Groningen, Nijenborgh 4, 9747 AG Groningen, The Netherlands.}

\emailAdd{g.orlando@rug.nl}

\abstract{It is well known that non-trivial squeezed tensor bispectra can lead to anisotropies in the inflationary stochastic gravitational wave (GW) background, providing us with an alternative and complementary window to primordial non-Gaussianities (NGs) with respect to the CMB. Previous works have highlighted the detection prospects of parity-even tensor NGs via the GW $I$-mode anisotropies. In this work we extend this by analysing for the first time the additional information carried by GW $V$-mode anisotropies due to squeezed NGs. We show that GW $V$ modes allow us to probe parity-odd squeezed $\langle \rm tts \rangle$ and $\langle \rm ttt \rangle$ bispectra. These bispectra break parity at the non-linear level and can be introduced by allowing alternative symmetry breaking patterns during inflation, like those comprised in solid inflation. Considering a BBO-like experiment, we find that a non-zero detection of squeezed $\langle \rm tts \rangle$ parity-odd bispectra in the $V$ modes dipole is possible without requiring any short-scale enhancement of the GW power spectrum amplitude over the constraints set by the CMB. We also briefly discuss the role of $V$-CMB cross-correlations. Our work can be extended in several directions and motivates a systematic search for polarized GW anisotropies in the next generations of GW experiments.}

\begin{document}

\maketitle
\flushbottom

\section{Introduction}

The Cosmic Microwave Background (CMB) has been crucial in shaping the current cosmological model. However, recently the first detection of a gravitational wave (GW) event on September 14, 2015 \cite{LIGOScientific:2016aoc} has lead to the rise of the GW cosmology. In fact, even if currently GW experiments mostly detect GWs belonging to individual astrophysical sources, while they can only put upper bounds on stochastic gravitational wave backgrounds (SGWB) \cite{LIGOScientific:2019vic,KAGRA:2021kbb}, the next generations of interferometers should possess the sensitivity necessary to detect backgrounds of GWs, relevant to cosmology. In particular, among the SGWB we can distinguish between two types of backgrounds: the astrophysical gravitational wave background (AGWB) which consists of the superposition of GWs belonging to individual unresolved sources; and the cosmological gravitational wave background (CGWB) which presents a background of gravitational waves that originated from some early universe mechanism (see e.g. \cite{Caprini:2018mtu} for a review of possibilities). The AGWB is the main aim of planned next generation interferometers that will become operative within the next decade: we have for instance the ground-based interferometers Cosmic Explorer (CE) \cite{Reitze:2019iox} and the Einstein Telescope (ET) \cite{Punturo:2010zz}, the space-based interferometers Laser Interferometer Space Antenna (LISA) \cite{LISA:2017pwj,LISACosmologyWorkingGroup:2022jok} and Taiji \cite{Hu:2017mde}, together with their networks (see e.g.~\cite{Ruan:2020smc,Wang:2021uih}). More futuristic, there are already designs for experiments like DECIGO \cite{Kawamura:2006up,Kawamura:2020pcg} and the Big Bang Observer (BBO) \cite{Crowder:2005nr} whose ultimate aim is the detection of the CGWB, in particular the background of gravitational waves from inflation, which is the topic of this work.

The inflationary background of gravitational waves differs from other backgrounds as it affects the Universe and can in principle be detected across many different scales \cite{Campeti:2020xwn}, from the very small frequencies of the CMB experiments ($f \approx 8 \times 10^{-17}\, \mbox{Hz}$ corresponding to the wave-number $k =  0.05 \, \mbox{Mpc}^{-1}$) up to e.g. the interferometric frequencies $f \approx 1$ Hz (for the BBO/DECIGO experiments). In particular, at the low frequencies of CMB, hopes to detect the primordial background of GWs rely on the so-called $B$-mode, whose detection represent the main target of the next generation CMB experiments, such as LiteBIRD \cite{LiteBIRD:2022cnt}, the Simons Observatory \cite{SimonsObservatory:2018koc} and CMB-S4 \cite{CMB-S4:2020lpa}. So far, the tightest constraints on the power spectrum of GWs is given by the last combined BICEP/Keck+Planck data constraint on the tensor-to-scalar ratio, $r< 0.032$ at $k =  0.05 \, \mbox{Mpc}^{-1}$ \cite{Tristram:2021tvh}\footnote{Similar constraints are given by the recent Refs. \cite{Galloni:2022mok,Paoletti:2022anb}.}. Given the uncertainty towards a possible detection of $B$ modes, it is important to develop the theoretical and experimental tools to search for primordial gravitational waves at scales different from the CMB. This is even more important considering the large class of models comprising growth mechanisms of the GWs amplitude at interferometric scales, such as those with extra degrees of freedom (see e.g. \cite{Cook:2011hg,Barnaby:2011qe,Maleknejad:2011jw,Dimastrogiovanni:2016fuu,Garcia-Bellido:2016dkw,Thorne:2017jft,Domcke:2018eki,Bordin:2018pca,Iacconi:2019vgc,Iacconi:2020yxn,Watanabe:2020ctz,Fumagalli:2020nvq,Fumagalli:2020adf}), models comprising non-attractor phases (see e.g.~\cite{Mylova:2018yap,Byrnes:2018txb,Carrilho:2019oqg,Ozsoy:2019lyy,Ozsoy:2019slf,Tasinato:2020vdk}),  models with excited non-Bunch Davies initial states (see e.g. \cite{Holman:2007na,Agarwal:2012mq,Akama:2020jko,Fumagalli:2021mpc}), and models with alternative symmetry breaking patterns (see e.g. ~\cite{Endlich:2012pz,Endlich:2013jia,Bartolo:2015qvr,Bartolo:2018elp,Ricciardone:2016lym,Ricciardone:2017kre,Mirzagholi:2020irt,Celoria:2020diz,Bordin:2020eui,Cabass:2021iii,Celoria:2021cxq}). For all these models, a detection of inflationary CGWB could be possible even if no detection is made on CMB scales. 

While the inflationary CGWB is expected to be predominantly isotropic, in a way analogous to the formation of CMB anisotropies we can have tiny fluctuations in the GW background due to its propagation in a perturbed universe. These have been studied e.g. in Refs.~\cite{Alba:2015cms,Contaldi:2016koz,Bartolo:2019oiq,Bartolo:2019yeu,Domcke:2020xmn,Dimastrogiovanni:2022eir}. In addition to these we can have anisotropies sourced by some primordial production mechanism that will be present at the formation of the inflationary CGWB. One example of such anisotropies are those induced by primordial non-Gaussianities (NGs) via a non-zero correlation between one very large and two very short scales, corresponding to the so-called squeezed configuration of primordial bispectra. In fact, it is well established that long-mode perturbations in squeezed bispectra induce local modulations in primordial power spectra (see e.g. Refs. \cite{Jeong:2012df,Dai:2013kra,Dimastrogiovanni:2014ina}). As the fractional energy density per logarithmic wave-number of the CGWB from inflation is by definition linearly related to the tensor power spectrum, these inhomogeneities result in intrinsic anisotropies already present at the formation of the inflationary CGWB. Previous works have mostly studied the theoretical predictions and the detection prospects for the anisotropies induced by squeezed parity invariant tensor-tensor-scalar ($\langle \rm tts \rangle$) and tensor-tensor-tensor ($\langle \rm ttt \rangle$) bispectra~\cite{Dimastrogiovanni:2019bfl,Adshead:2020bji,Malhotra:2020ket,Dimastrogiovanni:2021mfs,Dimastrogiovanni:2022afr}\footnote{For other probes of primordial tensor NGs complementary to the CMB, see e.g. \cite{Jeong:2012df, Dimastrogiovanni:2014ina,Bartolo:2019yeu, Bartolo:2019oiq,Biagetti:2020lpx,Orlando:2021nkv}.}. It was shown that they give non-zero $I$(Intensity)-mode anisotropies to the GW background from inflation. The main contamination for a possible detection are the presence of astrophysical foregrounds, characterized by the AGWB anisotropies.

In this work we will extend these studies by considering for the first time the effect of parity violation in the primordial tensor bispectra. Searching for signatures of parity violation from inflation is highly motivated in light of the recent observational evidence of parity-odd 4-th point correlation function of galaxies \cite{Hou:2022wfj,Philcox:2022hkh,Cabass:2022rhr,Cabass:2022oap}.  In particular, here we will focus on parity-odd squeezed $\langle \rm tts \rangle$ and $\langle \rm ttt \rangle$ bispectra which, as shown in \cite{Cabass:2021fnw}\footnote{See \cite{Bordin:2020eui,Cabass:2021iii,Cabass:2022jda} for further literature on graviton non-Gaussianities from symmetry principles.}, break the parity symmetry at the non-linear level and can be introduced at tree level and without necessarily requiring modification in the primordial power spectra in models of inflation with alternative symmetry breaking patters, such as solid inflation \cite{Endlich:2012pz}\footnote{These bispectra can appear also in ~\cite{Bartolo:2017szm,Bartolo:2020gsh}.}. Moreover, the strength of these parity-odd bispectra can be sizeable, potentially providing a quite large parameter-space available for detection.

We will show that these bispectra source anisotropies in GW $V$ modes, the GW circular polarization, without necessarily sourcing $I$-mode anisotropies or a $V$-mode monopole. Differently from GW $I$ modes, GW $V$ modes are expected to be free from the contamination of the astrophysical foregrounds \cite{Yunes:2013dva,Berti:2015itd,Kostelecky:2016kfm,Nishizawa:2018srh,ColemanMiller:2019tqn,Wang:2020cub}, and therefore detecting a $V$-mode GW signal can be crucial to disentangle it from a non-primordial origin. Previous studies have been focused mostly on the detection of the $V$ modes monopole \cite{Seto:2006hf,Seto:2006dz,Smith:2016jqs,Domcke:2019zls,Seto:2020zxw,Orlando:2020oko}, which is sensitive to the so-called chirality of gravitational waves, i.e. the asymmetry between the right (R)- and left (L)-handed primordial tensor power spectrum. Here, for the first time we look to $V$-mode anisotropies sourced by primordial parity-odd NGs.

In our investigation we first derive more general formulas for the $I$- and $V$-mode anisotropies in terms of the full chiral GW basis, accounting for primordial tensor bispectra with mixed chirality. We provide the expressions for the $I$ and $V$ modes power-spectra together with their cross-correlation with CMB anisotropies for tensor squeezed bispectra that are commonly found in the literature. Next, we focus on the case of parity-odd bispectra, assuming no parity violation in the tensor power spectrum. We find that only the GW $I$-mode monopole, $V$-mode anisotropies and their cross-correlations with CMB $T$ and $E$ modes are non-zero, while the $V$-mode monopole and $I$-mode anisotropies and their cross-correlations with CMB vanish. Therefore, finding evidence for a non-zero signal compatible with these characteristics could be a strong indicator of parity violation in the primordial universe via non-linear mechanisms.
Moreover, by considering a BBO-like experiment and the formalism developed in Ref.~\cite{Alonso:2020rar} we will explore the detection prospects of GW $V$-mode anisotropies due to squeezed primordial tensor NGs. In particular, we will take the coplanar BBO-star configuration. This is in apparent contradiction with what is well-known since a study in Ref. \cite{Smith:2016jqs}, where it was shown that only networks of interferometers can measure the GW $V$-mode monopole. The physical reason is that a planar interferometer responds identical to a L-handed GW arriving perpendicular to the plane of the detector and to a R-handed GW of the same amplitude coming from the opposite direction. However, an exception arises in the presence of anisotropies. In that case, as shown by e.g. Refs. \cite{Seto:2006hf,Seto:2006dz,Domcke:2019zls}, a planar detector can be sensitive to GW $V$-mode anisotropies.

We find that in 5 years a non-zero detection of a $V$-mode dipole due to squeezed $\langle \rm tts \rangle$ parity-odd bispectra is possible for $f_{\rm NL}^{\rm odd, tts} \sim 3 \times 10^3$ without requiring any short-scale enhancement of the gravitational waves power spectrum amplitude with respect to the constraint in place at CMB scales. This result is already relevant considering that CMB experiments are forecasted to place poor constraints on squeezed parity-odd $\langle \rm tts \rangle$ bispectra, even in case of a primordial $B$-mode detection (see e.g.~\cite{Bartolo:2018elp}). We also briefly discuss the role of $V$-CMB cross-correlations to improve detection prospects and confirm a primordial detection. We find that in the case of a noise-dominated detection of $V$-mode anisotropies, cross-correlations with CMB anisotropies are unable to improve the parameter-space probed. On the contrary, we find that a signal-dominated detection of $V$ modes can always be confirmed to be of primordial origin by exploiting the $V$-CMB cross-correlations. Our work can be extended in several directions as explained in the conclusions and motivate the development of further tools for the systematic search of polarized GW anisotropies in the next generations of GW experiments.

The paper is organized as follows. In Sec. \ref{sec:basics} we introduce some basics about inflation, introducing general parametrizations of squeezed bispectra at leading order and review the topic of squeezed modulations of primordial power spectra. In Sec. \ref{sec:anisotropies_rev} we review the mechanisms leading to CMB and GW anisotropies. In Sec. \ref{sec:parity-odd_inv} we derive general expressions for the auto GW and cross GW-CMB correlations due to primordial (squeezed) tensor NGs. We make a match with the literature for the parity-even case, and we study the detection prospects of parity-odd NGs in a BBO-like experiment, considering also the cross-correlations with CMB anisotropies. Finally, in Sec. \ref{sec:conclusions} we present our conclusions and discuss possible extensions of this work. We also include appendixes where we derive explicitly some technical results of this work.

\section{Basics} \label{sec:basics}

In this section we provide the reader with a brief review of the fundamental elements and formal conventions adopted to describe primordial perturbations from inflation. We also provide a general parametrization of the leading order value of squeezed primordial bispectra assuming rotational invariance, and review the topic of squeezed modulations of primordial tensor power spectra.

\subsection{Primordial perturbations from inflation}

Here, we provide the conventions used to describe primordial perturbations from inflation. First, we define the Fourier transform decomposition of scalar and tensor perturbations as
\be
\zeta(\bx) = \int \frac{d^3 k}{(2 \pi)^3} \,  e^{i  \bk \cdot \bx} \, \zeta_{\bk} \, ,
\ee
and
\be
\gamma_{ij}(\bx) = \int \frac{d^3 k}{(2 \pi)^3} \,  e^{i \bk \cdot \bx} \, \sum_{\lambda = R/L} \, \left[\gamma_{\bk}^\lambda \, \epsilon_{ij}^\lambda(\hat k) \right] \, .
\ee
For the purpose of what follows, we are decomposing tensor perturbations in terms of the chiral polarization basis defined through\footnote{While here we are introducing these definitions to describe primordial gravitational waves, these can also refer to GWs of alternative origin.} 
\ba
\epsilon_{ij}^{R, L} &= \frac{1}{2} \left[\epsilon_{ij}^+ \pm i \, \epsilon_{ij}^\times \right] \, ,\\
\gamma^{R, L} &= \frac{1}{2} \left[ \gamma_+ \pm i \, \gamma_\times \right] \, ,
\ea
where $\gamma_{+,\times}$ and $\epsilon_{ij}^{+, \times}$ are the usual linear polarizations of tensor perturbations. 

We remind that, if the tensor wave-vector is written in polar coordinates as 
\be
\hat k = (\sin\theta\cos\phi,\sin\theta\sin\phi,\cos\theta)\, ,
\ee
we can define the linear polarization tensors in terms of two unit vectors perpendicular to $\hat k$ as
\ba
\epsilon_{ij}^{+} &= (u_1)_i (u_1)_j - (u_2)_i (u_2)_j \, ,\\
\epsilon_{ij}^{\times} &= (u_1)_i (u_2)_j + (u_2)_i (u_1)_j \, ,
\ea
where 
\be
u_1 = \left(\sin \phi , - \cos \phi, 0\right)\, ,  \qquad u_2 = \left(\cos \theta \cos \phi , \cos \theta \sin \phi, - \sin \theta \right) \, .
\ee
The chiral polarization basis introduced is normalized such that it satisfies the following identities
\begin{align}
\epsilon_{ij}^{L}(\hat k)\epsilon_{L}^{ij}(\hat k)&=\epsilon_{ij}^{R}(\hat k)\epsilon_{R}^{ij}(\hat k) = 0 \, , \nonumber\\
\epsilon_{ij}^L(\hat k)\epsilon_R^{ij}(\hat k)&= 1 \, ,\nonumber\\
\epsilon_{ij}^{R}(-\hat k)&=\epsilon_{ij}^{L}(\hat k) \, ,\nonumber\\
\epsilon^{R *}_{ij}(\hat k)&=\epsilon^{L}_{ij}(\hat k),\nonumber\\
\gamma^{R*}_{\bk}&= \gamma_{\bk}^{L}  \, ,\nonumber\\
k_l \, \epsilon^{mlj} \, {\epsilon_j^{(\lambda) i}}(\hat k)&= -i \alpha_\lambda \, k \, \epsilon^{(\lambda) im}(\hat k) \, ,\label{eq:circ_identities}
\end{align}
where $\alpha_R=+1$ and $\alpha_L=-1$, and $\epsilon^{mlj}$ denotes the Levi-Civita anti-symmetric symbol.

We define the primordial power spectra as
\ba
\langle \zeta_{\bk_1} \zeta^*_{\bk_2}\rangle &= (2 \pi)^3 \delta(\bk_1 - \bk_2) \, P_s(\bk_1) \, , \\
\langle \gamma_{ij}(\bk_1) \gamma^{ij *}(\bk_2)\rangle &= (2 \pi)^3 \delta(\bk_1 - \bk_2) \, P_t(\bk_1) \, ,
\ea
where 
\be
\gamma_{ij}(\bk) = \sum_{\lambda = R/L} \, \gamma_{ij}^\lambda(\bk)   = \sum_{\lambda = R/L} \, \left[\gamma_{\bk}^\lambda \, \epsilon_{ij}^\lambda(\hat k) \right] \, .
\ee
In these definitions, we are implicitly assuming invariance under translations during inflation. Assuming invariance under rotations, the scalar and tensor power spectra only depend on the magnitude of the momentum and are typically expressed as
\be \label{eq:power_inflation}
P_{s}(k) = \frac{2 \pi^2}{k^3}  \mathcal A_s(k) \, ,\qquad P_{t}(k) = \frac{2 \pi^2}{k^3} \mathcal A_t(k) \, , \qquad r(k) = \frac{P_{t}(k)}{P_{s}(k)} \, ,
\ee
where $\mathcal A_s(k)$ and $\mathcal A_t(k)$ are dimensionless amplitudes and $r$ denotes the tensor-to-scalar ratio.

By introducing the spatial modulations generated by soft momenta (see Sec.~\ref{sec:squeezed}), we can redefine the previous quantities as
\be \label{eq:power_inflation_loc}
P_{s}(\vk, \vx_c) = \frac{2 \pi^2}{k^3}  \mathcal A_s(\vk, \vx_c) \, ,\qquad P_{t}(\vk, \vx_c) = \frac{2 \pi^2}{k^3} \mathcal A_t(\vk, \vx_c) \, .
\ee
Here, $\vx_c$ denotes the midpoint of the spatial separation $\vx_1 - \vx_2$ corresponding to the momentum $\vk$.

Notice that we can also define the polarized-power spectra of tensor perturbations
\ba
\langle  \gamma_{ij}^R(\bk_1) \, \gamma^{ij, \, R*}(\bk_2)\rangle &= (2 \pi)^3 \delta(\bk_1 - \bk_2) \, P_t^R(\bk_1) \, , \\
\langle \gamma_{ij}^L(\bk_1) \, \gamma^{ij, \, L*}(\bk_2)\rangle &= (2 \pi)^3 \delta(\bk_1 - \bk_2) \, P_t^L(\bk_1) \, .
\ea
These power spectra can be used to define the quantity $\chi$
\be \label{eq:chi}
\chi(k) = \frac{P_t^R(k) - P_t^L(k)}{P_t^R(k) + P_t^L(k)} \, ,
\ee
which is referred to as the chirality of gravitational waves. It measures the asymmetry between the $R$- and $L$-handed power spectra caused by parity violation arising in the primordial universe\footnote{For examples of parity violation mechanisms see e.g. \cite{Lue:1998mq,Alexander:2004wk,Takahashi:2009wc,Satoh:2010ep,Sorbo:2011rz,Maleknejad:2011jw,Wang:2012fi,Adshead:2013nka,Adshead:2013qp,Maleknejad:2016qjz,Peloso:2016gqs,Dimastrogiovanni:2016fuu,Domcke:2018rvv,Maleknejad:2018nxz,Mylova:2019jrj,Qiao:2019hkz,Papageorgiou:2019ecb,Watanabe:2020ctz}.}. Assuming parity is a symmetry of the theory, $P_t^{R, L}$ are related to $P_t$ as
\be
P_t^{R, L} = \frac{P_t}{2} \, .
\ee
Again, we can define the spatial-modulated $\lambda$-handed dimensionless tensor power spectrum through
\be \label{eq:power_inflation_loc_RL}
 P^\lambda_{t}(\vk, \vx_c) = \frac{2 \pi^2}{k^3} \mathcal A^\lambda_t(\vk, \vx_c) \, .
\ee
Finally, we define the primordial bispectra
\ba \label{eq:def_bispectra}
\langle \zeta_{\bk_1} \zeta_{\bk_2} \zeta_{\bk_3}\rangle &= (2 \pi)^3 \delta(\bk_1+ \bk_2+ \bk_3) \, B_{\rm sss}(\bk_1, \bk_2, \bk_3) \nonumber\\
\langle \zeta_{\bk_1} \zeta_{\bk_2} \gamma^{\lambda_3}_{\bk_3}\rangle &= (2 \pi)^3 \delta(\bk_1+ \bk_2+ \bk_3) \, B^{\lambda_3}_{\rm sst}(\bk_1, \bk_2,\bk_3) \nonumber\\
\langle \gamma^{\lambda_1}_{\bk_1} \gamma^{\lambda_2}_{\bk_2} \zeta_{\bk_3} \rangle &= (2 \pi)^3 \delta(\bk_1+ \bk_2+ \bk_3) \, B^{\lambda_1 \lambda_2}_{\rm tts}(\bk_1, \bk_2, \bk_3) \nonumber\\
\langle \gamma^{\lambda_1}_{\bk_1} \gamma^{\lambda_2}_{\bk_2} \gamma^{\lambda_3}_{\bk_3}\rangle &= (2 \pi)^3 \delta(\bk_1+ \bk_2+\bk_3) \, B^{\lambda_1 \lambda_2 \lambda_3}_{\rm ttt}(\bk_1,\bk_2,\bk_3) \, ,
\ea
where we assumed invariance under translations. If we account for the invariance under rotations, the bispectra would depend only on the magnitude of the momenta.

\subsection{Squeezed bispectra and power spectra modulation} \label{sec:squeezed}

In squeezed bispectra the maximum of the signal arises in momenta configurations when one of the three momenta is much smaller than the other two. Physically, these configurations correspond to a significant correlation between very long and very short scales. Here, we will develop a phenomenological approach to introduce the leading order value of squeezed bispectra, solely relying on rotational invariance. However, it has been shown\footnote{See the earliest classical investigations \cite{Maldacena:2002vr,Tanaka:2011aj,Creminelli:2013cga,Pajer:2013ana}, but also the more recent Refs. \cite{Sreenath:2014nca,Sreenath:2014nka,Bordin:2017ozj,Bravo:2017wyw,Finelli:2017fml,Cai:2018dkf,Jazayeri:2019nbi,Bravo:2020hde,Suyama:2021adn} which investigated soft theorems in more general scenarios.} that a series of soft theorems constrains the power of squeezed configurations in single-field models of inflation. Typically, we can evade these theorems and have significant squeezed tensor and mixed scalar-tensor NGs in models of inflation with isocurvature fields \cite{Chen:2009zp,Baumann:2017jvh,Bordin:2018pca,Dimastrogiovanni:2018gkl,Iacconi:2019vgc,Iacconi:2020yxn,Dimastrogiovanni:2021mfs}, models with non-Bunch Davies initial states \cite{Holman:2007na,Agarwal:2012mq,Akama:2020jko}, and in models with alternative symmetry breaking patterns \cite{Endlich:2012pz,Endlich:2013jia,Bartolo:2015qvr,Bartolo:2017szm,Bartolo:2018elp,Ricciardone:2016lym,Ricciardone:2017kre,Mirzagholi:2020irt,Celoria:2020diz,Bartolo:2020gsh,Bordin:2020eui,Cabass:2021iii,Celoria:2021cxq,Cabass:2021fnw}. These are the models to which our phenomenological parametrizations apply. 

We start by considering a generic bispectrum 
\ba
B_{\Phi \Phi \Psi}(\bk_1, \bk_2, \bq) \, .
\ea
In the squeezed limit $q \rightarrow 0$ it is convenient to re-write the short momenta as
\begin{equation} \label{eq:redef_momenta_main}
\bk_1 = \bk - \bq/2  \, , \qquad \bk_2 = -\bk - \bq/2  \, .
\end{equation}
Since we are assuming rotational invariance, our bispectrum will depend only on the magnitude of the momenta. These values can be expressed in terms of two momenta and their scalar product. By using the parametrization of Eq. \eqref{eq:redef_momenta_main}, we need only $k$, $q$ and $\hat k \cdot \hat q$. We can opt for the following parametrization in terms of short- and long-modes power spectra
\ba \label{eq:def_terms_F}
B_{\Phi \Phi \Psi}(\bk_1, \bk_2, \bq)|_{q \rightarrow 0} = F_{\rm NL}^{\Phi \Phi \Psi}(\bk, \bq) \, P_\Phi(k) P_\Psi(q) \, .
\ea
Here the product of the two short- and long-scale power spectra is required to provide the bispectrum with the correct dimensional scaling. The function $F_{\rm NL}^{\Phi \Phi \Psi}(\bk, \bq)$ may depend on the magnitude of the momenta $k,q$ and their scalar product which provides the bispectrum with an angular dependence in terms of $\hat k \cdot \hat q$. In general, depending on the nature of the long-mode, we can expand this angular dependence in terms of spin-weighted spherical harmonics. This can be expressed as
\ba
 F_{\rm NL}^{\Phi \Phi \Psi}(\bk, \bq) = \sum_{J= {\rm even}}  F_{J}^{\Phi \Phi \Psi}(k, q) \,\, {}_{s} Y_{J 0}(\hat{k} \cdot \hat q) \, ,
\ea
where $s = 0$ when the long-mode is a scalar, $s = -2$ when the long-mode is a $R$-handed tensor, and $s = +2$ when the long-mode is a $L$-handed tensor. Here, the sum over even values of $J$ is a consequence of the assumed invariance under rotational symmetry. Moreover, if primordial bispectra depend on the polarization state of one or more tensor fields $\lambda_i$, we can re-absorb these inside $F_{\rm L}^{\Phi \Phi \Psi}(k, q)$
\ba
 F^{\lambda_1...\lambda_n, \, \Phi \Phi \Psi}_{\rm NL}(\bk, \bq) = \sum_{J= {\rm even}}  F_{J}^{\lambda_1...\lambda_n, \, \Phi \Phi \Psi}(k, q) \,\, {}_{s} Y_{J 0}(\hat{k} \cdot \hat q) \, .
\ea
Therefore, following this prescription, we can parametrize the squeezed limit primordial bispectra as
\ba 
B_{\rm sss}(\bk_1, \bk_2, \bq)|_{q \rightarrow 0}  &= F^{\rm sss}_{\rm NL}(\bk, \bq) \, P_{s}(k) P_{s}(q) = \sum_{J= {\rm even}} F_J^{sss}(k,q) \, Y_{J 0}(\hat{k} \cdot \hat q) \, P_{s}(k) P_{s}(q) \, , \label{eq:ansatzsss} 
\ea
\ba
B^{\lambda'}_{\rm sst}(\bk_1, \bk_2, \bq)|_{q \rightarrow 0}  &=  F^{\lambda', \, \rm sst}_{\rm NL}(\bk, \bq) \, P_{s}(k) P^{\lambda'}_{t}(q) = \sum_{J= {\rm even}} F_J^{\lambda', \, \rm sst}(k,q) \,\, {}_{\pm 2} Y_{J 0}(\hat{k} \cdot \hat q) \, P_{s}(k) P^{\lambda'}_{t}(q) \, , \label{eq:ansatzsst} 
\ea
\ba
B^{\lambda \lambda}_{\rm tts}(\bk_1, \bk_2, \bq)|_{q \rightarrow 0}  &=  F^{\lambda \lambda, \, \rm tts}_{\rm NL}(\bk, \bq) \, P^{\lambda}_{t}(k) P_{s}(q) = \sum_{J= {\rm even}} F_J^{\lambda \lambda, \, \rm tts}(k,q) \, Y_{J 0}(\hat{k} \cdot \hat q) \, P^{\lambda}_{t}(k) P_{s}(q) \, , \label{eq:ansatztts} 
\ea
\ba
B^{\lambda \lambda \lambda'}_{\rm ttt}(\bk_1, \bk_2, \bq)|_{q \rightarrow 0}  &=  F^{\lambda \lambda \lambda', \, \rm ttt}_{\rm NL}(\bk, \bq) \, P^{\lambda}_{t}(k) P^{\lambda'}_{t}(q) \nonumber\\
&= \sum_{J= {\rm even}} F_J^{\lambda \lambda \lambda', \, \rm ttt}(k,q) \,\, {}_{\pm 2} Y_{J 0}(\hat{k} \cdot \hat q) \, P^{\lambda}_{t}(k) P^{\lambda'}_{t}(q)  \, , \label{eq:ansatzttt}
\ea
where we have introduced the quantities
\begin{align}
F_{\rm NL}^{\rm sss}(\vk,\vq) =& \frac{B_{\rm sss}(\vk-\vq/2,-\vk-\vq/2,\vq)|_{q \rightarrow 0}}{P_{s}(k) \, P_{s}(q)} = \sum_{J= {\rm even}} F_J^{sss}(k,q) \, Y_{J 0}(\hat{k} \cdot \hat q) \, , \\
F_{\rm NL}^{\lambda', \rm sst}(\vk,\vq) =& \frac{B^{\lambda'}_{\rm sst}(\vk-\vq/2,-\vk-\vq/2,\vq)|_{q \rightarrow 0}}{P_{s}(k) \, P^{\lambda'}_{t}(q)}  = \sum_{J= {\rm even}} F_J^{\lambda', \, \rm sst}(k,q) \,\, {}_{\pm 2} Y_{J 0}(\hat{k} \cdot \hat q) \, , \\
F_{\rm NL}^{\lambda \lambda, \rm tts}(\vk,\vq) =& \frac{B^{\lambda \lambda}_{\rm tts}(\vk-\vq/2,-\vk-\vq/2,\vq)|_{q \rightarrow 0}}{P^{\lambda}_{t}(k) \, P_{s}(q)}  = \sum_{J= {\rm even}} F_J^{\lambda \lambda, \, \rm tts}(k,q) \, Y_{J 0}(\hat{k} \cdot \hat q) \, , \label{eq:defNtts_main} \\
F_{\rm NL}^{\lambda\lambda \lambda', \rm ttt}(\vk,\vq) =& \frac{B^{\lambda \lambda \lambda'}_{\rm ttt}(\vk-\vq/2,-\vk-\vq/2,\vq)|_{q \rightarrow 0}}{P^{\lambda}_{t}(k) \, P^{\lambda'}_{t}(q) }  = \sum_{J= {\rm even}} F_J^{\lambda \lambda \lambda', \, \rm ttt}(k,q) \,\, {}_{\pm 2} Y_{J 0}(\hat{k} \cdot \hat q) \, . \label{eq:defNttt_main}
\end{align}
Depending on the underlying inflationary model, we can have different angular dependencies where one or more $F_J^{\lambda_1...\lambda_n, \, \rm x_1x_2x_3}(k,q)$ are non-zero. It is worth to mention that while in a rotationally invariant environment a pure scalar bispectrum is insensitive to parity violation (see e.g.~\cite{Shiraishi:2016mok}), we can have parity violation in bispectra involving tensors. In this regard, we expect three different scenarios:
\begin{itemize}
    \item Parity-even bispectra: in this case the bispectra are invariant under parity transformation and therefore they do not depend on the polarization states that are cross-correlated. In terms of the $F_{\rm NL}$'s quantities above the following relations hold
    \ba \label{eq:parityeven}
F^{R, \, \rm sst}_{\rm NL} &= F^{L, \, \rm sst}_{\rm NL} \,  , \nonumber \\
F^{RR, \, \rm tts}_{\rm NL} &= F^{LL, \, \rm tts}_{\rm NL} \, , \nonumber \\
F^{RRR, \, \rm ttt}_{\rm NL} &= F^{LLL, \, \rm ttt}_{\rm NL} = F^{RRL, \, \rm ttt}_{\rm NL} = F^{LLR, \, \rm ttt}_{\rm NL} \, .
\ea
\item Parity-odd bispectra: in this case the bispectra switch sign under parity transformation. In terms of the $F_{\rm NL}$'s quantities above the following relations hold
    \ba \label{eq:parityodd}
F^{R, \, \rm sst}_{\rm NL} &= - F^{L, \, \rm sst}_{\rm NL} \,  , \nonumber \\
F^{RR, \, \rm tts}_{\rm NL} &= - F^{LL, \, \rm tts}_{\rm NL} \, , \nonumber \\
F^{RRR, \, \rm ttt}_{\rm NL} &= F^{RRL, \, \rm ttt}_{\rm NL} = - F^{LLL, \, \rm ttt}_{\rm NL} =  - F^{LLR, \, \rm ttt}_{\rm NL} \, .
\ea
\item Maximum parity violation: in this case parity is maximally violated, i.e. the correlators involving only one of the two polarization states dominate over the others. If we assume tensor perturbations with a predominant $R$-handed polarization we expect the following relations to hold
\ba \label{eq:noparity}
B^{R}_{\rm sst} & \gg B^{L}_{\rm sst} \,  , \\
B^{RR}_{\rm tts} & \gg B^{LL}_{\rm tts} \, , \\
B^{RRR}_{\rm ttt} & \gg B^{RRL}_{\rm ttt}, \, B^{LLL}_{\rm ttt}, \, B^{LLR}_{\rm ttt} \, .
\ea
In other words, only the $R$-handed correlators will leave significant imprints on the cosmological observables, while correlators involving the $L$-handed polarization state will leave either vanishing or highly suppressed signals. 
\end{itemize}
Now that we have clarified our conventions and introduced squeezed bispectra, we can move on and briefly discuss their physical impact on tensor power spectra. In fact, it is well-known (see e.g. Refs. \cite{Jeong:2012df,Dai:2013kra,Dimastrogiovanni:2014ina}) that non-zero squeezed bispectra can provide spatial modulations to either scalar and tensor primordial power spectra. Here we are interested in the modulations to R- and L-handed tensor power spectra. In that case, spatial modulations are provided by non-zero $\langle \rm tts \rangle$ and $\langle \rm ttt \rangle$ squeezed bispectra only. In terms of the parametrizations introduced, these modulations read (for an explicit derivation, see App. \ref{appen:squeez})
\begin{align}
\label{eq:final_mod_scalar_main}
\mathcal A^{\lambda, \, \rm tot}_{t}(\vk,\vx_c) = \mathcal A^\lambda_{t}(k)\left[1+\int \frac{d^3q}{(2\pi)^3} \, e^{i\vq\cdot\vx_c} \,  F^{\lambda \lambda, \rm tts}_{\rm NL}(\vk,\vq) \, \zeta(\vq) \right] \, ,
\end{align}
and
\begin{align}
\label{eq:final_mod_tensor_main}
\mathcal A^{\lambda, \, \rm tot}_{t}(\vk,\vx_c) = \mathcal A^\lambda_{t}(k)\left[1+\int \frac{d^3q}{(2\pi)^3}\,e^{i\vq\cdot\vx_c} \, \sum_{\lambda' = R/L} F^{\lambda \lambda \lambda', \rm ttt}_{\rm NL}(\vk,\vq) \, \gamma^{\lambda'}(\vq) \right] \, .
\end{align}
Taking the small spatial separation $\vx_1 - \vx_2$ limit, corresponding to the (short) Fourier mode scale $k$, $\vx_c$ represents the point in the middle of $\vx_1 - \vx_2$. This local modulation of the power spectrum, as a function of $\bx_c$, is meaningful provided that the correlation scale $\vx_1 - \vx_2$ is small compared to the smallest wavelength of the modulating field which, translated in Fourier space, sets the upper bound of the $q$-integration, $q_{\rm max} < k$ for a given $k$. What about the lower bound $q_{\rm min}$? This will correspond to the largest scale at which perturbations are produced during inflation (therefore, it is related to the duration of inflation).

\section{CMB and GW anisotropies} \label{sec:anisotropies_rev}

\subsection{Review of CMB anisotropies}

Here, we give a brief overview of the physics of the CMB and how we characterize CMB anisotropies. In general, the CMB fluctuation field includes four different polarization states, the so-called Stokes parameters (see e.g.  \cite{Kosowsky:1994cy}), which are functions of the position and direction on the sky ${\hat n}$. We have the so-called $T$ modes, representing anisotropies in unpolarized radiation, $Q$ and $U$ modes representing anisotropies in the linear polarization field, and $V$ modes representing anisotropies in the CMB circular polarization. These can be expanded on the sky in terms of a spin-weighted basis as\footnote{When we refer to the Stokes parameters, we take only the relative fluctuations over the respective mean value, e.g. $\Delta_T = (\Delta T - T_0)/T_0$ and so on.} \cite{Hu:1997hp}
\be
\Delta_T({\hat{n}})=\sum_{\ell, m}a^{T}_{\ell m} \, Y_{\ell m}({\hat{n}})\, ,
\ee
\be
\Delta_V({\hat{n}})=\sum_{\ell, m}a^{V}_{\ell m} \, Y_{\ell m}({\hat{n}})\, ,
\ee
\be
\Delta_P^{\pm}({\hat{n}})=(\Delta_Q\pm i \, \Delta_U)({\hat{n}})=\sum_{\ell, m}a^{\pm 2}_{\ell m} \,\, _{\pm 2\, }\!Y_{\ell m}({\hat{n}})\, . \label{P_def}
\ee
This decomposition is possible since the $\Delta_T$ and $\Delta_V$ polarization fields turn out to be spin-0 fields, while the $(\Delta_Q\pm i \, \Delta_U)$ combination is a spin $\pm2$ field \cite{Hu:1997hp}. In particular, this last feature implies that $\Delta_Q$ and $\Delta_U$ polarization modes are not invariant under a rotation on the polarization plane (while $\Delta_T$ and $\Delta_V$ modes are). In general, a description of the CMB polarization in terms of quantities invariant under rotations is preferable. In order to define these quantities, we need to act on $\Delta_P^{\pm}$ with the spin raising and lowering operators\footnote{See App. \ref{appen:spin_operators} for the definition of these operators and an example.} $\eth$ and $\bar{\eth}$ as
\be \label{eq:defE}
\Delta_{E}({\hat{n}})=-\frac{1}{2}\left[\bar{\eth}^2 \Delta_P^{+}({\hat{n}})+\eth^2 \Delta_P^{-}({\hat{n}})\right] \, ,
\ee
\be \label{eq:defB}
\Delta_{B}({\hat{n}})=\frac{i}{2}\left[\bar{\eth}^2 \Delta_P^{+}({\hat{n}})-\eth^2 \Delta_P^{-}({\hat{n}}) \right] \, .
\ee
Here, we have introduced the so-called $E$ and $B$ polarization modes. These modes offer an alternative description of CMB linear polarization which, differently from the description using $Q$ and $U$ modes, is invariant under a rotation on the polarization plane. $E$ and $B$ modes are the preferred choice to characterize the CMB linear polarization field.

The connection between primordial perturbations from inflation and CMB anisotropies is made through a set of Boltzmann equations (see e.g. \cite{Hu:1997hp,Zaldarriaga:1996xe,Dodelson:2003ft}), which describe the time evolution of CMB temperature and polarization modes at linear level and predict the expected amount of each mode today, given certain initial conditions. These equations take care of two main contributions: the Compton scattering between CMB photons and electrons and the propagation through a background space made inhomogeneous by scalar and tensor perturbations from inflation. These lead to the formation of CMB anisotropies immediately after the decoupling with baryons through the so-called Sachs-Wolfe (SW) effect, and source mechanisms that arise during the propagation of the CMB from the last scattering surface to our detectors, the Integrated Sachs-Wolfe (ISW) effect.

We can define the spherical harmonic coefficients of each (rotationally invariant) CMB mode on the sky as
\be \label{harmonic_dec}
a^{X}_{\ell m} =\int d^2 {\hat{n}} \, Y_{\ell m}({\hat{n}}) \, \Delta_X({\hat{n}})\, ,
\ee
where $X= T, E, B, V$.

The coefficients of the unpolarized ($X = T$) and $E, B$-mode polarization ($X = E, B$) anisotropies given by the scalar ($\zeta$) and the tensor perturbations ($\gamma^{R, L}$) from inflation, are expressed, respectively, as \cite{Shiraishi:2010sm,Shiraishi:2010kd} \footnote{Notice the factor $i^{\ell}$ instead of $(-i)^{\ell}$ as it is commonly used in literature. This is due to the fact that, according to our convention, $\hat{n}$ denotes the direction to the sky, while in the literature it usually refers to the CMB photons direction.}
\begin{align} 
a_{\ell m}^{(s) X} &=
4\pi \, i^{\ell} \, \int \frac{d^3 p}{(2\pi)^{3}}
{\cal T}_{\ell(s)}^{X}(p) \, Y_{\ell m}^*(\hat{p}) \,\zeta_{\mathbf{p}}  \, , \label{eq:a_CMB_scalar} \\
a_{\ell m}^{(t) X} &=
4\pi \, i^{\ell} \, \int \frac{d^3 p}{(2\pi)^{3}}
{\cal T}_{\ell(t)}^{X}(p)  \left[ {}_{-2} Y_{\ell m}^*(\hat{p})  \, \gamma_{\mathbf{p}}^{R} + \left(-1\right)^x  {}_{+ 2} Y_{\ell m}^*(\hat p) \, \gamma_{\mathbf{p}}^{L}  \right]\, ,  \label{eq:a_CMB_tensor}
\end{align}
where ${\cal T}_{\ell (s)}^{X}(p)$ and ${\cal T}_{\ell (t)}^{X}(p)$ are the scalar and tensor CMB transfer functions obtained by solving the aforementioned Boltzmann equations, and $x = 0 \,(1)$ for $X= T, E \,(B)$. Conventional physics implies the CMB is invariant under parity transformation. For that reason $V$ modes are neglected and we set $a_{\ell m}^{V} = 0$. 

It is evident from these equations that CMB fluctuations are closely related to initial primordial perturbations. In this work we will evaluate the spherical harmonics coefficients of CMB anisotropies using Eqs. \eqref{eq:a_CMB_scalar} and \eqref{eq:a_CMB_tensor} and evaluate the CMB transfer functions using the publicly available Boltzmann numerical code \texttt{CAMB} \cite{camb_notes}. We adopt the best-fit \textit{Planck} 2018 flat $\Lambda$CDM cosmology \cite{Aghanim:2018eyx}, whose main cosmological parameters are summarized in Tab. \ref{tab:cosm_par}. 
\begin{table}[h!]
    \begingroup
    \setlength{\tabcolsep}{8pt} 
    \renewcommand{\arraystretch}{1.5} 
    \centering
    \begin{tabular}{c  c  c}
     \toprule
     &  \textbf{Parameters input in \texttt{CAMB}} &\\
     \midrule
        $H_0= 67.32 \, \mbox{km}/\mbox{s} \, \mbox{Mpc}^{-1}$ &  $\Omega_\mathrm{b} h^2 = 0.022383$ & $\Omega_\mathrm{c} h^2 = 0.12011$ \\
        $\Omega_\mathrm{k} = 0$ &  $\Omega_\mathrm{c} h^2 = 0.12011$  & $\tau = 0.0543$\\
        \bottomrule
    \end{tabular}
    \endgroup
    \caption{Best-fit \textit{Planck} parameters obtained combining $TT$, $TE$, $EE$+low$E$+lensing (see the \textit{Plik} best-fit of Tab.~1 of Ref.~\cite{Aghanim:2018eyx}).}
    \label{tab:cosm_par}
\end{table}
\FloatBarrier

\subsection{Review of SGWB anisotropies} \label{sec:self_corr}

There are two types of SGWB that in principle can be observed: the cosmological background (CGWB) which consist of a background of gravitational waves produced during some early universe mechanism, such as inflation, and the astrophysical background (AGWB), which consists of the superposition of gravitational waves emitted by unresolved astrophysical sources. In both cases a direct detection is typically characterized statistically in terms of their fractional energy density per logarithmic wave-number at a given conformal time $\eta$ (see e.g. \cite{Maggiore:1999vm,Watanabe:2006qe,Caprini:2018mtu})
\begin{align} \label{def:Omega_I}
\Omegagw(k, \eta) = \frac{1}{\rho_{\rm{cr}}}\frac{d \rho^I_{\rm GW}}{d \ln k} \, ,
\end{align}
where the GW overall energy density is given by
\ba
\rho^I_{\rm GW} = \frac{\langle \gamma'_{ij}(\eta, \bx) \, \gamma^{ij '}(\eta, \bx) \rangle}{32 \pi G a^2(\eta)} = \frac{\langle \gamma^{' R}_{ij}(\eta, \bx) \, \gamma^{ij ', \, R}(\eta, \bx) \rangle}{32 \pi G a^2(\eta)} + \frac{\langle \gamma^{' L}_{ij}(\eta, \bx) \, \gamma^{ij ', \, L}(\eta, \bx) \rangle}{32 \pi G a^2(\eta)} \, ,
\ea
and $\rho_{\rm cr}$ is the critical energy density. Typically, the quantity in Eq. \eqref{def:Omega_I} is expressed in terms of the frequency domain, with a given frequency $f$ linked to the wave-number $k$ through $k = 2 \pi f a$, with $a$ being the scale factor. 

Notice the superscript "$I$" in the definition of Eq.  \eqref{def:Omega_I}. This is used to refer to the $I$ modes of SGWB, i.e. unpolarized backgrounds obtained by summing over the two independent $R$- and $L$-handed polarizations. We can also define the circular polarization of SGWB, the $V$ modes, as 
\begin{align} \label{def:Omega_V}
\OmegagwV(k, \eta) = \frac{1}{\rho_{\rm{cr}}}\frac{d \rho^V_{\rm GW}}{d \ln k} \, ,
\end{align}
where the SGWB $V$-mode energy density is given by
\ba
\rho^V_{\rm GW}  = \frac{\langle \gamma^{' R}_{ij}(\eta, \bx) \, \gamma^{ij ', \, R}(\eta, \bx) \rangle}{32 \pi G a^2(\eta)} - \frac{\langle \gamma^{' L}_{ij}(\eta, \bx) \, \gamma^{ij ', \, L}(\eta, \bx) \rangle}{32 \pi G a^2(\eta)} \, ,
\ea
which measures the asymmetry in the amount of $R$- and $L$-handed gravitational waves. $\OmegagwV(k, \eta)$ is related to $\Omegagw(k, \eta)$ through
\ba
\OmegagwV(k, \eta) = \chi(k) \, \Omegagw(k, \eta) \, ,
\ea
where $\chi$ is defined as in Eq. \eqref{eq:chi}. This last quantity is sensitive to violation of the parity symmetry in the gravitational sector.

The quantities defined so far are homogeneous, i.e. they are defined globally for a given wave-number $k$ and do not possess any spatial modulation. However, there are physical mechanisms capable to make them not-uniform across the sky. Allowing for anisotropies we can decompose the $I$ and $V$ modes polarizations in the homogeneous and inhomogeneous components as 
\begin{align}  \label{eq:OmegaI_dec}
  \Omegagw(k,\hat{n}) =  \Big[\overline{\Omega}^I_{\text{\tiny GW}}(k)+ \tilde{\Omega}^I_{\text{\tiny GW}}(k,\hat{n})\Big] \,,
\end{align}
and
\begin{align} \label{eq:OmegaV_dec}
  \OmegagwV(k,\hat{n}) = \Big[\overline{\Omega}^V_{\text{\tiny GW}}(k)+ \tilde{\Omega}^V_{\text{\tiny GW}}(k,\hat{n})\Big] \,.
\end{align}
We can normalize the inhomogeneous components over the unpolarized homogeneous component $\overline{\Omega}^I_{\text{\tiny GW}}(k)$  as 
\begin{align} 
  \delta^I_{\rm GW}(k,\hat{n}) =  \frac{\tilde{\Omega}^I_{\text{\tiny GW}}(k,\hat{n})}{\overline{\Omega}^I_{\text{\tiny GW}}(k)} \, , \qquad \qquad \delta^V_{\rm GW}(k,\hat{n}) =  \frac{\tilde{\Omega}^V_{\text{\tiny GW}}(k,\hat{n})}{\overline{\Omega}^I_{\text{\tiny GW}}(k)} \, .
\end{align}
In terms of these normalized anisotropies we can rewrite Eqs. \eqref{eq:OmegaI_dec} and \eqref{eq:OmegaV_dec} as 
\begin{align} \label{eq:OmegaI_anis}
  \Omegagw(k,\hat{n}) =  \overline{\Omega}^I_{\text{\tiny GW}}(k)\Big[1+ \delta^I_{\rm GW}(k,\hat{n})\Big] \,,
\end{align}
and 
\begin{align} \label{eq:OmegaV_anis}
  \OmegagwV(k,\hat{n}) =  \overline{\Omega}^I_{\text{\tiny GW}}(k)\Big[\chi(k) +  \delta^V_{\rm GW}(k,\hat{n})\Big] \,,
\end{align}
where $\chi(k)$ denotes the homogeneous component of the chirality parameter defined in Eq.~\eqref{eq:chi}.
In the following we will review the physical mechanisms that give rise to anisotropies in the CGWB from inflation and the state-of-art modeling of anisotropies in the AGWB.

\subsubsection{Inflationary CGWB}
\label{sec:CGWB}

For inflationary gravitational waves we can link the $\Omegagw(k, \eta_0)$ quantity evaluated today to the dimensionless primordial tensor power spectrum from inflation via \cite{Watanabe:2006qe}
\begin{align} \label{eq:OmegaI}
\Omegagw(k, \eta_0, \hat{n})  = \frac{\left[\mathcal{T}'(k,\eta_0)\right]^2}{12 a_0^2 H_0^2}  \, \Big[\mathcal A^R_{t}(k, \vx_c) + \mathcal A^L_{t}(k, \vx_c) \Big]  = \frac{\left[\mathcal{T}'(k,\eta_0)\right]^2}{12 a_0^2 H_0^2} \, \mathcal A_{t}(k, \vx_c) \, ,
\end{align}
where $\hat n = \vx_c/d$ with $d=\eta_0-\eta_{\rm entry}$ being the conformal time elapsed from horizon re-entry of the mode $k$ to the present, and $\mathcal{T}'(k,\eta_0)$ is the tensor transfer function evaluated today which describes the evolution of tensor perturbations after their horizon re-entry\footnote{In this time evolution we are not considering effects yielding to the formation of late-time inhomogeneities, like the induced anisotropies discussed later on at the end of Sec. \ref{sec:CGWB}.}. For very short scales that have re-entered the horizon early on ($k \eta_0 \gg 1$), after performing an oscillation-averaging procedure, the transfer function reads \cite{Caprini:2018mtu}
\begin{align} 
\left[\mathcal{T}'(k,\eta_0)\right]^2 = 
\begin{cases}
\eta_*/(2 \eta^4_0) \qquad   k > k_* \, ,\\
9/(2 \eta^4_0 k^2) \qquad  k < k_* \, ,
\end{cases}    
\end{align}
where $\eta_*$ is the conformal time at which the radiation and matter solutions for the scale factor of the Universe cross. The scale $k_*$ is the wave-number entering the horizon at $\eta_*$, i.e. $k_* \equiv 1/\eta_*$. In Ref. \cite{Caprini:2018mtu} a very compact approximation of $\Omegagw(k, \eta_0)$ was derived in terms of the radiation energy density $\Omega_r \simeq 10^{-4}$ and $k_{\rm eq} \simeq 10^{-2}$, the comoving wave-number entering the horizon at the time of matter-radiation equality $\eta_{\rm eq}$. This approximation reads 
\ba \label{eq:OmegaI_prim}
\Omegagw(k, \eta_0, \hat{n}) = \frac{3}{128} \Omega_r \left[ \frac{1}{2} \left(\frac{k_{\rm eq}}{k}\right)^2 + \frac{16}{9} \right] \, \Big[\mathcal A^R_{t}(k, \vx_c) + \mathcal A^L_{t}(k, \vx_c) \Big] \, .
\ea
This was shown to work well for the very short modes that re-enter the horizon deep inside the radiation dominated epoch, which is true for all the modes we can observe at interferometer scales. Similarly, the circular polarization of primordial gravitational waves can be well-approximated as
\begin{align} \label{eq:OmegaV_prim}
\OmegagwV(k, \eta_0, \hat{n})  =  \frac{3}{128} \Omega_r \left[ \frac{1}{2} \left(\frac{k_{\rm eq}}{k}\right)^2 + \frac{16}{9} \right] \, \Big[\mathcal A^R_{t}(k, \vx_c) - \mathcal A^L_{t}(k, \vx_c) \Big] \, .
\end{align}
The unpolarized homogeneous component $\overline{\Omega}^I_{\text{\tiny GW}}(k)$ is linearly related to the homogeneous component of tensor power spectrum in Eq.~\eqref{eq:power_inflation} as 
\begin{align}
\overline{\Omega}^I_{\text{\tiny GW}}(k,\eta_0) = \frac{3}{128} \Omega_r \left[ \frac{1}{2} \left(\frac{k_{\rm eq}}{k}\right)^2 + \frac{16}{9} \right] \,  \mathcal A_{t}(k) \,.
\end{align}
From the fundamental definitions of Eqs. \eqref{eq:OmegaI_prim} and \eqref{eq:OmegaV_prim} physics generating inhomogeneities in the primordial tensor power spectra, will inevitably cause anisotropies in the CGWB. Here, we are interested in mechanisms able to provide anisotropies despite assuming rotational invariance during inflation. These are for instance the squeezed modulations considered in Sec. \ref{sec:squeezed}, which will generate \textit{intrinsic} anisotropies. However, primordial gravitational waves on short scales propagate over very large distance before reaching our detectors. Propagation through the perturbed universe leads to some level of anisotropies as well, referred to as \textit{induced} anisotropies. In the following we will review these independent sources of anisotropy.

\subsubsection*{Intrinsic anisotropies from squeeezed bispectra}

These anisotropies arise from squeezed primordial bispectra of the form $\langle \gamma^\lambda_{\vk_1}\gamma^\lambda_{\vk_2} \, \Psi_\vq \rangle|_{\vq\to 0}$, where $\Psi_\vq$ denotes a long wavelength mode of either a scalar or tensor perturbation while $\gamma^\lambda_{\vk_{1,2}}$ are the short wavelength tensor modes of polarization $\lambda$. In Sec. \ref{sec:squeezed} we have seen that when $\Psi_\vq = \zeta_\vq$, we get the following squeezed modulation of tensor power spectra
\begin{align} \label{eq:modPz}
\mathcal A^{\lambda, \, \rm tot}_{t}(\vk,\vx_c) = \mathcal A^\lambda_{t}(k)\left[1+\int \frac{d^3q}{(2\pi)^3} \, e^{i\vq\cdot\vx_c} \,  F^{\lambda \lambda, \rm tts}_{\rm NL}(\vk,\vq) \, \zeta_{\vq} \right] \, .
\end{align}
 By inserting this equation into Eqs. \eqref{eq:OmegaI_prim} and \eqref{eq:OmegaV_prim}, and exploiting the definitions in Eqs. \eqref{eq:OmegaI_anis} and \eqref{eq:OmegaV_anis}, we get the following anisotropic components of the GW $I$ and $V$ modes in terms of the R- and L-handed primordial tensor power spectra
 \begin{align} \label{eq_deltaI_tts}
\deltagw_{\rm tts}(k, \hat{n}) = \frac{1}{\mathcal A_{t}(k)}\int \frac{d^3q}{(2\pi)^3} \, e^{i d \, \vq \cdot \hat{n}} \left[\mathcal A^R_{t}(k) \, F_{\rm NL}^{RR,\rm tts}(\vk,\vq) + \mathcal A^L_{t}(k) \, F_{\rm NL}^{LL,\rm tts}(\vk,\vq) \right] \zeta_{\vq} \,,
\end{align}
and
 \begin{align} \label{eq_deltaV_tts}
\deltagwV_{\rm tts}(k, \hat{n}) = \frac{1}{\mathcal A_{t}(k)}\int \frac{d^3q}{(2\pi)^3} \, e^{i d \, \vq \cdot \hat{n}} \left[\mathcal A^R_{t}(k) \, F_{\rm NL}^{RR,\rm tts}(\vk,\vq) - \mathcal A^L_{t}(k) \, F_{\rm NL}^{LL,\rm tts}(\vk,\vq) \right] \zeta_{\vq} \,,
\end{align}
 where we evaluated the local modulation provided by Eq.  \eqref{eq:modPz} at the position $\vx_c$ where the CGWB of wave-number $k$ starts propagating after the horizon re-entry. Therefore, by denoting $\hat n = - \vk/k$ a certain propagation direction in the sky, $\vx_c = d \, \hat n$, with $d=\eta_0-\eta_{\rm entry}$. Physically, Eqs. \eqref{eq_deltaI_tts} and \eqref{eq_deltaV_tts} provide anisotropies already present at the CGWB formation. 

For the squeezed modulation sourced by a tensor perturbation in Eq. \eqref{eq:final_mod_tensor_main}, we obtain the following anisotropies generated by a long wavelength tensor mode 
\begin{align}\label{eq:deltagw_ttt}
\deltagw_{\rm ttt}(k,\hat{n}) = \frac{1}{\mathcal A_{t}(k)}\int\frac{d^3q}{(2\pi)^3}\,e^{id \vq \cdot \hat{n}} \sum_{\lambda' = R/L} \left[\mathcal A^R_{t}(k) \, F_{\rm NL}^{RR\lambda',\rm ttt}(\vk,\vq) + \mathcal A^L_{t}(k) \, F_{\rm NL}^{LL\lambda',\rm ttt}(\vk,\vq) \right] \gamma^{\lambda'}_{\vq} \,,
\end{align}
 and
\begin{align}\label{eq:deltagwV_ttt}
\deltagwV_{\rm ttt}(k,\hat{n}) = \frac{1}{\mathcal A_{t}(k)} \int\frac{d^3q}{(2\pi)^3}\,e^{id \vq \cdot \hat{n}} \sum_{\lambda' = R/L} \left[\mathcal A^R_{t}(k) \, F_{\rm NL}^{RR\lambda',\rm ttt}(\vk,\vq) - \mathcal A^L_{t}(k) \, F_{\rm NL}^{LL\lambda',\rm ttt}(\vk,\vq) \right] \gamma^{\lambda'}_{\vq} \,.
\end{align}
The anisotropies just introduced can be expanded in spherical harmonic coefficients in the same way as CMB anisotropies, i.e. we define the wave-number dependent coefficients as
\be \label{eq:def_sphe_I}
\delta^{{\rm GW}, \, I}_{\ell m}(k) =  \int d^2 \hat n \, Y^*_{\ell m}(\hat n) \, \deltagw(k,\hat n) \, ,
\ee
and
\be
\delta^{{\rm GW}, \, V}_{\ell m}(k) =  \int d^2 \hat n \, Y^*_{\ell m}(\hat n) \, \deltagwV(k,\hat n) \, .
\ee
The only difference with CMB anisotropies is that GW anisotropies carry information on the wave-number (frequency) $k$.
In order to perform the angular integrations inside the spherical harmonics coefficients definition we need to specify a given angular dependence in the quantities  $F_{\rm NL}^{\lambda_1...\lambda_n, \, \rm x_1x_2x_3}(\vk,\vq)$. In the following, we provide the spherical harmonic coefficients obtained by considering angular dependencies that typically appear in the literature, referring to the original Refs. \cite{Malhotra:2020ket, Dimastrogiovanni:2021mfs} and App. \ref{app:Wigner} for computational details.

\subsubsection*{Monopolar $\langle \gamma \gamma \zeta \rangle$}

Let us begin with the case where the parameter $\fnls$ defined in Eq.~\eqref{eq:defNtts_main} has a monopolar dependence in terms of $\hat q\cdot \hat k$, i.e.\footnote{Here and in the following we will re-normalize the $F_{\rm NL}^{\lambda_1...\lambda_n, \, \rm x_1x_2x_3}(\vk,\vq)$ coefficients in order to be coherent with the normalization chosen in literature \cite{Dimastrogiovanni:2021mfs}. This allows for better comparison of these results with previous works.}
\begin{align}
    F_{\rm NL}^{\lambda\lambda,\rm tts}(\vk,\vq) = \, \sqrt{4 \pi} \, Y_{0 0}(\hat k \cdot \hat q)  \, f_{\rm NL}^{\lambda\lambda, \rm tts}(k,q) = f_{\rm NL}^{\lambda\lambda, \rm tts}(k,q) \,.
    \label{eq:tts_mono}
\end{align}
In this case, the spherical harmonic coefficients of GW anisotropies become
\begin{align}\label{eq:deltagw_tts_mono}
\delta_{\ell m}^{\rm GW, I, \, tts}(k)  = \frac{4 \pi}{\mathcal A_t(k)} \, i^\ell \, \int \frac{d^3q}{(2\pi)^3}\, Y^*_{\ell m}(\hat q) \, j_\ell(q d) \,\left[ \mathcal A^R_t(k) \, f_{\rm NL}^{RR, \rm tts}(k,q) + \mathcal A^L_t(k) \, f_{\rm NL}^{LL, \rm tts}(k,q) \right]\, \zeta_{\vq}  \, ,
\end{align}
and 
\begin{align}\label{eq:deltagw_tts_mono_V}
\delta_{\ell m}^{\rm GW, V, \, tts}(k)  = \frac{4 \pi}{\mathcal A_t(k)} \, i^\ell \, \int \frac{d^3q}{(2\pi)^3}\, Y^*_{\ell m}(\hat q) \, j_\ell(q d) \,\left[ \mathcal A^R_t(k) \, f_{\rm NL}^{RR, \rm tts}(k,q) - \mathcal A^L_t(k) \, f_{\rm NL}^{LL, \rm tts}(k,q) \right]\, \zeta_{\vq} \, .
\end{align}

\subsubsection*{Quadrupolar $\langle \gamma \gamma \zeta \rangle$}
\noindent Next, we consider the case where $\fnls$ has a quadrupolar angular dependence in $\hat q\cdot \hat k$, therefore the case $J = 2$ of Eq. \eqref{eq:defNtts_main}. We have
\begin{align}
    F_{\rm NL}^{\lambda\lambda,\rm tts}(\vk,\vq) = \sqrt{\frac{4 \pi}{ 5}} \, Y_{2 0}(\hat k \cdot \hat q) \, f_{\rm NL}^{\lambda\lambda, \rm tts}(k,q)  \,.
    \label{eq:tts_quad}
\end{align}
In this case, the spherical harmonic coefficients of GW anisotropies read 
\begin{align}\label{eq:deltagw_tts_quad}
\delta_{\ell m}^{\rm GW, I, \, tts}(k)  = &\frac{1}{\mathcal A_t(k)} \frac{16 \pi^2}{5} \sqrt{\frac{5}{4 \pi}} \, \sum_{J, M, m'} \, i^J \, \sqrt{(2 J + 1)(2 \ell + 1)} \, \, \begin{pmatrix}
	\ell & J & 2 \\
	0 & 0 & 0
	\end{pmatrix}\begin{pmatrix}
	\ell & J & 2 \\
	m & M & m'
	\end{pmatrix} \,  \times \nonumber \\
&\times \int \frac{d^3q}{(2\pi)^3}\, Y_{J M}(\hat q) \, Y_{2 m'}(\hat q) \, j_J(q d) \, \left[ \mathcal A^R_t(k) \, f_{\rm NL}^{RR, \rm tts}(k,q) + \mathcal A^L_t(k) \, f_{\rm NL}^{LL, \rm tts}(k,q) \right] \, \zeta_{\vq}  \, ,
\end{align}
and 
\begin{align}\label{eq:deltagwV_tts_quad}
\delta_{\ell m}^{\rm GW, V, \, tts}(k)  = &\frac{1}{\mathcal A_t(k)} \frac{16 \pi^2}{5} \sqrt{\frac{5}{4 \pi}} \, \sum_{J, M, m'} \, i^J \, \sqrt{(2 J + 1)(2 \ell + 1)} \, \, \begin{pmatrix}
	\ell & J & 2 \\
	0 & 0 & 0
	\end{pmatrix}\begin{pmatrix}
	\ell & J & 2 \\
	m & M & m'
	\end{pmatrix} \,  \times \nonumber \\
&\times \int \frac{d^3q}{(2\pi)^3}\, Y_{J M}(\hat q) \, Y_{2 m'}(\hat q) \, j_J(q d) \, \left[ \mathcal A^R_t(k) \, f_{\rm NL}^{RR, \rm tts}(k,q) - \mathcal A^L_t(k) \, f_{\rm NL}^{LL, \rm tts}(k,q) \right] \, \zeta_{\vq} \, .
\end{align}
Here
\begin{equation}
\begin{pmatrix}
	\ell_1 & \ell_2 & \ell_3 \\
	m_1 & m_2 & - m_3
	\end{pmatrix} \, ,
\end{equation}
denotes the Wigner 3-j symbol as defined in Eq. \eqref{eq:3jsymbols-CG}.  

\subsubsection*{Quadrupolar $\langle \gamma \gamma \gamma \rangle$}

Here we consider the case where $\fnlt$ has a quadrupolar dependence in $\hat q\cdot \hat k$. This corresponds to the case $J = 2$ of Eq. \eqref{eq:defNttt_main}.  We can parametrise this as 
\begin{align}
F_{\rm NL}^{\lambda\lambda \lambda',\rm ttt}(\vk,\vq) = - \frac{2}{3} \, \sqrt{\frac{6 \pi}{5}} \, {}_{\pm 2}Y_{20}(\hat k \cdot \hat q)  \, f_{\rm NL}^{\lambda\lambda \lambda', \rm ttt}(k,q) = - \epsilon_{ij}^{\lambda'}(q)\,k^i k^j \, f_{\rm NL}^{\lambda\lambda \lambda', \rm ttt}(k,q)   \, .
\end{align}
The spherical harmonic coefficients become
\begin{align}\label{eq:deltagw_ttt_quad}
    \delta^{\rm GW, I, ttt}_{\ell m}(k)  = \frac{2 \pi}{\mathcal A_t(k)} \,  i^\ell \, &\sqrt{\frac{(\ell+2)!}{(\ell-2)!}} \, \int  \frac{d^3q}{(2\pi)^3} \, \frac{j_\ell(q d)}{(qd)^2} \times \nonumber \\
   & \times \left\{ {}_{- 2}Y_{\ell m}^*(\hat q) \,
\left[\mathcal A^R_{t}(k) \, f_{\rm NL}^{RRR,\rm ttt}(k,q) + \mathcal A^L_{t}(k) \, f_{\rm NL}^{LLR,\rm ttt}(k,q) \right]  \gamma_{\vq}^{R} + \right. \nonumber  \\
&\left. \qquad + {}_{+2}Y_{\ell m}^*(\hat q) \, \left[\mathcal A^L_{t}(k) \, f_{\rm NL}^{LLL,\rm ttt}(k,q) + \mathcal A^R_{t}(k) \, f_{\rm NL}^{RRL,\rm ttt}(k, q) \right] \gamma_{\vq}^{L} \right\}\,,
\end{align}
and
\begin{align}\label{eq:deltagwV_ttt_quad}
    \delta^{\rm GW, V, ttt}_{\ell m}(k)  = \frac{2 \pi}{\mathcal A_t(k)} \,  i^\ell \, &\sqrt{\frac{(\ell+2)!}{(\ell-2)!}} \, \int  \frac{d^3q}{(2\pi)^3} \, \frac{j_\ell(q d)}{(qd)^2} \times \nonumber \\
   & \times \left\{ {}_{- 2}Y_{\ell m}^*(\hat q) \,
\left[\mathcal A^R_{t}(k) \, f_{\rm NL}^{RRR,\rm ttt}(k,q) - \mathcal A^L_{t}(k) \, f_{\rm NL}^{LLR,\rm ttt}(k,q) \right]  \gamma_{\vq}^{R} - \right. \nonumber  \\
&\left. \qquad - {}_{+2}Y_{\ell m}^*(\hat q) \, \left[\mathcal A^L_{t}(k) \, f_{\rm NL}^{LLL,\rm ttt}(k,q) - \mathcal A^R_{t}(k) \, f_{\rm NL}^{RRL,\rm ttt}(k,q) \right] \gamma_{\vq}^{L} \right\}\,.
\end{align}
As a final remark, notice that our spherical harmonic coefficients have been written in terms of angular-independent quantities $f_{\rm NL}^{\lambda_1...\lambda_n, \, \rm x_1x_2x_3}(k,q)$, which can be considered as coefficents that label the strength of NGs. These quantities give information about the strength of the cross-correlation between the large and small scales $q$ and $k$ in a given inflationary scenario.

\subsubsection*{Induced anisotropies}

These anisotropies arise from the propagation of GW through a background made inhomogeneous by scalar and tensor perturbations of the metric tensor. Their formal and computational description is very similar to the formation of CMB temperature and polarization anisotropies as a consequence of the SW and ISW effects. The detailed study of the associated Boltzmann equations and the prediction for the amount of induced anisotropies today for $I$ modes has been derived in Refs. \cite{Contaldi:2016koz,Bartolo:2019oiq,Bartolo:2019yeu,ValbusaDallArmi:2020ifo}. Here we are interested in the large-scale limit of the result for the anisotropies. In this limit, the dominant contribution is given by the SW term induced by scalar perturbations, which reads as 
\begin{align}
    \deltagw_{\rm ind}(k, \hat{n}) \simeq \left[4-\frac{\partial \ln{\overline{\Omega}_{\rm GW}(k)}}{\partial \ln{k}}\right] \frac{2}{3} \, \int \frac{d^3q}{(2\pi)^3} \, e^{id \, \vq \cdot \hat{n}} \, \zeta_{\vq} \, .
    \label{eq:deltagw_ind}
\end{align}
There is also an analogous contribution due to tensor perturbations which is however negligible with respect to the effect of scalar perturbations. 

What about induced anisotropies in $V$ modes? As scalar perturbations at linear level cannot perceive parity violation (see e.g. \cite{Shiraishi:2016mok}), then they do not induce $V$ modes. In principle, we can have a non-zero effect from tensor perturbations assuming a primordial mechanism able to provide some asymmetry between R- and L-handed tensor power spectra, resulting in $\chi \neq 0$. In absence of such a mechanism, the $V$-mode induced anisotropies are expected to be vanishing 
\begin{align}
    \deltagwV_{\rm ind}(k, \hat{n}) = 0\, .
    \label{eq:deltagw_ind_V}
\end{align}
In this study we will assume $\chi = 0$, and Eq. \eqref{eq:deltagw_ind_V} will hold. Needless to say, studying the effect of tensor-induced $V$ modes from models of inflation with significant parity violation in tensor power spectra can be considered in a separate analysis.

In a way similar to the spherical harmonic coefficients of monopolar $\langle \gamma \gamma \zeta \rangle$, we can compute the spherical harmonics coefficients of Eqs. \eqref{eq:deltagw_ind} and \eqref{eq:deltagw_ind_V}. We find
\be
\label{eq:delta_GW_ind}
\deltagw_{{\rm ind}, \, \ell m}(k)  \simeq \frac{8 \pi}{3} \, i^\ell \, \left[4-\frac{\partial \ln{\overline{\Omega}_{\rm GW}(k)}}{\partial \ln{k}}\right] \, \int \frac{d^3q}{(2\pi)^3} \, Y^*_{\ell m}(\hat q)  \, j_\ell(q d) \, \zeta_{\vq} \, ,
\ee
and 
\be
\label{eq:delta_GW_ind_V}
\deltagwV_{{\rm ind}, \, \ell m}(k) = 0 \, .
\ee
These induced contributions are of primordial origin similar to the intrinsic contributions introduced above. These are linearly related to primordial perturbations from inflation, and contain information on the power spectrum statistics of the underlying inflationary model, specifically on the tensor tilt due to the $4-\partial\ln{\overline{\Omega}_{\rm GW}(k)}/\partial \ln{k}$ prefactor. Given this prefactor, these anisotropies will be relevant in an early-universe mechanism able to provide a sharp peak in the amplitude of primordial gravitational waves (see e.g. \cite{Bartolo:2019zvb,Dimastrogiovanni:2022eir}). As in the present work we will not introduce such a mechanism and we will focus on $V$-mode anisotropies, the induced anisotropies will be subdominant to the intrinsic anisotropies in the regime of sizeable non-Gaussianities.

\subsubsection{AGWB}

The modeling of the astrophysical background of gravitational waves is one of the challenges of the gravitational wave physics. As this background is formed by the superposition of un-resolved individual GW sources, such as merging binary systems, it will be strongly related to the distribution of the large scale structure of the Universe. In addition, this background of gravitational waves is expected to be anisotropic.  Many literature studies have proposed realistic modelization of the AGWB monopole and anisotropies in different frequency windows, see e.g. Refs. \cite{Regimbau:2011rp,Cusin:2017fwz,Cusin:2018ump,Cusin:2018rsq,Cusin:2019jhg,Cusin:2019jpv,Jenkins:2018kxc,Jenkins:2018uac,Jenkins:2019nks,Jenkins:2019uzp,Bertacca:2019fnt,Pitrou:2019rjz,Capurri:2021zli,DallArmi:2022wnq,Pizzuti:2022nnj}. Of specific interest is the recent code \cite{Bellomo:2021mer} that compute anisotropies of the AGWB.  

Most importantly the literature has pointed out that the amount of GW signal from the expected astrophysical background can make the detection of the primordial background of gravitational waves challenging. While methods to separate the monopoles of the cosmological background from the astrophysical one have been proposed (see e.g. \cite{Regimbau:2016ike,Pan:2019uyn,Mukherjee:2019oma,Sharma:2020btq,Pieroni:2020rob,Biscoveanu:2020gds,Martinovic:2020hru,Poletti:2021ytu,Boileau:2020rpg,Boileau:2021sni}), it has not convincingly shown to determine the nature of the detected anisotropies. Recently one direction has been investigated in Ref. \cite{Ricciardone:2021kel}, considering the cross-correlation of GW anisotropies with CMB $T$ modes as a way to distinguish between astrophysical and primordial background-anisotropies. 
On the other hand, according to Ref. \cite{Dimastrogiovanni:2021mfs}, a net detection of a GW-$T$ signal of primordial origin is still made difficult by its cosmic variance limited uncertainty in the regime where AGWB-anisotropies significantly dominate over CGWB-anisotropies. 

In this work we will mainly be focused on the information provided by GW $V$ modes. As we have seen above these are sourced by parity violation. While there are no useful constraints on parity violation of the CGWB from inflation due to the lack of detection of inflationary gravitational waves, parity violation mechanisms for AGWB are strongly constrained and disfavoured by the current observations of astrophysical GW-events (see e.g. \cite{Yunes:2013dva,Berti:2015itd,Kostelecky:2016kfm,Nishizawa:2018srh,ColemanMiller:2019tqn,Wang:2020cub}). Therefore, as we will assume, the amount of GW $V$ modes of astrophysical origin is expected to be negligible, and detecting a net parity violation signal in a SGWB is a powerful tool to determine the primordial origin of this background.

\section{Probing parity-odd bispectra with GW $V$ modes} \label{sec:parity-odd_inv}

In this section we first provide general expressions for auto GW-GW and cross GW-CMB correlations induced by the primordial mechanisms considered above and we will compare our results with the literature considering parity even bispectra. Next, we will study $V$-mode anisotropies generated by parity-odd squeezed bispectra. We will determine their detection prospects by considering a BBO-like experiment. 

\subsection{GW-GW and GW-CMB cross-correlations} \label{sec:correlators_self_cross}

We start defining a general angular cross-correlation between the quantities $X$ and $Y$ as
\begin{align} \label{eq:def_corr}
C_{\ell}^{X Y} &= \frac{1}{2 \ell+1} \, \sum_m \langle X_{\ell m} \, Y_{\ell m} \rangle \, ,
\end{align}
where $X = I, V$ corresponding to $I$- and $V$-mode anisotropies of the CGWB, and $Y = I, V, T, E, B$, where $T$, $E$ and $B$ refer to the corresponding CMB anisotropic modes. Using the results for the GW and CMB spherical harmonics coefficients derived in the previous section (for technical details see App. \ref{app:Wigner}), we get the following expressions we will provide, these separately for different angular dependence in $\hat k \cdot \hat q$.

\subsubsection*{Monopolar TTS}

\begin{align}\label{eq:spectrafirst}
C_{\ell}^{I T}(k)  =  G_{\rm cross}^{\rm tts}(k)  \int \, \frac{dq}{q}  \, j_\ell(q d) \, {\cal T}_{\ell(s)}^{T}(q) \, \mathcal A_s(q) \, \left[ \mathcal A^R_t(k) \, f_{\rm NL}^{R, \rm tts}(k,q) + \mathcal A^L_t(k) \, f_{\rm NL}^{L, \rm tts}(k,q)\right] \, ,
\end{align}
\begin{align}
C_{\ell}^{I E}(k)  =  G_{\rm cross}^{\rm tts}(k)  \int \,  \frac{dq}{q}  \, j_\ell(q d) \, {\cal T}_{\ell(s)}^{E}(q) \, \mathcal A_s(q) \, \left[ \mathcal A^R_t(k) \, f_{\rm NL}^{R, \rm tts}(k,q) + \mathcal A^L_t(k) \, f_{\rm NL}^{L, \rm tts}(k,q)\right] \, ,
\end{align}
\begin{align}
C_{\ell}^{V T}(k)  =  G_{\rm cross}^{\rm tts}(k)  \int \, \frac{dq}{q} \, j_\ell(q d) \, {\cal T}_{\ell(s)}^{T}(q) \, \mathcal A_s(q) \, \left[ \mathcal A^R_t(k) \, f_{\rm NL}^{R, \rm tts}(k,q) - \mathcal A^L_t(k) \, f_{\rm NL}^{L, \rm tts}(k,q)\right] \, ,
\end{align}
\begin{align}
C_{\ell}^{V E}(k)  =  G_{\rm cross}^{\rm tts}(k) \int \, \frac{dq}{q} \, j_\ell(q d) \, {\cal T}_{\ell(s)}^{E}(q) \, \mathcal A_s(q) \, \left[ \mathcal A^R_t(k) \, f_{\rm NL}^{R, \rm tts}(k,q) - \mathcal A^L_t(k) \, f_{\rm NL}^{L, \rm tts}(k,q)\right] \, ,
\end{align}
\begin{align}
C_{\ell}^{I B}(k)  =  0 \, ,
\end{align}
\begin{align}
C_{\ell}^{V B}(k)  =  0 \, ,
\end{align}
\begin{align}
C_{\ell}^{I I}(k)  =  G_{\rm auto}^{\rm tts}(k)  \int \, \frac{dq}{q} \, \left[j_\ell(q d)\right]^2 \, \mathcal A_s(q) \, \left[ \mathcal A^R_t(k) \, f_{\rm NL}^{R, \rm tts}(k,q) + \mathcal A^L_t(k) \, f_{\rm NL}^{L, \rm tts}(k,q)\right]^2 \, ,
\end{align}
\begin{align}
C_{\ell}^{V V}(k)  =  G_{\rm auto}^{\rm tts}(k)  \int \, \frac{dq}{q} \, \left[j_\ell(q d)\right]^2 \, \mathcal A_s(q) \, \left[ \mathcal A^R_t(k) \, f_{\rm NL}^{R, \rm tts}(k,q) - \mathcal A^L_t(k) \, f_{\rm NL}^{L, \rm tts}(k,q)\right]^2 \, ,
\end{align}
\begin{align}
C_{\ell}^{I V}(k)  =  G_{\rm auto}^{\rm tts}(k)  \int \, \frac{dq}{q} \, \left[j_\ell(q d)\right]^2 \, \mathcal A_s(q) \, \left[ \left(\mathcal A^R_t(k) \, f_{\rm NL}^{R, \rm tts}(k,q) \right)^2 - \left(\mathcal A^L_t(k) \, f_{\rm NL}^{L, \rm tts}(k,q)\right)^2\right] \, .
\end{align}

\subsubsection*{Quadrupolar TTS}

\begin{align}
C_{\ell}^{I T}(k)  =  G_{\rm cross}^{\rm tts}(k) & \, \sum_J \, i^{J - \ell} \, (2 J +1) \, \left[\begin{pmatrix}
	\ell & J & 2 \\
	0 & 0 & 0
	\end{pmatrix}\right]^2 \times \nonumber \\
	& \times \int \, \frac{dq}{q} \, j_J(q d) \, {\cal T}_{\ell(s)}^{T}(q) \, \mathcal A_s(q) \, \left[ \mathcal A^R_t(k) \, f_{\rm NL}^{R, \rm tts}(k,q) + \mathcal A^L_t(k) \, f_{\rm NL}^{L, \rm tts}(k,q)\right] \, ,
\end{align}
\begin{align}
C_{\ell}^{I E}(k)  =  G_{\rm cross}^{\rm tts}(k) & \, \sum_J \, i^{J - \ell} \, (2 J +1) \, \left[\begin{pmatrix}
	\ell & J & 2 \\
	0 & 0 & 0
	\end{pmatrix}\right]^2 \times \nonumber \\
	& \times \int \, \frac{dq}{q} \, j_J(q d) \, {\cal T}_{\ell(s)}^{E}(q) \, \mathcal A_s(q) \, \left[ \mathcal A^R_t(k) \, f_{\rm NL}^{R, \rm tts}(k,q) + \mathcal A^L_t(k) \, f_{\rm NL}^{L, \rm tts}(k,q)\right] \, ,
\end{align}
\begin{align}
C_{\ell}^{V T}(k)  = G_{\rm cross}^{\rm tts}(k) & \, \sum_J \, i^{J - \ell} \, (2 J +1) \, \left[\begin{pmatrix}
	\ell & J & 2 \\
	0 & 0 & 0
	\end{pmatrix}\right]^2 \times \nonumber \\
	& \times \int \, \frac{dq}{q} \, j_J(q d) \, {\cal T}_{\ell(s)}^{T}(q) \, \mathcal A_s(q) \, \left[ \mathcal A^R_t(k) \, f_{\rm NL}^{R, \rm tts}(k,q) - \mathcal A^L_t(k) \, f_{\rm NL}^{L, \rm tts}(k,q)\right] \, ,
\end{align}
\begin{align}
C_{\ell}^{V E}(k)  =  G_{\rm cross}^{\rm tts}(k) & \, \sum_J \, i^{J - \ell} \, (2 J +1) \, \left[\begin{pmatrix}
	\ell & J & 2 \\
	0 & 0 & 0
	\end{pmatrix}\right]^2 \times \nonumber \\
	& \times \int \, \frac{dq}{q} \, j_J(q d) \, {\cal T}_{\ell(s)}^{E}(q) \, \mathcal A_s(q) \, \left[ \mathcal A^R_t(k) \, f_{\rm NL}^{R, \rm tts}(k,q) - \mathcal A^L_t(k) \, f_{\rm NL}^{L, \rm tts}(k,q)\right] \, ,
\end{align}
\begin{align}
C_{\ell}^{I B}(k)  =  0 \, ,
\end{align}
\begin{align}
C_{\ell}^{V B}(k)  =  0 \, ,
\end{align}
\begin{align}
C_{\ell}^{I I}(k)  =  G_{\rm auto}^{\rm tts}(k) & \, \sum_{J, J'} \, i^{J - J'} \, (2 J +1) \, (2 J' +1) \, \left[\begin{pmatrix}
	\ell & J & 2 \\
	0 & 0 & 0
	\end{pmatrix}\right]^2 \, \left[\begin{pmatrix}
	\ell & J' & 2 \\
	0 & 0 & 0
	\end{pmatrix}\right]^2 \times  \nonumber \\
	& \times \int \, \frac{dq}{q} \, j_{J}(q d) \, j_{J'}(q d) \, \mathcal A_s(q) \, \left[ \mathcal A^R_t(k) \, f_{\rm NL}^{R, \rm tts}(k,q) + \mathcal A^L_t(k) \, f_{\rm NL}^{L, \rm tts}(k,q)\right]^2 \, ,
\end{align}
\begin{align}
C_{\ell}^{V V}(k)  =  G_{\rm auto}^{\rm tts}(k) & \, \sum_{J, J'} \, i^{J - J'} \, (2 J +1) \, (2 J' +1) \, \left[\begin{pmatrix}
	\ell & J & 2 \\
	0 & 0 & 0
	\end{pmatrix}\right]^2 \, \left[\begin{pmatrix}
	\ell & J' & 2 \\
	0 & 0 & 0
	\end{pmatrix}\right]^2 \times \nonumber \\
	& \times \int \, \frac{dq}{q} \, j_{J}(q d) \, j_{J'}(q d) \, \mathcal A_s(q) \, \left[ \mathcal A^R_t(k) \, f_{\rm NL}^{R, \rm tts}(k,q) - \mathcal A^L_t(k) \, f_{\rm NL}^{L, \rm tts}(k,q)\right]^2 \, ,
\end{align}
\begin{align}
C_{\ell}^{I V}(k)  =  G_{\rm auto}^{\rm tts}(k) & \, \sum_{J, J'} \, i^{J - J'} \, (2 J +1) \, (2 J' +1) \, \left[\begin{pmatrix}
	\ell & J & 2 \\
	0 & 0 & 0
	\end{pmatrix}\right]^2 \, \left[\begin{pmatrix}
	\ell & J' & 2 \\
	0 & 0 & 0
	\end{pmatrix}\right]^2 \times \nonumber \\
	& \times \int \, \frac{dq}{q} \, j_{J}(q d) \, j_{J'}(q d) \, \mathcal A_s(q) \, \left[ \left(\mathcal A^R_t(k) \, f_{\rm NL}^{R, \rm tts}(k,q) \right)^2 - \left(\mathcal A^L_t(k) \, f_{\rm NL}^{L, \rm tts}(k,q)\right)^2\right] \, .
\end{align}

\subsubsection*{Quadrupolar TTT}

\begin{align}
C_{\ell}^{I T}(k)  =  G_{\ell, \, \rm cross}^{\rm ttt}(k) \int \, \frac{dq}{q} \, & \frac{j_\ell(q d)}{(q d)^2} \, {\cal T}_{\ell(t)}^{T}(q) \Big\{ \mathcal A^R_t(q) \left[ \mathcal A^R_t(k) \, f_{\rm NL}^{RR, \rm ttt}(k,q) + \mathcal A^L_t(k) \, f_{\rm NL}^{LR, \rm ttt}(k,q)\right] +  \nonumber \\
& \qquad\qquad  + \mathcal A^L_t(q) \left[ \mathcal A^L_t(k) \, f_{\rm NL}^{LL, \rm ttt}(k,q) + \mathcal A^R_t(k) \, f_{\rm NL}^{RL, \rm ttt}(k,q)\right] \Big\} \, ,
\end{align}
\begin{align}
C_{\ell}^{I E}(k)  =  G_{\ell, \, \rm cross}^{\rm ttt}(k) \int \, \frac{dq}{q} \, & \frac{j_\ell(q d)}{(q d)^2} \, {\cal T}_{\ell(t)}^{E}(q) \Big\{ \mathcal A^R_t(q) \left[ \mathcal A^R_t(k) \, f_{\rm NL}^{RR, \rm ttt}(k,q) + \mathcal A^L_t(k) \, f_{\rm NL}^{LR, \rm ttt}(k,q)\right] + \nonumber \\
&\qquad\qquad + \mathcal A^L_t(q)\left[ \mathcal A^L_t(k) \, f_{\rm NL}^{LL, \rm ttt}(k,q) + \mathcal A^R_t(k) \, f_{\rm NL}^{RL, \rm ttt}(k,q)\right] \Big\} \, ,
\end{align}
\begin{align}
C_{\ell}^{V T}(k)  =  G_{\ell, \, \rm cross}^{\rm ttt}(k) \int \, \frac{dq}{q} \, & \frac{j_\ell(q d)}{(q d)^2} \, {\cal T}_{\ell(t)}^{T}(q) \Big\{ \mathcal A^R_t(q) \left[ \mathcal A^R_t(k) \, f_{\rm NL}^{RR, \rm ttt}(k,q) - \mathcal A^L_t(k) \, f_{\rm NL}^{LR, \rm ttt}(k,q)\right] - \nonumber \\
&\qquad\qquad - \mathcal A^L_t(q) \left[ \mathcal A^L_t(k) \, f_{\rm NL}^{LL, \rm ttt}(k,q) - \mathcal A^R_t(k) \, f_{\rm NL}^{RL, \rm ttt}(k,q)\right] \Big\} \, ,
\end{align}
\begin{align}
C_{\ell}^{V E}(k)  =  G_{\ell, \, \rm cross}^{\rm ttt}(k) \int \, \frac{dq}{q} \, & \frac{j_\ell(q d)}{(q d)^2} \, {\cal T}_{\ell(t)}^{E}(q) \Big\{ \mathcal A^R_t(q) \left[ \mathcal A^R_t(k) \, f_{\rm NL}^{RR, \rm ttt}(k,q) - \mathcal A^L_t(k) \, f_{\rm NL}^{LR, \rm ttt}(k,q)\right] - \nonumber \\
&\qquad\qquad - \mathcal A^L_t(q) \left[ \mathcal A^L_t(k) \, f_{\rm NL}^{LL, \rm ttt}(k,q) - \mathcal A^R_t(k) \, f_{\rm NL}^{RL, \rm ttt}(k,q)\right] \Big\} \, ,
\end{align}
\begin{align}
C_{\ell}^{I B}(k)  =  G_{\ell, \, \rm cross}^{\rm ttt}(k) \int \, \frac{dq}{q} \, & \frac{j_\ell(q d)}{(q d)^2} \, {\cal T}_{\ell(t)}^{B}(q) \Big\{ \mathcal A^R_t(q) \left[ \mathcal A^R_t(k) \, f_{\rm NL}^{RR, \rm ttt}(k,q) + \mathcal A^L_t(k) \, f_{\rm NL}^{LR, \rm ttt}(k,q)\right] - \nonumber \\
&\qquad\qquad - \mathcal A^L_t(q) \left[ \mathcal A^L_t(k) \, f_{\rm NL}^{LL, \rm ttt}(k,q) + \mathcal A^R_t(k) \, f_{\rm NL}^{RL, \rm ttt}(k,q)\right] \Big\} \, ,
\end{align} 
\begin{align}
C_{\ell}^{V B}(k)  =  G_{\ell, \, \rm cross}^{\rm ttt}(k) \int \, \frac{dq}{q} \, & \frac{j_\ell(q d)}{(q d)^2} \, {\cal T}_{\ell(t)}^{B}(q) \Big\{ \mathcal A^R_t(q) \left[ \mathcal A^R_t(k) \, f_{\rm NL}^{RR, \rm ttt}(k,q) - \mathcal A^L_t(k) \, f_{\rm NL}^{LR, \rm ttt}(k,q)\right] + \nonumber \\
&\qquad\qquad + \mathcal A^L_t(q) \left[ \mathcal A^L_t(k) \, f_{\rm NL}^{LL, \rm ttt}(k,q) - \mathcal A^R_t(k) \, f_{\rm NL}^{RL, \rm ttt}(k,q)\right] \Big\} \, ,
\end{align}
\begin{align}
C_{\ell}^{I I}(k)  =  G_{\ell, \, \rm auto}^{\rm ttt}(k) \int \, \frac{dq}{q} \, & \left[\frac{j_\ell(q d)}{(q d)^2}\right]^2 \, \Big\{ \mathcal A^R_t(q) \left[ \mathcal A^R_t(k) \, f_{\rm NL}^{RR, \rm ttt}(k,q) + \mathcal A^L_t(k) \, f_{\rm NL}^{LR, \rm ttt}(k,q)\right]^2 + \nonumber \\
&\qquad\qquad + \mathcal A^L_t(q) \left[ \mathcal A^L_t(k) \, f_{\rm NL}^{LL, \rm ttt}(k,q) + \mathcal A^R_t(k) \, f_{\rm NL}^{RL, \rm ttt}(k,q)\right]^2 \Big\} \, ,
\end{align}
\begin{align}
C_{\ell}^{V V}(k)  =  G_{\ell, \, \rm auto}^{\rm ttt}(k) \int \, \frac{dq}{q} \, & \left[\frac{j_\ell(q d)}{(q d)^2}\right]^2 \, \Big\{ \mathcal A^R_t(q) \left[ \mathcal A^R_t(k) \, f_{\rm NL}^{RR, \rm ttt}(k,q) - \mathcal A^L_t(k) \, f_{\rm NL}^{LR, \rm ttt}(k,q)\right]^2 + \nonumber \\
&\qquad\qquad + \mathcal A^L_t(q) \left[ \mathcal A^L_t(k) \, f_{\rm NL}^{LL, \rm ttt}(k,q) - \mathcal A^R_t(k) \, f_{\rm NL}^{RL, \rm ttt}(k,q)\right]^2 \Big\} \, ,
\end{align}
\begin{align} \label{eq:spectralast}
C_{\ell}^{I V}(k)  =  G_{\ell, \, \rm auto}^{\rm ttt}(k) \int \, \frac{dq}{q} \, & \left[\frac{j_\ell(q d)}{(q d)^2}\right]^2 \, \Big\{ \mathcal A^R_t(q) \left[ \left(\mathcal A^R_t(k) \, f_{\rm NL}^{RR, \rm ttt}(k,q)\right)^2 - \left(\mathcal A^L_t(k) \, f_{\rm NL}^{LR, \rm ttt}(k,q)\right)^2 \right] - \nonumber \\
&\qquad - \mathcal A^L_t(q) \left[ \left(\mathcal A^L_t(k) \, f_{\rm NL}^{LL, \rm ttt}(k,q) \right)^2 - \left(\mathcal A^R_t(k) \, f_{\rm NL}^{RL, \rm ttt}(k,q)\right)^2 \right] \Big\} \, .
\end{align}
In the above expressions we have applied the following for simplicity of notation,
\begin{align}
G_{\rm cross}^{\rm tts}(k) & \equiv  \frac{4 \pi}{\mathcal A_t(k)}  \, , \nonumber \\
G_{\rm auto}^{\rm tts}(k) &  \equiv  \frac{4 \pi}{\mathcal A^2_t(k)}  \, , \nonumber \\
G_{\ell, \, \rm cross}^{\rm ttt}(k) &  \equiv   \frac{2 \pi}{\mathcal A_t(k)} \, \sqrt{\frac{(\ell + 2)!}{(\ell - 2)!}} \, , \nonumber \\
G_{\ell, \, \rm auto}^{\rm ttt}(k) &  \equiv   \frac{\pi}{\mathcal A^2_t(k)} \, \frac{(\ell + 2)!}{(\ell - 2)!} \, .
\end{align}

\subsection{Parity-even bispectra: matching with previous literature}

Let us assume that primordial power spectra and bispectra are invariant under parity transformations. Therefore, according to Eq. \eqref{eq:parityeven}, amplitudes of non-Gaussianity do not depend on the polarization index, leading to the following identities 
\begin{align}
f_{\rm NL}^{R, \, \rm tts} = f_{\rm NL}^{L, \, \rm tts} = f_{\rm NL}^{\rm even, \, tts} \, ,
\end{align}
and
\begin{align} \label{eq:mixedeqnonmix}
f_{\rm NL}^{RR, \, \rm ttt} = f_{\rm NL}^{LL, \, \rm ttt}  =
f_{\rm NL}^{LR, \, \rm ttt} = f_{\rm NL}^{RL, \, \rm ttt} = f_{\rm NL}^{\rm even, \, ttt} \, .
\end{align}
In this case the expressions for the correlators given above will simplify and become as specified in the following.

\subsubsection*{Monopolar TTS}

\begin{align}
C_{\ell}^{I T}(k)  =  4 \pi \, \int \, \frac{dq}{q} \, j_\ell(q d) \, {\cal T}_{\ell(s)}^{T}(q) \, \mathcal A_s(q) \,  \, f_{\rm NL}^{\rm even, \, tts}(k,q)  \, ,
\end{align}
\begin{align}
C_{\ell}^{I E}(k)  =  4 \pi \, \int \, \frac{dq}{q} \, j_\ell(q d) \, {\cal T}_{\ell(s)}^{E}(q) \, \mathcal A_s(q) \, f_{\rm NL}^{\rm even, \, tts}(k,q)  \, ,
\end{align}
\begin{align}
C_{\ell}^{V T}(k)  =  0 \, ,
\end{align}
\begin{align}
C_{\ell}^{V E}(k)  =  0 \, ,
\end{align}
\begin{align}
C_{\ell}^{I B}(k)  =  0 \, ,
\end{align}
\begin{align}
C_{\ell}^{V B}(k)  =  0 \, ,
\end{align}
\begin{align}
C_{\ell}^{I I}(k)  =  4 \pi \, \int \, \frac{dq}{q} \, \left[j_\ell(q d)\right]^2 \, \mathcal A_s(q) \, \left[f_{\rm NL}^{\rm even, \, tts}(k,q)\right]^2 \, ,
\end{align}
\begin{align}
C_{\ell}^{V V}(k)  =  0 \, ,
\end{align}
\begin{align}
C_{\ell}^{I V}(k)  = 0 \, .
\end{align}

\subsubsection*{Quadrupolar TTS}

\begin{align}
C_{\ell}^{I T}(k)  = 4 \pi  & \, \sum_J \, i^{J - \ell} \, (2 J +1) \, \left[\begin{pmatrix}
	\ell & J & 2 \\
	0 & 0 & 0
	\end{pmatrix}\right]^2 \times \int \, \frac{dq}{q} \, j_J(q d) \, {\cal T}_{\ell(s)}^{T}(q) \, \mathcal A_s(q) \, f_{\rm NL}^{\rm even, \, tts}(k,q) \, ,
\end{align}
\begin{align}
C_{\ell}^{I E}(k)  =  4 \pi  & \, \sum_J \, i^{J - \ell} \, (2 J +1) \, \left[\begin{pmatrix}
	\ell & J & 2 \\
	0 & 0 & 0
	\end{pmatrix}\right]^2  \times \int \, \frac{dq}{q} \, j_J(q d) \, {\cal T}_{\ell(s)}^{E}(q) \, \mathcal A_s(q) \, f_{\rm NL}^{\rm even, \, tts}(k,q) \, ,
\end{align}
\begin{align}
C_{\ell}^{V T}(k)  =  0 \, ,
\end{align}
\begin{align}
C_{\ell}^{V E}(k)  = 0 \, ,
\end{align}
\begin{align}
C_{\ell}^{I B}(k)  =  0 \, ,
\end{align}
\begin{align}
C_{\ell}^{V B}(k)  =  0 \, ,
\end{align}
\begin{align}
C_{\ell}^{I I}(k)  =  4 \pi \, \sum_{J, J'} \, i^{J - J'} \, & (2 J +1) \, (2 J' +1)  \, \left[\begin{pmatrix}
	\ell & J & 2 \\
	0 & 0 & 0
	\end{pmatrix}\right]^2 \, \left[\begin{pmatrix}
	\ell & J' & 2 \\
	0 & 0 & 0
	\end{pmatrix}\right]^2 \times  \nonumber \\
	& \times \int \, \frac{dq}{q} \, j_{J}(q d) \, j_{J'}(q d) \, \mathcal A_s(q) \, \left[ f_{\rm NL}^{\rm even, \, tts}(k,q) \right]^2 \, ,
\end{align}
\begin{align}
C_{\ell}^{V V}(k)  = 0 \, ,
\end{align}
\begin{align}
C_{\ell}^{I V}(k)  =  0 \, .
\end{align}

\subsubsection*{Quadrupolar TTT}

\begin{align}
C_{\ell}^{I T}(k)  =  2 \pi  \, \sqrt{\frac{(\ell + 2)!}{(\ell - 2)!}} \int \, \frac{dq}{q} \, & \frac{j_\ell(q d)}{(q d)^2} \, {\cal T}_{\ell(t)}^{T}(q) \, \mathcal A_t(q) \,  f_{\rm NL}^{\rm even, \, ttt}(k,q)  \, ,
\end{align}
\begin{align}
C_{\ell}^{I E}(k)  =  2 \pi  \, \sqrt{\frac{(\ell + 2)!}{(\ell - 2)!}} \int \, \frac{dq}{q} \, & \frac{j_\ell(q d)}{(q d)^2} \, {\cal T}_{\ell(t)}^{E}(q) \, \mathcal A_t(q) \, f_{\rm NL}^{\rm even, \, ttt}(k,q) \, ,
\end{align}
\begin{align}
C_{\ell}^{V T}(k)  =  0 \, ,
\end{align}
\begin{align}
C_{\ell}^{V E}(k)  =  0 \, ,
\end{align}
\begin{align}
C_{\ell}^{I B}(k)  =  0 \, ,
\end{align} 
\begin{align}
C_{\ell}^{V B}(k)  =  0 \, ,
\end{align}
\begin{align}
C_{\ell}^{I I}(k)  =  \pi \, \frac{(\ell + 2)!}{(\ell - 2)!} \int \, \frac{dq}{q} \, & \left[\frac{j_\ell(q d)}{(q d)^2}\right]^2 \, \mathcal A_t(q) \, \left[ f_{\rm NL}^{\rm even, \, ttt}(k,q) \right]^2 \, ,
\end{align}
\begin{align}
C_{\ell}^{V V}(k)  =  0 \, ,
\end{align}
\begin{align}
C_{\ell}^{I V}(k)  =  0 \, .
\end{align}
These agree with results presented in Ref. \cite{Dimastrogiovanni:2021mfs} apart for the use of the exact $T$- and $E$-mode CMB transfer functions. Also, here we have accounted for non-zero squeezed $\langle \rm ttt \rangle$ bispectra with mixed chiralities, which in \cite{Dimastrogiovanni:2021mfs} were assumed to be equal to $0$. However, according to the general equation \eqref{eq:mixedeqnonmix}, these should give additional contributions, resulting in a factor $2$ in the GW-CMB cross-correlations and a factor $4$ in the GW-GW auto-correlations with respect to \cite{Dimastrogiovanni:2021mfs}. 

From our results we see that only cross-correlators involving $I$, $T$ and $E$ modes are non-vanishing. This is not a surprise and it is strictly related to the parity-invariance of the signatures. We can understand this from the intuition by considering how each fundamental mode transforms under the parity transformation operator $\mathcal P$
\begin{align}
\mathcal P \left(a_{\ell m}^{T}\right) &= (-1)^\ell \, a_{\ell  m}^{T} \, , \\
\mathcal P \left(a_{\ell m}^{E}\right) &= (-1)^\ell \, a_{\ell  m}^{E} \, , \\
\mathcal P \left(a_{\ell m}^{B}\right) &= (-1)^{\ell+1} \, a_{\ell  m}^{B} \, , \\
\mathcal P \left(\delta^{{\rm GW}, \, I}_{\ell m}\right) &= (-1)^\ell \, \delta^{{\rm GW}, \, I}_{\ell m} \, ,
\end{align} 
and the fact that under parity symmetry
\begin{align}
\delta^{{\rm GW}, \, V}_{\ell m} = 0 \, .
\end{align}
Therefore, by employing the parity transformation to the definition \eqref{eq:def_corr}, we get the following transformation rules and results
\begin{align}
C_{\ell}^{I T} & \stackrel{\mathcal P }{\longrightarrow} C_{\ell}^{I T} \, , \\
C_{\ell}^{I E} & \stackrel{\mathcal P }{\longrightarrow}  C_{\ell}^{I E} \, , \\
C_{\ell}^{I B} & \stackrel{\mathcal P }{\longrightarrow}  - C_{\ell}^{I B} = 0\, , \\
C_{\ell}^{V T} & = 0 \, , \\
C_{\ell}^{V E} & = 0  \, , \\
C_{\ell}^{V B}  & = 0  \, , \\
C_{\ell}^{I I}  & \stackrel{\mathcal P }{\longrightarrow} C_{\ell}^{I I}  \, , \\
C_{\ell}^{V V}  &  = 0  \, , \\
C_{\ell}^{V I}  & = 0 \, .
\end{align}

\subsection{Parity-odd bispectra}
\label{sec:parity-odd}

We now switch to the case study focus in this work, i.e. the case of parity-odd bispectra. In particular, we assume no parity violation at the power spectrum level, so $\chi = 0$, and therefore the $V$-mode monopole is vanishing.  As noted in~\cite{Cabass:2021fnw}, parity-odd bispectra may appear at tree-level with no modification to tensor power spectra in inflationary models with alternative symmetry breaking patterns, like solid inflation \cite{Endlich:2012pz,Endlich:2013jia}. Solid inflation is an inflationary model where the acceleration is driven by a triplet of scalar fields that pick up spatial vevs, leading to the spontaneous breaking of spatial diffeomorphisms.  Internal symmetries of the scalar triplet then ensure that the background geometry is homogeneous and isotropic, and the effect of the symmetry breaking pattern is manifest only in cosmological correlators. We will not review in detail this inflationary model, but refer the reader to the original literature and Ref.~\cite{Cabass:2021fnw} for more details on how to construct the operators that give rise to parity-odd bispectra. Moreover, in~\cite{Cabass:2021fnw} it was also shown that the strength of these parity-odd NGs can be sizeable.

More general, according to~\cite{Cabass:2021fnw}, at tree level and assuming only locality, unitarity and the scale invariance of the theory, the shape functions of parity-odd bispectra are highly constrained, with only a few possibilities. By taking the squeezed limit of these, we find that either $\langle \rm tts \rangle$ and $\langle \rm ttt \rangle$ admit non-vanishing squeezed shapes, with $\langle \rm tts \rangle$/$\langle \rm ttt \rangle$ having a monopolar/quadrupolar angular dependence in terms of $\hat k \cdot \hat q$. Even if solid inflation admits mixed scalar-graviton interactions that are not manifestly local, Ref. \cite{Cabass:2021fnw} showed that we can still introduce cubic operators that lead to manifestly local scalar-graviton interactions. Therefore, we take the monopolar/quadrupolar angular dependence in $\langle \rm tts \rangle$/$\langle \rm ttt \rangle$ as well-motivated ansatz for our parity-odd bispectra. 

How are these bispectra constrained with current experiments? The lack of observation of a squeezed non-Gaussian signal in CMB $T$ and $E$ anisotropies disfavour an important level of parity-odd $\langle \rm ttt \rangle$ squeezed NGs at CMB scales (see e.g. \cite{Shiraishi:2013wua,Shiraishi:2014ila,Shiraishi:2017yrq,Planck:2019kim,Shiraishi:2019yux}). On the contrary, when referring to parity-odd $\langle \rm tts \rangle$ squeezed NGs, CMB experiments have been forecasted to provide very poor constraints when relying solely on $T$ and $E$ modes  (see e.g. Ref. \cite{Bartolo:2018elp}). Therefore, at present the strength of squeezed (parity-odd) $\langle \rm tts \rangle$ NGs can be considered experimentally unconstrained. 

Let us now switch to the computations of the imprints of parity-odd bispectra on GW anisotropies. We start by reminding that the parity-odd condition \eqref{eq:parityodd} determines the following relationships between the non-Gaussian coefficients 
\begin{align}
f_{\rm NL}^{R, \,\rm tts} = - f_{\rm NL}^{L, \,\rm tts} = f_{\rm NL}^{\rm odd, \, tts} \, ,
\end{align}
and
\begin{align}
f_{\rm NL}^{RR, \, \rm ttt} = f_{\rm NL}^{RL, \,\rm ttt} &= - f_{\rm NL}^{LR, \,\rm ttt} = -f_{\rm NL}^{LL, \,\rm ttt} = f_{\rm NL}^{\rm odd, \, ttt} \, .
\end{align}
By applying these last equations to the correlators given in Sec. \ref{sec:correlators_self_cross}, we obtain the following expressions.

\subsubsection*{Monopolar TTS}

\begin{align}
C_{\ell}^{I T}(k)  =  0 \, ,
\end{align}
\begin{align}
C_{\ell}^{I E}(k)  =  0 \, ,
\end{align}
\begin{align} \label{eq:CVT}
C_{\ell}^{V T}(k)  =  4 \pi \,  \int \, \frac{dq}{q} \, j_\ell(q d) \, {\cal T}_{\ell(s)}^{T}(q) \, \mathcal A_s(q) \,  \, f_{\rm NL}^{\rm odd, \, tts}(k,q) \, ,
\end{align}
\begin{align} \label{eq:CVE}
C_{\ell}^{V E}(k)  =  4 \pi \, \int \, \frac{dq}{q} \, j_\ell(q d) \, {\cal T}_{\ell(s)}^{E}(q) \, \mathcal A_s(q) \,  \, f_{\rm NL}^{\rm odd, \, tts}(k,q) \, ,
\end{align}
\begin{align}
C_{\ell}^{I B}(k)  =  0 \, ,
\end{align}
\begin{align}
C_{\ell}^{V B}(k)  =  0 \, ,
\end{align}
\begin{align}
C_{\ell}^{I I}(k)  =  0 \, ,
\end{align}
\begin{align}
C_{\ell}^{V V}(k)  =  4 \pi \, \int \, \frac{dq}{q} \, \left[j_\ell(q d)\right]^2 \, \mathcal A_s(q) \, \left[ f_{\rm NL}^{\rm odd, \, tts}(k,q) \right]^2 \, ,
\end{align}
\begin{align}
C_{\ell}^{I V}(k)  =  0 \, .
\end{align}

\begin{figure}[!htbp]
    \centering

        \centering
        \includegraphics[width=0.72\linewidth]{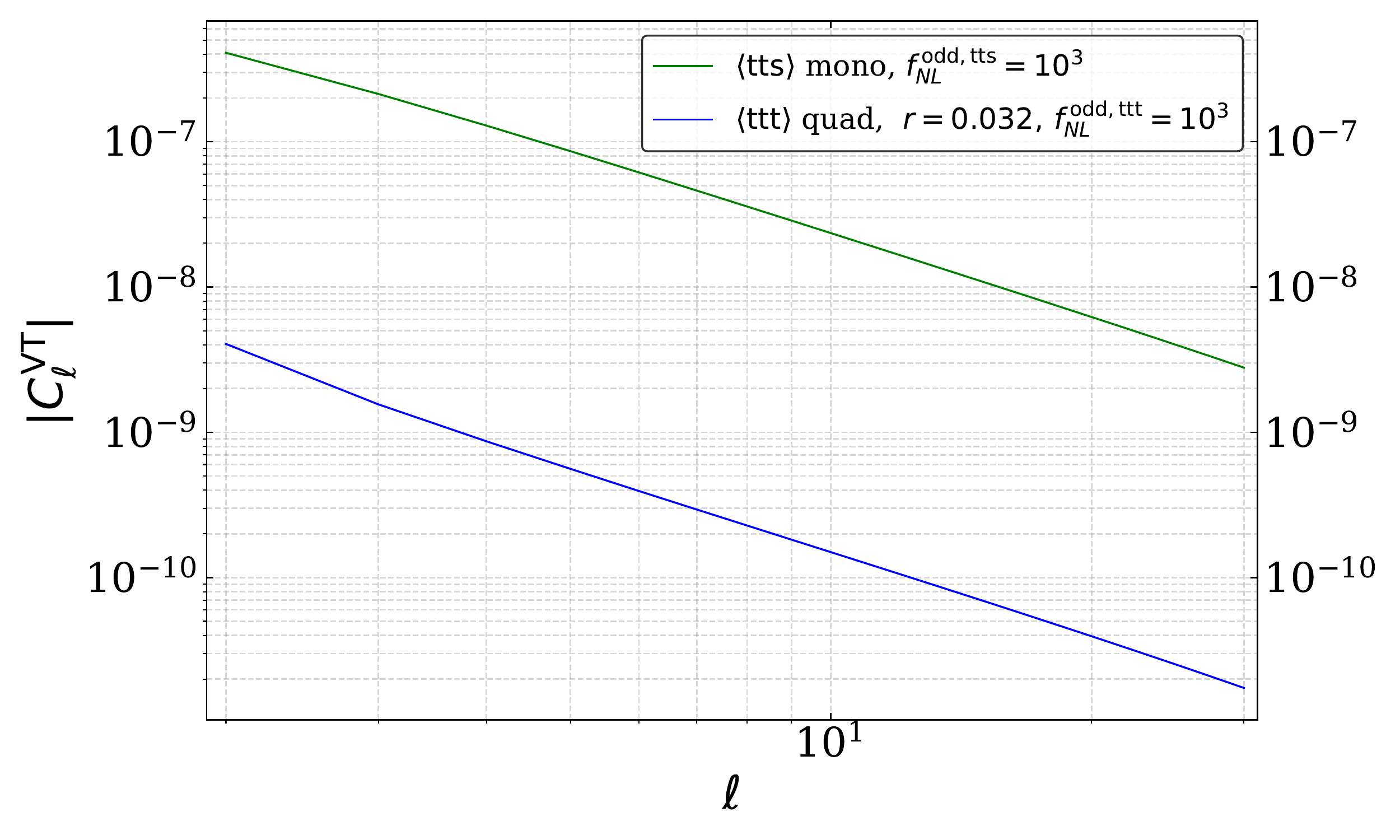} 

        \centering
        \includegraphics[width=0.72\linewidth]{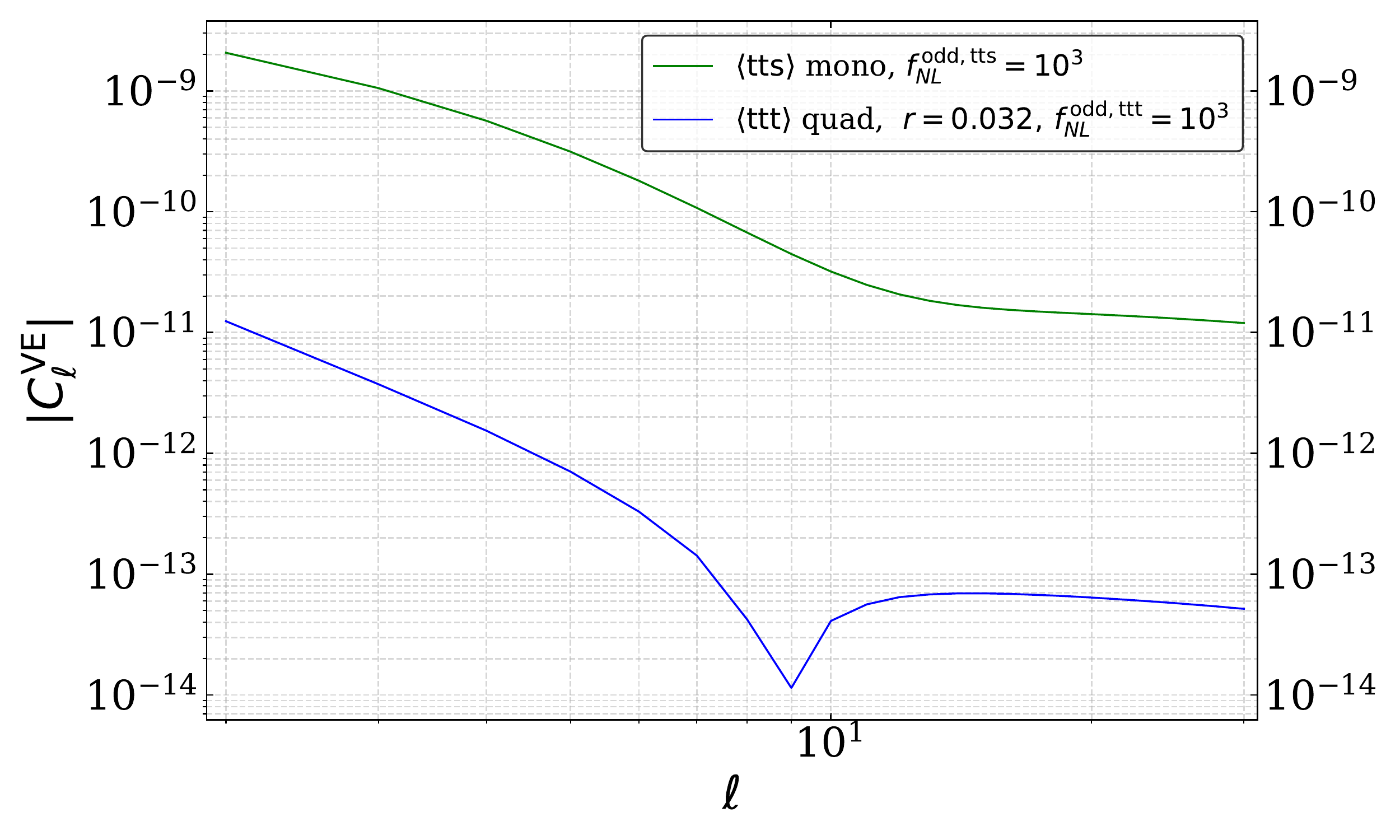} 

        \centering
        \includegraphics[width=0.72\linewidth]{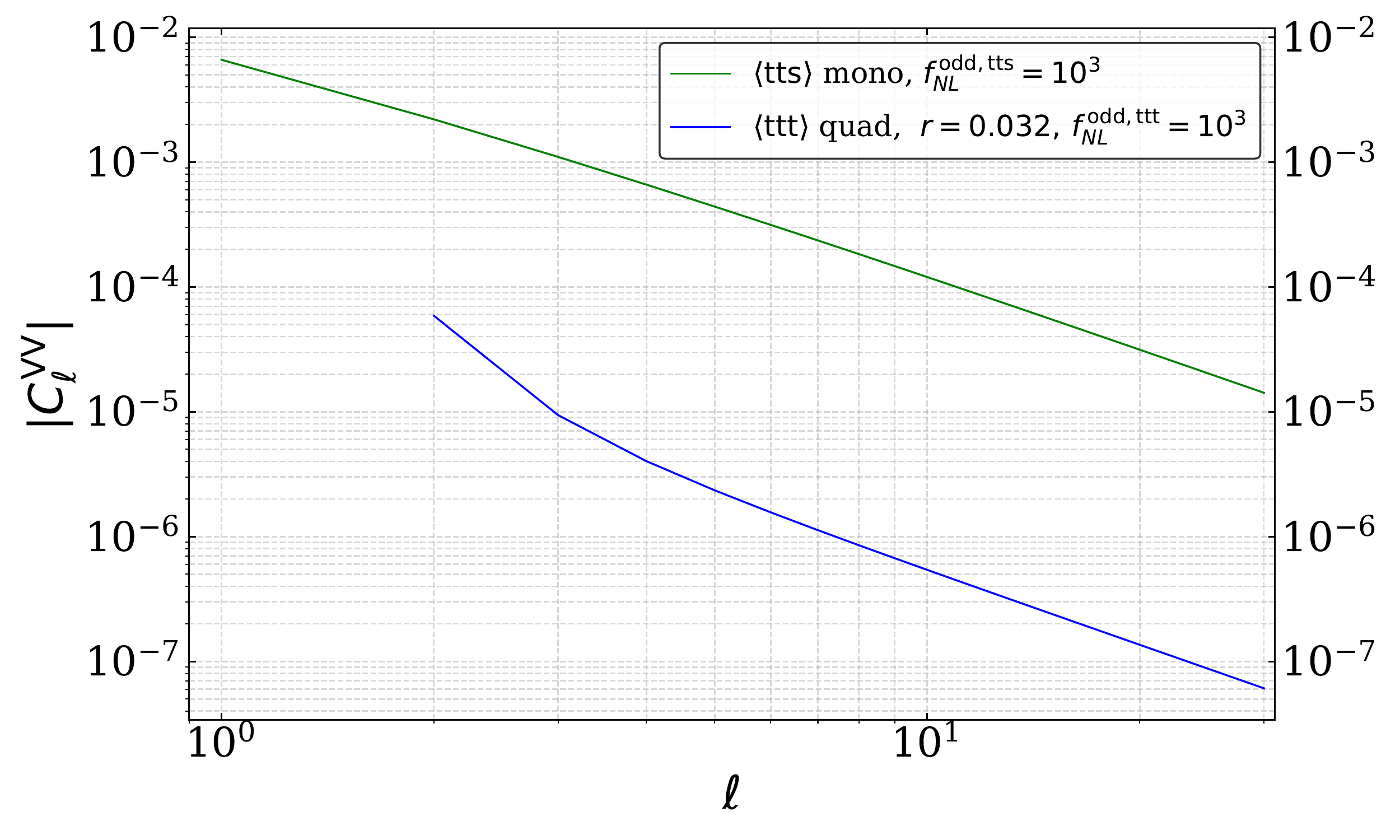} 

    \caption{The $V$-$V$ auto- and $V$-CMB cross-correlations generated by parity-odd bispectra for some fixed value of the parameters. We plotted only the absolute values. The $C_{\ell>8}^{\rm VE}$'s from $\langle \rm ttt \rangle$ take negative values. All the other values are positive. $C_{1}^{\rm VV}$ from $\langle \rm ttt \rangle$ is not shown as it is vanishing.} \label{fig:Cells} 
\end{figure}

\subsubsection*{Quadrupolar TTT}

\begin{align}
C_{\ell}^{I T}(k)  =  0 \, ,
\end{align}
\begin{align}
C_{\ell}^{I E}(k)  = 0 \, ,
\end{align}
\begin{align}
C_{\ell}^{V T}(k)  =  2 \pi \,  \, \sqrt{\frac{(\ell + 2)!}{(\ell - 2)!}} \int \, \frac{dq}{q} \, & \frac{j_\ell(q d)}{(q d)^2} \, {\cal T}_{\ell(t)}^{T}(q) \, \mathcal A_t(q) \, f_{\rm NL}^{\rm odd, \, ttt}(k,q)  \, ,
\end{align}
\begin{align}
C_{\ell}^{V E}(k)  =  2 \pi \, \sqrt{\frac{(\ell + 2)!}{(\ell - 2)!}} \int \, \frac{dq}{q} \, & \frac{j_\ell(q d)}{(q d)^2} \, {\cal T}_{\ell(t)}^{E}(q) \, \mathcal A_t(q) \,  f_{\rm NL}^{\rm odd, \, ttt}(k,q)  \, ,
\end{align}
\begin{align}
C_{\ell}^{I B}(k)  =  0 \, ,
\end{align} 
\begin{align}
C_{\ell}^{V B}(k)  =  0 \, ,
\end{align}
\begin{align}
C_{\ell}^{I I}(k)  =  0 \, ,
\end{align}
\begin{align}
C_{\ell}^{V V}(k) =   \pi \, \frac{(\ell + 2)!}{(\ell - 2)!} \int \, \frac{dq}{q} \, & \left[\frac{j_\ell(q d)}{(q d)^2}\right]^2 \, \mathcal A_t(q) \, \left[ f_{\rm NL}^{\rm odd, \, ttt}(k,q)\right]^2 \, ,
\end{align}
\begin{align}
C_{\ell}^{I V}(k) =  0 \, .
\end{align}
Here, we find that only correlators involving $V$, $T$ and $E$ modes are non-vanishing. Interestingly all spectra with $I$ modes are vanishing, therefore a net detection of GW $I$ modes of primordial origin or their cross-correlation with CMB anisotropies as considered in previous literature would rule out scenarios of inflation with parity-odd bispectra, but no in the tensor power spectrum. In Fig. \ref{fig:Cells} we plot these quantities fixing constant values of the non-Gaussian parameters $f^{\rm odd, tts}_{\rm NL} = f^{\rm odd, ttt}_{\rm NL} = 10^3$ and $r = 0.032$, which corresponds to the current constraint on the tensor-to-scalar-ratio~\cite{Planck:2018jri,Tristram:2021tvh}. As we are considering anisotropies of GW that re-enter the horizon very early, we can fix $d=\eta_0-\eta_{\rm entry} \approx \eta_0$ without any significant modification in the final results. We assume constant amplitude for the large scale primordial power spectra, with $\mathcal A_\mathrm{s}(q) = 2.100549\times 10^{-9}$ for the dimensionless scalar power spectrum, while the dimensionless tensor power spectrum is taken as $\mathcal A_\mathrm{t}(q) \simeq r A_\mathrm{s}$. We verified that accounting for the slight red-tilted dependence of scalar perturbations as measured by CMB experiments~\cite{Planck:2018jri} has a negligible effect. We can fix $q_{\rm min} = 0$ to perform the $q$-integrations as the integrands in the equations for the angular power spectra are vanishing in the $q \rightarrow 0$ limit, erasing any dependence on the exact duration of inflation. There is only one exception: the $\ell =2$ auto-$V$ modes power spectrum generated by the parity-odd $\langle \rm ttt \rangle$ bispectrum  possesses a logarithmic divergence in the $q \rightarrow 0$ limit if we assume that $\mathcal A_t(q)$ and $f^{\rm odd, ttt}_{\rm NL}(k,q)$ are scale-invariant. In the $V$-mode power spectrum shown in Fig. \ref{fig:Cells} we cured this divergence by fixing $q_{\rm min} = 10^{-5} \, \mbox{Mpc}^{-1}$, corresponding to the order of magnitude of the largest CMB scale. However, we stress that this choice is just indicative and smaller values of $q_{\rm min}$ would cause a logarithmic increase of this $\ell = 2$ auto-$V$ modes power spectrum.

From the figures we conclude that the $\langle \rm tts \rangle$ bispectrum systematically leads to a signal that is at least an order of magnitude stronger than $\langle \rm ttt \rangle$ for the same level of NGs. Also, $\langle \rm tts \rangle$ is the only bispectrum that affects the GW $V$-mode dipole which, as we will see later on, represents the best observational channel in GW experiments searching for $V$-mode anisotropies. This together with the lack of any constraint from CMB experiments makes the $\langle \rm tts \rangle$ bispectrum a better candidate than $\langle \rm ttt \rangle$ for a possible detection with GW $V$ modes. For that reason, in the next section we will present a preliminary investigation of the detection prospects on the GW $V$ modes generated by the squeezed parity-odd $\langle \rm tts \rangle$ bispectrum.
\FloatBarrier

\subsection{Detection prospects of parity-odd bispectra from a BBO-like experiment}

In this section we study the detection prospects of the $V$-mode intrinsic anisotropies generated by parity-odd squeezed bispectra. We first estimate the minimum noise level in the $V$ modes detection from a BBO-like case study: the experiment we consider consists of the space-based BBO star configuration, with two LISA-like triangular constellations arranged in the same plane as a six-pointed star, i.e. with an angular phase difference of $60^\circ$ between the constellation planes \cite{Crowder:2005nr}. In a very advanced stage of the BBO mission there will also be $2$ outer constellations trailing and leading the star constellation by $120^\circ$ in an earth-like orbit around the sun. However, in this work we do not take these additional constellations into account. The individual arm lengths of each constellation are taken as the current planned value of $L = 5 \times 10^7 \mbox{ m}$.  As shown for the first time in \cite{Smith:2016jqs}, with the observatories lying on the same plane, the resulting experiment would be insensitive to the $V$-mode monopole, while instead this is not true when dealing with $V$-mode anisotropies (see e.g. \cite{Domcke:2019zls}). 

Here, we will not go into details defining the quantities involved in the characterization of LISA-like networks of interferometers and the final results for the noise angular power spectra, but we will give only the fundamental details, referring to the literature and previous references on this topic for in depth explanations. 

\begin{figure}[!htbp]

        \centering
        \includegraphics[width=0.8\linewidth]{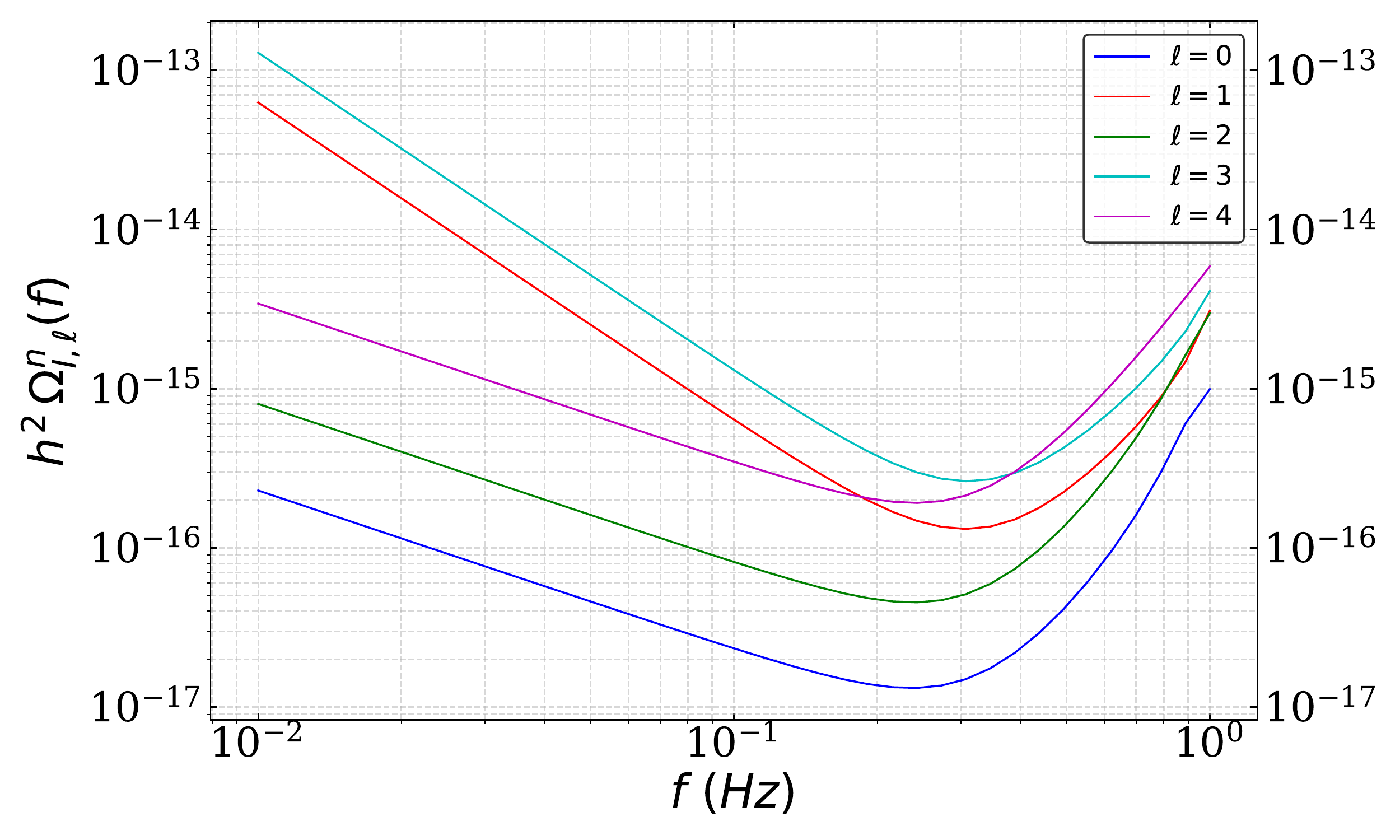} 

        \centering
        \includegraphics[width=0.8\linewidth]{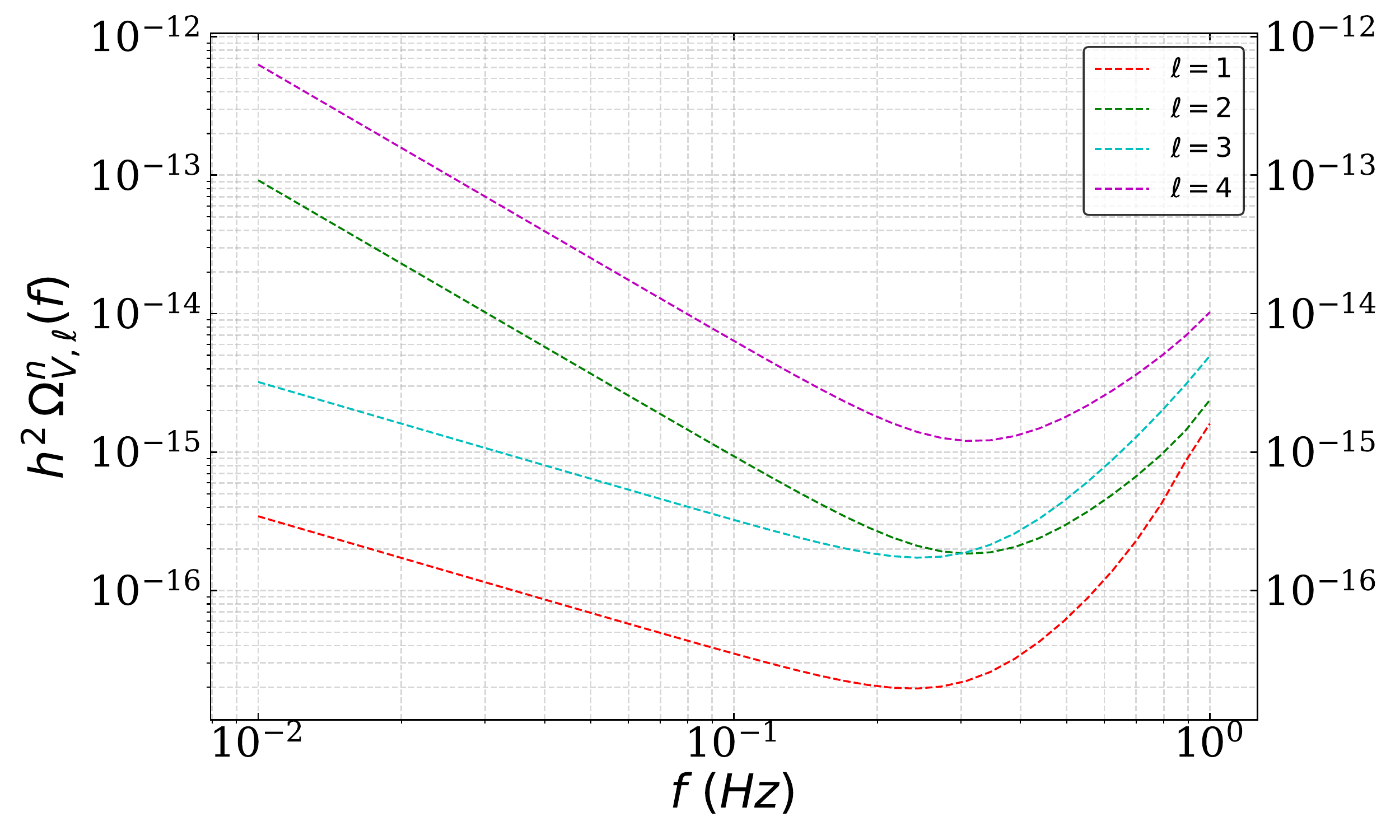} 

    \caption{Figures showing the angular sensitivity curves of the first four $\ell$-th multipoles. Only cross-correlations between different constellations of the BBO-star configuration are considered. The noise power spectral densities on the BBO channels are taken by Ref. \cite{Smith:2016jqs}. $1$ year of observation is assumed.} \label{fig:Nell}
\end{figure}

In general, for each of the two constellations we can define three observational channels, the so-called Time Domain Interferometry (TDI) $X$, $Y$ and $Z$ variables \cite{Flauger:2020qyi} \footnote{In practise, it is common to switch to the so called AET TDI channels basis, which are obtained by diagonalizing the co-variance matrix of the $X$, $Y$ and $Z$ channels, in a way that the cross-correlated noise between the different channels of the same constellation is zero \cite{Hogan:2001jn,Adams:2010vc}. In our case we consider only cross-correlations between different constellations, therefore employing the $X, Y, Z$ or the $A, E, T$ basis will not make any difference in our final statements.}. In the frequency domain, the cross-correlations between the generic channels $A$ and $B$ can be written as~\cite{Flauger:2020qyi}
\begin{equation} \label{eq:signalsGW}
\langle s_A(f) \, s_B(f')\rangle = \frac{\delta_T(f-f')}{2} \, \left[ \mathcal R^I_{AB}(f) \, P^I_t(f) +  \mathcal R^V_{AB}(f) \, P^V_t(f) \right] \, ,
\end{equation}
where $\delta_T(f-f') = T \sinc(f - f')$ \footnote{In the limit of infinite observation time the $\delta_T$ can be replaced with a proper Dirac delta.}. Here, the quantities $\mathcal R^X_{AB}(f)$ are the monopole overlap reduction functions of the GW $X$-mode, where we defined 
\begin{equation}
P^I_t(f) = P^R_t(f) + P^L_t(f) \, , \qquad \qquad P^V_t(f) = P^R_t(f) - P^L_t(f) \, .
\end{equation}
In Eq. \eqref{eq:signalsGW} the $\hat n$ angular dependence of the signal power spectrum is integrated out. By accounting for this angular dependence, we can define the angular overlap reduction functions as~\cite{Alonso:2020rar}
\begin{equation}
\mathcal R^X_{AB, \, \ell m}(f) = \int d^2 \hat n \, \mathcal R^X_{AB}(f,\hat n) \, Y^*_{\ell m}(\hat n) \, . 
\end{equation}
Here we consider only cross-correlations between channels of different constellations. In this case, and working in the rigid detector approximation (time independent antenna patterns), the $X$-mode noise power spectrum reads~\cite{Alonso:2020rar}
\begin{equation}
\left(N^X_\ell\right)^{-1} =  T_{\rm obs} \sum_{A, B > A} \int df \, \left( \frac{2 \xi(f)}{5}\right)^2 \, \left(\frac{3 H_0^2}{4 \pi^2 f^3} \right)^2 \frac{\sum_m |\mathcal R^X_{AB, \, \ell m}(f)|^2}{(2 \ell+1) N^A(f) \, N^B(f)}  \, ,
\end{equation}
where $T_{\rm obs}$ denotes the total observation time, 
\begin{equation}
\xi(f) = \left(\frac{f}{f_*}\right)^{n_t} \, ,
\end{equation}
is a quantity sensitive to the spectral dependence of the signal under consideration\footnote{Here a power-law dependence with tensor tilt $n_t$ and reference frequency $f_*$ is assumed.}, and $N^A(f), N^B(f)$ are the noise power spectral densities of the channels $A$ and $B$, respectively. For an almost scale invariant background of GW we have $\xi(f) \simeq 1$. According to our normalization conventions of GW anisotropies, assuming a background of GW with $\overline{\Omega}^I_{\text{\tiny GW}}(f) = \overline{\Omega}^I_{\text{\tiny GW}}$, the noise just introduced can be related to $C_\ell^{II}$ and $C_\ell^{VV}$ as
\begin{equation} \label{eq:rescaling}
N_\ell^{II} = \frac{N^I_\ell}{\left(\overline{\Omega}^I_{\text{\tiny GW}}\right)^2} \, , \qquad \qquad N_\ell^{VV} = \frac{N^V_\ell}{\left(\overline{\Omega}^I_{\text{\tiny GW}}\right)^2} \, .
\end{equation}
Alternatively, for a background with a relevant scale dependence, this last equation should be applied locally at each separate frequency $f$. It makes sense to define the multipole squared-SNR as \cite{LISACosmologyWorkingGroup:2022kbp}
\begin{equation} \label{eq:SNR_local}
\left({\rm SNR}_\ell^X\right)^2 = \int \, df \, (2 \ell+1) \, \frac{\left|C_\ell^{XX}(f)\right| \, \left[\overline{\Omega}^I_{\text{\tiny GW}}(f)\right]^2}{N_\ell^{X}(f)} \, ,
\end{equation}
where\footnote{Note the additional $\left(2/5\right)^2$ factor with respect to what given in Ref. \cite{LISACosmologyWorkingGroup:2022kbp}. This is because we are following the conventions in Ref. \cite{Alonso:2020rar} in defining the detector angular overlap reduction functions which differ from Ref. \cite{LISACosmologyWorkingGroup:2022kbp} by a factor $5/2$. The authors of Ref. \cite{Alonso:2020rar} adopted a normalization convention that is typically intended for L-shaped detectors rather than triangular detectors.}
\begin{equation} \label{eq:noise_def}
N^X_\ell(f) =  \left(T_{\rm obs} \sum_{A, B > A}  \, \left( \frac{2}{5}\right)^2 \, \left(\frac{3 H_0^2}{4 \pi^2 f^3} \right)^2 \, \frac{\sum_m |\mathcal R^X_{AB, \, \ell m}(f)|^2}{(2 \ell+1) N^A(f) \, N^B(f)} \right)^{-1} \, .
\end{equation}
Eq.~\eqref{eq:SNR_local} gives us the frequency-integrated squared-SNR at a given multipole $\ell$. $\left({\rm SNR}_\ell^X\right)^2>1$ implies evidence for a non-zero signal in the $\ell$-th pole anisotropy of the $X$-mode. In analogy with Ref. \cite{Dimastrogiovanni:2022eir}, we can define the following effective angular sensitivity of the detector to the $\ell$−th multipole
\begin{equation} \label{eq:Omega_X}
\Omega^n_{X, \ell}(f) =  \left(T_{\rm obs} \sum_{A, B > A}  \, \left( \frac{2}{5}\right)^2 \, \left(\frac{3 H_0^2}{4 \pi^2 f^3} \right)^2 \, \frac{\sum_m |\mathcal R^X_{AB, \, \ell m}(f)|^2}{(2 \ell+1) N^A(f) \, N^B(f)} \right)^{-1/2} \, .
\end{equation}
In Fig.~\ref{fig:Nell} we plot this quantity as function of the frequency for the $I$ and $V$-mode first four multipoles\footnote{As shown e.g.~in Ref. ~\cite{Malhotra:2020ket}, when taking cross-correlations between the two constellation of the BBO-star configuration, the noise level starts to increase fast for $\ell > 4$, making any detection of a signal challenging.} assuming $1$ year of observation. For each channel of the two BBO-like constellations we have taken the noise power spectral densities as given in Ref. \cite{Smith:2016jqs}\footnote{Specifically, for each channel "$A$" we assume a noise of the form
\begin{equation}
N^A(f) = 4 \left(S_{n, s} + 2 S_{n,a} \left[ 1 + \cos^2(f/f_{\rm BBO})\right] \right) \, ,  
\end{equation}
where $f_{\rm BBO}$ is the BBO-characteristic frequency $f_{\rm BBO} = \left(2 \pi L/c \right)^{-1} \simeq 1$ Hz, and we defined the shot-noise and acceleration-noise power spectra
\begin{align}
S_{n, s} &= 8 \times 10^{-50} \, \mbox{Hz}^{-1}  \, , \\
S_{n, a} &= 2.3 \times 10^{-52} \, \left(\frac{\mbox{Hz}}{f} \right)^4 \, \mbox{Hz}^{-1} \, .
\end{align}
}. 
From Fig. \ref{fig:Nell} we find that $f_{\rm min} \approx 0.3$ Hz represents the frequency of maximum sensitivity of the experiment for all the multipole-anisotropies, with a slight variation of the exact frequency of the minimum noise depending on the multipole and the $X$-mode considered. 

In particular, we can understand the frequency-scaling and the general behaviour of the $V$-mode angular sensitivity curves for reflection with the $I$-mode case: in fact, as explained e.g. in Ref. \cite{Alonso:2020rar}, in LISA-like interferometers the overall behaviour of these curves and their strength is linked to the parity-state of the corresponding angular overlap reduction functions. In the case of cross-correlations between channels of close constellations (as happens for the BBO star configuration) $\ell =$even(odd) angular $I$-mode overlap reduction functions have even(odd)-parity, respectively. Angular overlap functions with the same parity-state have very similar frequency-scaling and the minimum of the sensitivity function at similar frequency (here $f_{\rm min} \simeq 0.25(0.3)$ Hz for even(odd)-parity). Also, parity-even angular overlap functions have overall better sensitivity than the parity-odd ones. With this in mind, we can easily predict the behaviour of the $V$-mode angular sensitivity curves: as they have opposite parity-states with respect to the $I$-mode ones, we expect the $\ell=$even(odd) $V$-mode angular sensitivity curves to be similar to the $\ell=$odd(even) $I$-mode ones. This coincides with what quantitatively displayed in Fig. \ref{fig:Nell}.

\begin{figure}[!htbp]
        \centering
        \includegraphics[width=0.8\textwidth]{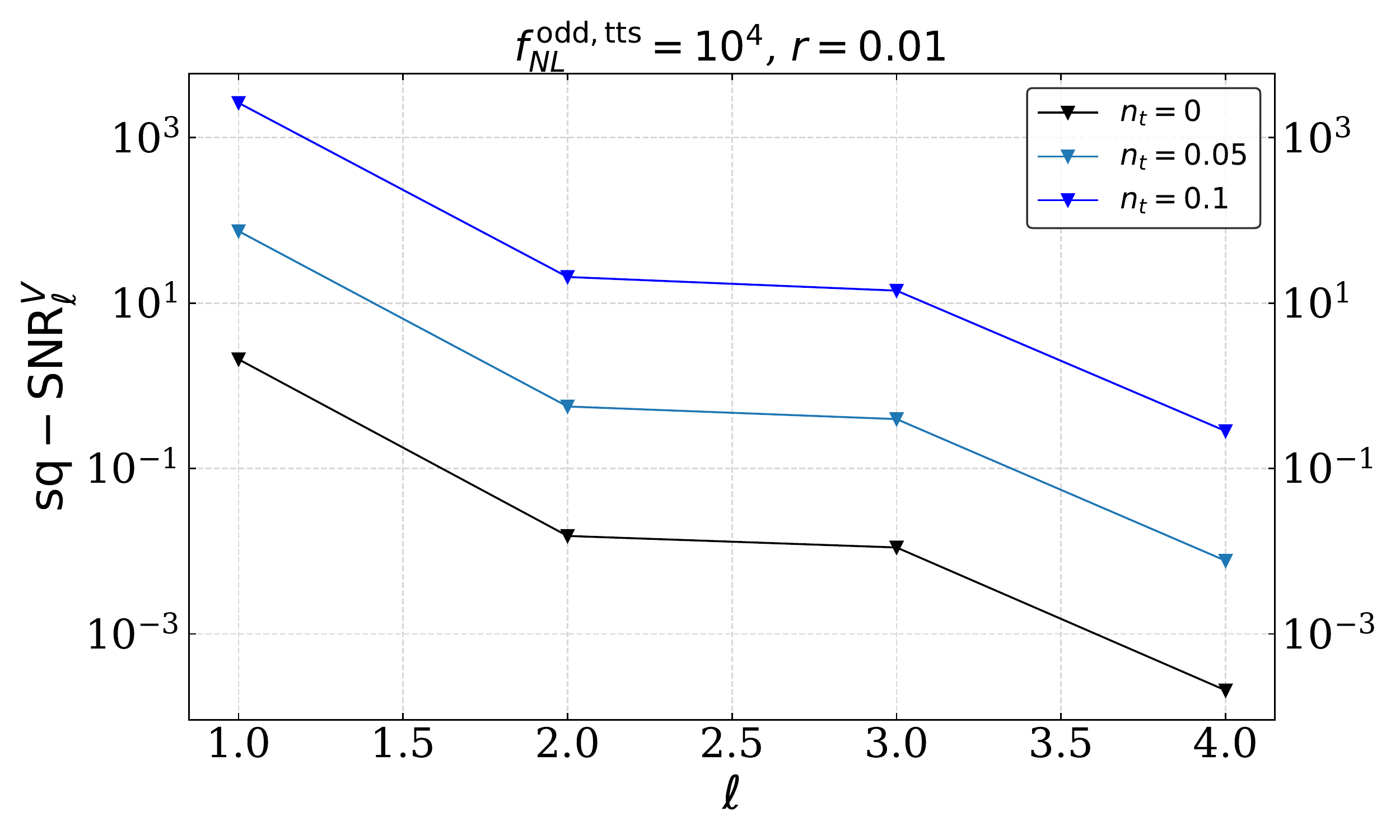} 
    \caption{Figure showing the squared-${\rm SNR}^V_\ell$ for the $V$-mode anisotropies sourced by squeezed parity-odd $\langle \rm tts \rangle$. A power-law of tensor-tilt $n_t$ is assumed for the tensor power spectrum. The value of the tensor-to-scalar-ratio refers to the CMB frequency $f \simeq 7.7 \times 10^{17}$ Hz (or $k = 0.05 \,\, \mbox{Mpc}^{-1}$).} \label{fig:VoverN}
\end{figure}

By assuming an almost constant $f^{\rm odd, tts}_{\rm NL} = 10^4$ coefficient and a power-law tensor power spectrum, in Fig. \ref{fig:VoverN} we show the $\left({\rm SNR}_\ell^V\right)^2$ for three different choices of the tensor spectral index taking $r = 0.01$ at $k = 0.05 \,\, \mbox{Mpc}^{-1}$. Using this figure, in Tab. \ref{tab:SNRV} we derive scaling formulas for the $\left({\rm SNR}_\ell^V\right)^2$. We see that the $V$-mode dipole is the best channel to probe parity-odd $\langle \rm tts\rangle$ bispectra. In particular by taking a scale invariant amplitude of GWs with $r = 0.01$, in $5$ years the $V$-mode dipole would provide evidence for a signal of $f^{\rm odd, tts}_{\rm NL} \sim 3 \times 10^3$. By taking a generic $r$ this value would become $f^{\rm odd, tts}_{\rm NL} \sim  3 \times 10^3 \times (0.01/r)$. On the contrary, higher multipoles are much less sensitive to a primordial signal, demanding a blue-tilted dependence in the tensor power spectrum to improve the detection prospects. Such a tilt-dependence can be obtained for instance within solid inflation (see e.g. \cite{Malhotra:2020ket}) which, as pointed out above, contains symmetry breaking patterns that allow the parity-odd NGs considered here. However, in this inflationary scenario large values of $n_t$ are disallowed, in a way that a detection of a $\ell >1$ $V$-mode anisotropy would still remain difficult. As squeezed $\langle \rm ttt\rangle$ bispectra do not leave signatures in the $V$-mode dipole, our analysis enforces the fact that $\langle \rm tts\rangle$ is in general a better candidate to make an observation rather than $\langle \rm ttt\rangle$.
\begin{table}[!htbp]
\centering
\begingroup
\setlength{\tabcolsep}{8pt} 
\renewcommand{\arraystretch}{2} 
     
    \begin{tabular}{| c | c | c | c | c | c}
       \cline{2-5}  \multicolumn{1}{c|}{$\times \left( \frac{r}{0.01}\right)^2 \, \Big( \frac{f^{\rm odd, tts}_{\rm NL}}{10^4}\Big)^2 \left(\frac{T_{\rm obs}}{1 \rm yr}\right)$}
         & $\ell = 1$ & $\ell = 2$ & $\ell = 3$ & $\ell = 4$ \\
       \cline{1-5}
       \rule{0pt}{1.5\normalbaselineskip}
  $\langle \rm tts \rangle: \quad$ $\left({\rm SNR}_\ell^V\right)^2|_{n_t = 0}$ = & $2.09$  & $0.015$ &  $0.011$ & $0.00021$ & \\
       \cline{1-5}
     $\,\,\,\,\,\,\,\langle \rm tts \rangle: \quad$   $\left({\rm SNR}_\ell^V\right)^2|_{n_t = 0.05}$ = & $74$  & $0.56$ & $0.4$ & $0.0077$ & \\
        \cline{1-5}
      $\,\,\,\,\langle \rm tts \rangle: \quad$  $\left({\rm SNR}_\ell^V\right)^2|_{n_t = 0.1}$ = & $2.64 \times 10^{3}$  & $21$ &  $14$ & $0.28$  \\
        \cline{1-5}
    \end{tabular}
    \endgroup
    \caption{Scaling formulas for $\left({\rm SNR}_\ell^V\right)^2$ for different values of the tensor tilt.} 
    \label{tab:SNRV}
\end{table}
\FloatBarrier

At this point we want to address the following question: is the $V$-mode observational channel here considered to measure parity-odd $\langle \rm tts \rangle$ competitive (as well as complementary) with CMB experiments? Even if squeezed primordial NGs probed by GW anisotropies refer to squeezed triangular configurations much different than those probed by CMB experiments, assuming scale invariance we can match the expected sensitivity to $f^{\rm odd, tts}_{\rm NL}$ of our case-study with that of the future generations of CMB experiments, assuming that a primordial $B$-mode detection is made. Using the results of \cite{Bartolo:2018elp} and in particular noting that the parameter $\Pi$ of this reference is linked to our $f^{\rm odd, tts}_{\rm NL}$ through
\begin{equation} 
f^{\rm odd, tts}_{\rm NL} \simeq 6.5 \times 10^{-2} \, r \, \Pi \, , 
\end{equation}
we derive the following CMB forecasts on the $1 \sigma$ error on $f^{\rm odd, tts}_{\rm NL}$ by using $BBT$ and $BBE$ bispectra for different values of $r$
\begin{equation} \label{eq:tts_CMB}
\Delta  f^{\rm odd, tts}_{\rm NL}\Big|^{r = 10^{-2}}_{\rm CMB} \gtrsim  10^4 \, , \qquad \Delta  f^{\rm odd, tts}_{\rm NL}\Big|^{r = 10^{-3}}_{\rm CMB} \gtrsim  10^5 \, , \qquad \Delta  f^{\rm odd, tts}_{\rm NL}\Big|^{r = 10^{-4}}_{\rm CMB} \gtrsim  5 \times 10^5  \, .
\end{equation}
These lower bounds on the 1-$\sigma$ errors are derived assuming a full-sky cosmic variance-limited detection of primordial $B$ modes accounting for the lensing contamination. By matching these values with
\begin{equation} \label{eq:tts_GW}
f^{\rm odd, tts}_{\rm NL}\Big|^{\rm 5 yrs}_{\rm GW} \sim  3 \times 10^3 \times \left(\frac{0.01}{r}\right)  \, , 
\end{equation}
we realize that for $r \gtrsim 10^{-4}$ our $V$-mode dipole can in principle provide stronger or commensurate constraints compared to those of hypothetical CMB experiments that are able to detect cosmic-variance limited primordial $B$ modes. 

It is important to mention that in this comparison we did not consider the information coming from $\ell>2$ $V$-mode anisotropies as the level of noise increases relatively rapidly by increasing $\ell$ (see Fig. \ref{fig:Nell}). As shown e.g. in Ref. \cite{Dimastrogiovanni:2021mfs} for the $I$-mode case, considering the full BBO-configuration might reduce the level of noise of higher $\ell$-poles anisotropies in a way that the ultimate detectable value of $f^{\rm odd, tts}_{\rm NL}$ is smaller than what presented in Eq. \eqref{eq:tts_GW}. Therefore, the forecast in Eq. \eqref{eq:tts_GW} should be considered as a first estimate.

As a final remark, we stress that in our investigation we have not considered the motion of our BBO-experiment with respect to the cosmic reference frame. By accounting for this, we would expect a non-zero $V$-mode dipole induced by a non-zero $V$-mode monopole \cite{Seto:2006hf,Seto:2006dz,Domcke:2019zls}. While here we have assumed no asymmetry in $R$- and $L$-handed tensor power spectra in a way that the $V$-mode monopole is vanishing, a non-zero $V$-mode monopole due to some other primordial parity-violation mechanism could contaminate the dipole channel. We will leave the implications of this contamination for the search of parity-odd squeezed primordial NGs in GW experiments for future research.

In the following we will see which role the GW-CMB cross-correlations play in providing additional evidence claiming a primordial detection of a GW $V$-mode signal.

\subsection{The role of GW-CMB cross-correlations}

\begin{figure}[!htbp]
        \centering
        \includegraphics[width=0.8\textwidth]{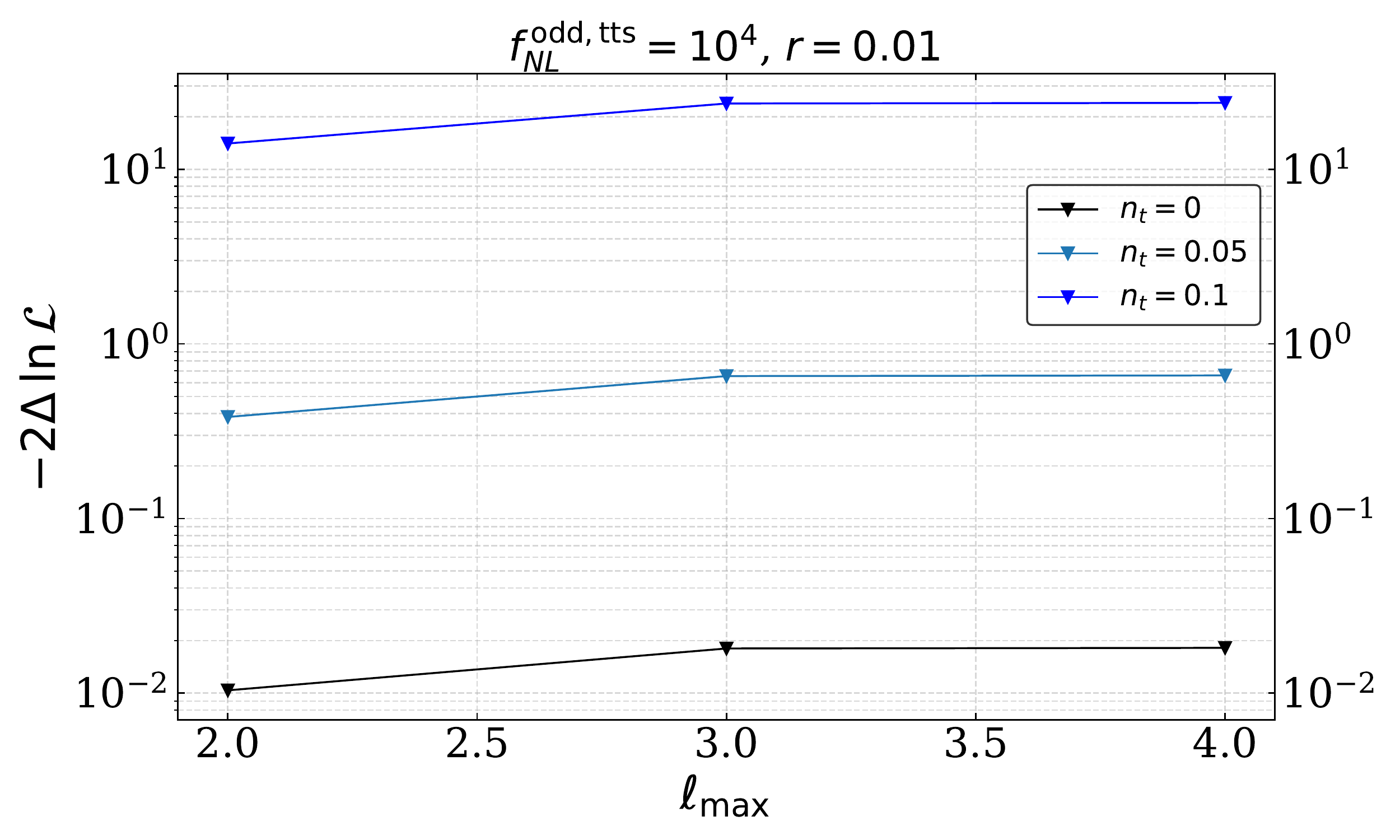} 
    \caption{Plot of the likelihood ratio in Eq. \eqref{eq:likel_N}.} \label{fig:likeli_N}
\end{figure}

Here we discuss the cross-correlations between GW $V$ modes and CMB $T$ and $E$ modes to improve the detection prospects of a parity-odd $\langle \rm tts \rangle$ bispectrum or to confirm a detection of primordial origin. For this purpose, let us consider the following vector of observables\footnote{In what follows we are assuming that the non-Gaussian amplitudes $f_{\rm NL}^{\rm odd, tts}$ do not depend on the short scale $k (f)$, in a way that the signals $C_\ell^{XY}$ do not depend on the frequency $f$.}
\begin{equation}
O_\ell = (C_\ell^{V T}, C_\ell^{V E}) \, .
\end{equation}
Assuming that the cross-correlations are Gaussian distributed\footnote{In principle these cross-correlations should follow a $\chi^2$ distribution with $2\ell+1$ degrees of freedom which approaches to a Gaussian in the large $\ell$ limit. As here we want to make an order-of-magnitude forecast, for simplicity we will approximate the distribution as a Gaussian also for small $\ell$'s.}, the vector of some given measurements $\tilde O_\ell$ will follow the multivariate Gaussian likelihood
\begin{equation}
\mathcal L \propto \exp\left[-\frac{1}{2} \sum_{\ell =2}^{\ell_{\rm max}} \, (\tilde O_\ell-\bar O_\ell)^T \cdot\Sigma^{-1} \cdot (\tilde O_\ell-\bar O_\ell)\right] \, ,
\end{equation} 
where $\Sigma$ denotes the covariance matrix
\begin{equation} \label{eq:sigma_N}
\Sigma = \frac{1}{2 \ell +1} \, \begin{pmatrix}
	 \left(N^{VV}_{\ell} + \bar C^{VV}_{\ell}\right)  C^{TT}_{\ell} + \left( \bar C^{VT}_{\ell}\right)^2 & \,\,\,\,\left(N^{VV}_{\ell} + \bar C^{VV}_{\ell}\right) C^{TE}_{\ell} + \bar C^{VT}_{\ell} \bar C^{VE}_{\ell}  \\
                                          \\
	 \left(N^{VV}_{\ell} + \bar C^{VV}_{\ell}\right)  C^{TE}_{\ell} + \bar C^{VT}_{\ell} \bar C^{VE}_{\ell} & \,\,\,\,\left(N^{VV}_{\ell} + \bar C^{VV}_{\ell}\right)  C^{EE}_{\ell} + \left(\bar C^{VE}_{\ell}\right)^2 
	\end{pmatrix} \, ,
\end{equation}
and
\begin{equation}
\left(N^{VV}_{\ell}\right)^{-1} = \int \, df \, \frac{\left[\overline{\Omega}^I_{\text{\tiny GW}}(f)\right]^2}{N_\ell^{V}(f)} \, .
\end{equation}
In Eq. \eqref{eq:sigma_N} the bar denotes quantities evaluated for some fiducial model where we fix the early-universe parameters, corresponding to Eqs. \eqref{eq:CVT} and \eqref{eq:CVE}. Now, let us assume that we have been unable to extract a signal out of the noise level in the $V$ modes power spectrum as $({\rm SNR}_\ell^V)^2 < 1$. From a measurement $\tilde O_\ell$ we want to reject the $H_0$ hypothesis of the absence of a primordial signal in the auto- and cross-correlations, i.e. $\bar C^{VV}_{\ell} = \bar O_\ell = 0$. The alternative hypothesis $H_1$ will be the presence of a primordial signal in the cross-correlation, i.e. $\bar O_\ell \neq 0$, consistent with Eqs. \eqref{eq:CVT} and \eqref{eq:CVE}. Assuming that we measure non-zero cross-correlations $\tilde O_\ell$ that are compatible with what is expected from a given fiducial theory (so $\tilde O_\ell \simeq \bar O_\ell$)  we can define the following log-likelihood ratio
\begin{equation} \label{eq:likel_N}
- 2 \, \Delta \ln \mathcal L = \sum_{\ell =2}^{\ell_{\rm max}} \,  \bar O_\ell^T \cdot \Sigma_N^{-1} \cdot \bar O_\ell \, ,
\end{equation}
where 
\begin{equation}
\Sigma_N = \frac{1}{2 \ell +1} \, \begin{pmatrix}
	 N^{VV}_{\ell}   C^{TT}_{\ell}  & \,\,\,\, N^{VV}_{\ell}  C^{TE}_{\ell}  \\
                                          \\
	 N^{VV}_{\ell}  C^{TE}_{\ell}  & \,\,\,\,N^{VV}_{\ell}  C^{EE}_{\ell} 
	\end{pmatrix} \, .
\end{equation}
In general, the larger the quantity $- 2 \, \Delta \ln \mathcal L $ is, the larger is the statistical significance supporting a non-zero primordial detection. We start to have evidence of a non-zero primordial signal when $- 2 \Delta \ln \mathcal L>1$ \footnote{Usually, we should impose $- 2 \Delta \ln \mathcal L > p$, where $p$ is a parameter sensitive to the significance $\alpha$ selected to reject $H_0$ in favour of $H_1$. In order to do so, we should study the distribution of $- 2 \Delta \ln \mathcal L$ under the $H_0$ hypothesis. However, here we are not interested in such elaborated analysis, that is left for future research. The condition $- 2 \Delta \ln \mathcal L>1$ tells us the parameter-space for which we should start to have an evidence for a non-zero cross-correlation of primordial origin.}. In Fig. \ref{fig:likeli_N} we show the log likelihood ratio in Eq. \eqref{eq:likel_N} as a function of $\ell_{\rm max}$ taking  $f^{\rm odd, tts}_{\rm NL} = 10^4$ and a power-law tensor power spectrum with different values of the tensor tilt, as above. Again, using this figure, in Tab. \ref{tab:logL} we derive some scaling formulas for the log likelihood ratio. In general, by matching tables \ref{tab:SNRV} and \ref{tab:logL}, we find that for the parameter-space where the condition $\sum_\ell \, ({\rm SNR}_\ell^V)^2 < 1$ is met, we always get $- 2 \Delta \ln \mathcal L|_{\rm \ell_{\rm max} = 4}<1$ as well. This suggests that by cross-correlating GW $V$ modes with CMB anisotropies we are unable to extract a net signal unless a GW $V$ modes detection is already made.

\begin{table}[!htbp]
\centering
\begingroup
\setlength{\tabcolsep}{8pt} 
\renewcommand{\arraystretch}{2} 
    \begin{tabular}{| c | c | c | c | }
       \cline{2-4}  \multicolumn{1}{c|}{$\times \left( \frac{r}{0.01}\right)^2 \, \Big( \frac{f^{\rm odd, tts}_{\rm NL}}{10^4}\Big)^2 \left(\frac{T_{\rm obs}}{1 \rm yr}\right)$}
          & $\ell_{\rm max} = 2$ & $\ell_{\rm max} = 3$ & $\ell_{\rm max} = 4$ \\
       \cline{1-4}
       \rule{0pt}{1.5\normalbaselineskip}
   $\langle \rm tts \rangle: \quad$ $- 2 \, \Delta \ln \mathcal L|_{n_t = 0}$ = & $0.0104$ &  $0.0180$ & $0.0182$  \\
       \cline{1-4}
        $\,\,\,\,\,\,\,\langle \rm tts \rangle: \quad$ $- 2 \, \Delta \ln \mathcal L|_{n_t = 0.05}$ = & $0.38$ & $0.65$ & $0.66$  \\
        \cline{1-4}
        $\,\,\,\,\langle \rm tts \rangle: \quad$ $- 2 \, \Delta \ln \mathcal L|_{n_t = 0.1}$ = & $14.05$ &  $23.80$ & $24.00$  \\
        \cline{1-4}
    \end{tabular}
    \endgroup
    \caption{Scaling formulas for $- 2 \, \Delta \ln \mathcal L$ for different values of the tensor tilt.} 
    \label{tab:logL}
\end{table}
\FloatBarrier

We now want to answer a different question: assuming that we get a clear non-zero detection in the $V$ modes auto-correlations from $\ell = 2$ up to a certain $\ell_{\rm max}$, with what statistical significance we can confirm this detection using cross-correlations with CMB anisotropies? To answer this question we need the following log likelihood ratio
\begin{equation} \label{eq:likel_S}
- 2 \, \Delta \ln \mathcal L = \sum_{\ell =2}^{\ell_{\rm max}} \,  \bar O_\ell^T \cdot \Sigma_S^{-1} \cdot \bar O_\ell \, ,
\end{equation}
where this time
\begin{equation}
\Sigma_S = \frac{1}{2 \ell +1} \, \begin{pmatrix}
	 \left(N^{VV}_{\ell} +\bar C^{VV}_{\ell} \right)   C^{TT}_{\ell}  & \,\,\,\, \left(N^{VV}_{\ell} + \bar C^{VV}_{\ell} \right)  C^{TE}_{\ell}  \\
                                          \\
	 \left(N^{VV}_{\ell} + \bar C^{VV}_{\ell} \right)  C^{TE}_{\ell}  & \,\,\,\, \left(N^{VV}_{\ell} + \bar C^{VV}_{\ell} \right)  C^{EE}_{\ell} 
	\end{pmatrix} \, .
\end{equation}
This log likelihood ratio allows us to reject the $\tilde H_0$ hypothesis of zero signal in the cross-correlations, but assuming a significant non-zero detection in the self-correlations, i.e. $\bar C^{VV}_{\ell} \neq 0 > N^{VV}_{\ell}$. Under this assumption the quantity $- 2 \Delta \ln \mathcal L$ does not depend significantly on the fixed fiducial early universe parameters and we give its value in terms of $\ell_{\rm max}$ in Tab.~\ref{tab:conditions_cross_S}. We find that if we observe $V$-mode anisotropies with $\ell \geq 2$ attributed to squeezed parity-odd NGs, we should always have additional evidence of their primordial origin by cross-correlating GW $V$ modes with CMB anisotropies. This provides us with a fundamental tool to exclude an alternative origin of the $V$-mode signal, such as instrument systematics. 
\begin{table}[!htbp]
 \centering
\begingroup
\setlength{\tabcolsep}{15pt} 
\renewcommand{\arraystretch}{2} 
   \begin{tabular}{ | c | c | c | c |}
    \cline{2-4} \multicolumn{1}{c|}{}
            & $\ell_{\rm max}  = 2$ & $\ell_{\rm max}  = 3$ & $\ell_{\rm max}  = 4$ \\
       \hline
        $\langle \rm tts \rangle: \quad$   $- 2 \Delta \ln \mathcal L =$   & $5.38$ ($\gtrsim 2.69$) & $12.26$ &  $20.87$ \\
       \hline
    \end{tabular}
    \endgroup
    \caption{Significance of a primordial detection from the $VT$ and $VE$ cross-correlations assuming a signal-dominated measurement of the GW $V$ modes, with $C^{VV}_{\ell} \gg N^{VV}_{\ell}$ for $\ell \leq \ell_{\rm max}$. In parenthesis the case $C^{VV}_{2} \gtrsim N^{VV}_{2}$.}
    \label{tab:conditions_cross_S}
\end{table}
\FloatBarrier

\section{Discussion and conclusion}
\label{sec:conclusions}

In previous decades the main observational tool for cosmology were the CMB anisotropies. However, in recent years we experienced the rise of the multi-messenger cosmology, and in the next decades gravitational-wave cosmology is expected to impact in confirming and unrevealing new information about our Universe.

In this paper we have considered primordial processes which give rise to anisotropies in the stochastic background of gravitational waves expected from inflation. The amount of these anisotropies is related to the strength of squeezed $\langle \rm \rm ttt \rangle$ and $\langle \rm tts \rangle$ primordial NGs, which can be important in models of inflation with isocurvature fields, with
non-Bunch Davies initial states, and in models with alternative symmetry breaking patterns, like solid inflation. Therefore, being able to observe this stochastic background together with anisotropies would allow us to probe new physics during inflation. 

With respect to previous literature on this topic, here we focused on the polarization of GW, and imprints of parity violation. In fact, similarly with what happens for CMB anisotropies, also GW anisotropies can possess different states of polarization. So far, only $I$-mode anisotropies (the so-called unpolarized anisotropies) due to squeezed NGs have been considered, while here for the first time we considered the GW $V$ modes that are linearly related to the local asymmetry between $R$- and $L$-handed GW. 

For this purpose, we first included the chiral basis of GWs and derived more general formulas for the $I$ and $V$ GW modes power spectra and their cross-correlations with CMB anisotropies. Next, we specified observables for parity-odd $\langle \rm tts \rangle$ and $\langle \rm ttt \rangle$ squeezed bispectra that are sensitive to the violation of parity symmetry in the primordial interactions and, according to \cite{Cabass:2021fnw}, can be sizeable (without necessarily introducing parity violation in primordial power spectra) in models of inflation with alternative symmetry breaking patterns, like solid inflation. 

Similarly to what happens when dealing with GW $I$ modes, for the same level of non-Gaussianity parity-odd $\langle \rm tts \rangle$ bispectra in general provide stronger signatures than $\langle \rm ttt \rangle$ in $V$ modes (see e.g. Fig. \ref{fig:Cells}). In particular, $\langle \rm ttt \rangle$  bispectra do not leave signatures in the $V$-mode dipole, which we have forecasted to be the best observational channel. This suggests that parity-odd $\langle \rm tts \rangle$ bispectra are in general better candidates to make an observation in GW experiments. Moreover, differently from the $I$ modes, the induced and astrophysical contamination to the intrinsic (non-Gaussian induced) $V$-mode signal is expected to be negligible, leaving the experimental noise as the only barrier to get a non-zero detection. 

We studied the detection prospects of parity-odd $\langle \rm tts \rangle$ bispectra from the BBO-star configuration. Interestingly, we found that after $5$ years a net observation of a non-Gaussian amplitude $f^{\rm odd, tts}_{\rm NL} \sim 3 \times 10^3$ is possible in the $V$-mode dipole, given the current constraints to the tensor-to-scalar ratio $r$ and without requiring a blue-tilted scale dependence of the gravitational waves power spectrum from inflation. More interestingly, next generation CMB experiments are not expected to provide better constraints on $f^{\rm odd, tts}_{\rm NL}$ assuming a primordial $B$-mode detection with $r \gtrsim 10^{-4}$. This makes our observational channel particularly interesting to explore signatures of parity violation in the $\langle \rm tts \rangle$ bispectrum, which can reveal signatures of new physics during inflation.

Assuming that no parity violation is introduced in the tensor power spectra, our parity-odd NGs are expected not to source any GW $I$-mode anisotropies or $V$-mode monopole. Therefore, a detection of a primordial $I$-mode monopole and $V$-mode anisotropies together with a zero amount of $V$-mode monopole and $I$-mode anisotropies would be a strong indicator for parity-odd bispectra, allowing us to distinguish this parity violation mechanism from others introduced in the literature.

Finally, we found that cross-correlating the $V$-mode signal with CMB anisotropies in general can not increase the parameter space probed by the $V$ modes power spectrum in the case this signal is noise dominated. However, it can always provide an additional evidence of a primordial detection in case where $V$-mode anisotropies are signal-dominated. This finding confirms the utility of the GW-CMB cross-correlations for confirming a $V$-mode detection as found in previous literature for the $I$ modes.

This work aimed to give a first look to the ultimate potential of the GW $V$-mode anisotropies towards probing non-linear parity violation processes in the early universe. This opens up plenty of follow-up investigations. 
\begin{itemize}
    \item In our general results for GW-GW and GW-CMB (cross) correlations, Eqs. \eqref{eq:spectrafirst}-\eqref{eq:spectralast}, we derive formulas in terms of non-Gaussian coefficients and power spectra where we take into account the a-priori difference between a $L$-handed and a $R$-handed gravitational wave. While in this work we fixed $P_t^R = P_t^L$ and we studied parity violation through parity-odd bispectra only, in general an asymmetry $P_t^R \neq P_t^L$ with or without the presence of parity-violation in tensor squeezed bispectra can lead to non-zero GW $I$ or $V$ modes and their cross-correlations with the CMB as well. Studying this instance would be interesting.
    \item In this work we considered the so-called BBO star-configuration. It would be interesting to understand what happens when including in the picture the $2$ outer constellations trailing and leading the star constellation by $120^\circ$ in an earth-like orbit around the sun. It would also be compelling to find the optimal configuration of the BBO constellations that maximizes the sensitivity to $V$-mode anisotropies. 
    \item It would be intriguing to find symmetry breaking patterns that allow for parity-odd bispectra while introducing growth mechanisms for the tensor power spectrum on small scales. In this case, the value of $\overline{\Omega}^I_{\text{\tiny GW}}$ can be enhanced at interferometer scales, therefore allowing for a non-zero detection associated to smaller values of $f^{\rm odd, tts}_{\rm NL}$. 
    \item Enhanced values of $\overline{\Omega}^I_{\text{\tiny GW}}$ at interferometer scales would open up the possibility to observe non-zero $V$ modes anisotropies from parity-odd bispectra with forthcoming GW experiments, such as the LISA-Taiji space-based network or the ET-CE ground-based network. It would be interesting to study the detection prospects of $V$-mode anisotropies for these networks.
    \item In this work we have studied the detection prospects of monopolar parity-odd $\langle \rm tts \rangle$ NGs only. However, as already shown in e.g. \cite{Dimastrogiovanni:2021mfs} for the $I$ modes case, specific realizations of inflation might enhance (beyond tree-level) the quadrupolar parity-odd $\langle \rm tts \rangle$ and $\langle \rm ttt \rangle$ bipectra, motivating a separate study of the detection prospects of these signatures. 
       \item As the GW-CMB cross-correlations that are supposed to matter most in the squeezed PNGs search are those at very large angular scales, it would be interesting to study their statistical properties and detection prospects by considering their real distribution, which is supposed to be a $\chi$-squared distribution with $2 \ell +1$ degrees of freedom. This might provide appreciable differences with the assumption of Gaussian likelihoods. 
       \item Here we studied anisotropies arising in the GW $V$-mode polarization. In principle this analysis could be extended to study anisotropies in the $Q$- and $U$-mode polarizations that describe the GW linear polarization field. As the GW linear polarizations are spin$\pm4$ fields, only $\ell \geq 4$ anisotropies can be non-vanishing (see e.g. the discussion in Ref. \cite{Conneely:2018wis}). These are sourced by non-zero primordial cross-correlations between R- and L-handed tensor perturbations (see e.g. Ref. \cite{Gubitosi:2016yoq}). Also, as recently shown in Ref. \cite{Garoffolo:2022usx}, GW linear polarization anisotropies can be sourced by interaction with the cosmic structures along the line-of-sight.  
      \item Finally, in this work we have considered negligible the amount of induced and astrophysical $V$ modes. While this is justified respectively by the assumption $P_t^R = P_t^L$ and the modeling based on the current observed GW events of astrophysical origin, it is however worth a more detailed investigation, particularly when considering inflationary models where $P_t^R \neq P_t^L$.
    \end{itemize}
We leave all these follow-up investigations to future works.

\paragraph{Acknowledgements}

We want to thank Ameek Malhotra and P. Daniel Meerburg for useful comments on the draft. We acknowledge support from the Netherlands organization for scientific research (NWO) VIDI grant (dossier 639.042.730).

\newpage

\appendix

\section{Squeezed modulation} \label{appen:squeez}

In this appendix we briefly review the mechanism of squeezed modulation of the primordial power spectrum of a generic field $\Phi$ as a result of a non-zero squeezed mixed-bispectrum with the field $\Psi$, i.e. $B_{\Phi_{\bk_1} \Phi_{\bk_2} \Psi_\bq} \neq 0$, with $q\ll k_1, k_2$. As we have seen in Sec. \ref{sec:squeezed}, assuming invariance under rotations, this squeezed bispectrum can be expressed in terms of the absolute value of the short modes $k_1, k_2$, and their scalar product $\hat k_1 \cdot \hat k_2$. We can parametrize the short modes as
\begin{equation} \label{eq:redef_momenta}
\bk_1 = \bk - \bq/2  \, , \qquad \bk_2 = -\bk - \bq/2  \, .
\end{equation}
In this way, the squeezed limit bispectrum will be a function of $k, q$ and $\hat k \cdot \hat q$. Moreover, following the prescriptions of Sec. \ref{sec:squeezed} (in particular Eq. \eqref{eq:def_terms_F}) we define the quantity
\begin{align}
F_{\rm NL}^{\Phi\Phi \Psi}(\bk,\bq) = \frac{B_{\Phi \Phi \Psi}(\bk - \bq/2,-\bk - \bq/2, \bq)|_{q \rightarrow 0}}{P_\Phi(k) P_\Psi(q)} \, .
\end{align}
It is well-known (see e.g. \cite{Jeong:2012df,Dai:2013kra,Dimastrogiovanni:2014ina}) that the presence of a non-zero squeezed bispectrum 
\begin{align}
B_{\Phi \Phi \Psi}(\bk - \bq/2,-\bk - \bq/2, \bq)|_{q \rightarrow 0} \neq 0
\end{align}
implies that the (long) Fourier mode $\Psi_\bq$ induces the following correlation between the short-momenta fields  $\Phi_{\bk_1}$ and $\Phi_{\bk_2}$
\begin{align}
\langle \Phi_{\bk_1} \Phi_{\bk_2} \rangle|_{\Psi_\bq} = (2 \pi)^3 \, \delta(\bk_1 + \bk_2 + \bq) \, P_\Phi(k) \, F_{\rm NL}^{\Phi \Phi \Psi}(\bk,\bq) \, \Psi^*_\bq \, .
\end{align}
This squeezed-induced correlation integrated for all the possible long modes and added to the standard (homogeneous) power spectrum of the variable $\Phi$ gives the following total contribution
\begin{align}
\langle \Phi_{\bk_1} \Phi_{\bk_2} \rangle_{\rm tot} = (2 \pi)^3 \, \delta(\bk_1 + \bk_2) P_\Phi(k) \,  + \int \frac{d^3q}{(2 \pi)^3} \, (2 \pi)^3 \, \delta(\bk_1 + \bk_2 + \bq) P_\Phi(k) \, F_{\rm NL}^{\Phi \Phi \Psi}(\bk,\bq) \, \Psi^*_{\bq} \, .
\end{align}
The primordial two-point correlation function in real space between two points $\bx_1$ and $\bx_2$ for the variable $\Phi$ is given by the inverse Fourier-transform of this previous quantity. Therefore, we have
\begin{align}
\langle \Phi_{\bx_1} \Phi_{\bx_2} \rangle_{\rm tot} = \int \frac{d^3 k_1}{(2 \pi)^3}  \int \frac{d^3 k_2}{(2 \pi)^3} \, e^{i (\bk_1 \cdot \bx_1 + \bk_2 \cdot \bx_2)} \, (2 \pi)^3 \, \delta(\bk_1 + \bk_2) P_\Phi(k) + \nonumber\\
+ \int \frac{d^3 k_1}{(2 \pi)^3}  \int \frac{d^3 k_2}{(2 \pi)^3} \, e^{i (\bk_1 \cdot \bx_1 + \bk_2 \cdot \bx_2)} \, \int \frac{d^3q}{(2 \pi)^3} \, (2 \pi)^3 \, \delta(\bk_1 + \bk_2 + \bq) \, P_\Phi(k) \, F_{\rm NL}^{\Phi \Phi \Psi}(\bk,\bq) \, \Psi^*_{\bq}  \, .
\end{align}
By making the following variable re-definitions
\begin{align}
\mathbf{p_1} = \bk_1 + \bk_2 \, , \qquad \mathbf{p_2}  = \left(\bk_1 - \bk_2\right)/2 \, ,
\end{align}
and
\begin{align}
\bx = \bx_1 - \bx_2 \, , \qquad \bx_c = (\bx_1 + \bx_2)/2 \, ,
\end{align}
we get
\begin{align}
\langle \Phi_{\bx_1} \Phi_{\bx_2} \rangle_{\rm tot} = & \int \frac{d^3 p_2}{(2 \pi)^3}  \, e^{i \mathbf{p_2} \cdot \bx} \, P_\Phi(p_2) + \nonumber \\
& + \int \frac{d^3 p_1}{(2 \pi)^3} \int \frac{d^3 p_2}{(2 \pi)^3} \, e^{i (\mathbf{p_1} \cdot \bx_c + \mathbf{p_2} \cdot \bx)} \, P_\Phi(p_2) \, F_{\rm NL}^{\Phi \Phi \Psi}(\mathbf{p_2},-\mathbf{p_1}) \, \Psi^*_{-\mathbf{p_1}}  \, ,
\end{align}
where by exploiting the Dirac-delta's we have integrated the $\mathbf{p_1}$ momentum in the first term and the $\bq$ momentum in the second term. Now, we just perform the change of variables $\mathbf{p_2} =\bk$ and $\mathbf{p_1} = - \mathbf{q}$ to get the final result
\begin{align} \label{eq:final_mod}
\langle \Phi_{\bx_1} \Phi_{\bx_2} \rangle_{\rm tot} = & \int \frac{d^3 k}{(2 \pi)^3}  \,  e^{i \bk \cdot \bx} \, P_\Phi(k) \, \left[ 1 +  \int \frac{d^3 q}{(2 \pi)^3} \, e^{-i \mathbf{q} \cdot \bx_c}  \, F_{\rm NL}^{\Phi \Phi\Psi}(\bk,\mathbf{q})  \, \Psi^*_{\mathbf{q}} \right]  \nonumber \\
= & \int \frac{d^3 k}{(2 \pi)^3}  \,  e^{i \bk \cdot \bx} \, P_\Phi(k) \, \left[ 1 +  \int \frac{d^3 q}{(2 \pi)^3} \, e^{i \mathbf{q} \cdot \bx_c}  \, F_{\rm NL}^{\Phi \Phi \Psi}(\bk,\mathbf{q})  \, \Psi_{\mathbf{q}} \right] \, ,
\end{align}
where in the last step we have employed the fact that by definition $F_{\rm NL}^{\Phi \Phi \Psi}(\bk,\mathbf{q})$ is real. What we have just derived is a local modulation of the primordial power spectrum of the variable $\Phi$, arising in the midpoint at $\bx_c = (\bx_1 + \bx_2)/2 $. This local modulation, as a function of $\bx_c$, is meaningful provided that the correlation scale is small compared to the smallest scale of the modulating field, which sets the upper bound of the $q$-integration, $q_{\rm max} < k$ for a given $k$.

In this work we are interested in modulations of the tensor power spectra, so $\Phi_{\mathbf{k}} = \gamma^{\lambda}_{\mathbf{k}}$, and a modulation provided by either long scalar and tensor perturbations. By taking $\Psi_{\mathbf{q}} = \zeta_{\mathbf{q}}$ we get the following modulation of the $\lambda$-handed tensor power spectrum from a long scalar mode
\begin{align}
\label{eq:final_mod_scalar}
P^{\lambda, \, \rm tot}_{t}(\vk,\vx_c) = P^\lambda_{t}(k)\left[1+\int \frac{d^3q}{(2\pi)^3} \, e^{i\vq\cdot\vx_c} \,  F^{\lambda \lambda, \rm tts}_{\rm NL}(\vk,\vq) \, \zeta(\vq) \right] \, ,
\end{align}
where we defined
\begin{align}
\label{eq:fnltts_def_app}
   F_{\rm NL}^{\lambda \lambda, \rm tts}(\vk,\vq) = \frac{B^{\lambda \lambda}_{\rm tts}(\vk-\vq/2,-\vk-\vq/2,\vq)|_{q \rightarrow 0}}{P^{\lambda}_{t}(k) \, P_{s}(q)} \, .
\end{align}
On the other hand, by taking $\Psi_{\mathbf{q}} = \gamma^{\lambda'}_{\mathbf{q}}$, we get the following modulation of the $\lambda$-handed tensor power spectrum from a long $\lambda'$-handed tensor mode
\begin{align}
\label{eq:final_mod_tensor}
P^{\lambda, \, \rm tot}_{t}(\vk,\vx_c) = P^\lambda_{t}(k)\left[1+\int \frac{d^3q}{(2\pi)^3}\,e^{i\vq\cdot\vx_c} \, \sum_{\lambda' = R/L} F^{\lambda \lambda\lambda', \rm ttt}_{\rm NL}(\vk,\vq) \, \gamma^{\lambda'}(\vq) \right] \, ,
\end{align}
where we defined
\begin{align}
\label{eq:fnlttt_def_app}
   F_{\rm NL}^{\lambda\lambda \lambda', \rm ttt}(\vk,\vq) = \frac{B^{\lambda \lambda \lambda'}_{\rm ttt}(\vk-\vq/2,-\vk-\vq/2,\vq)|_{q \rightarrow 0}}{P^{\lambda}_{t}(k) \, P^{\lambda'}_{t}(q) } \, .
\end{align}
Eqs. \eqref{eq:final_mod_scalar_main} and \eqref{eq:final_mod_tensor_main} follow once we switch to the dimensionless tensor power spectra.

\section{Spin-raising and lowering operators} \label{appen:spin_operators}

Here, we briefly review the definitions of the spin-raising and lowering operators, giving an example on how we can use them to define the spin-weighted spherical harmonics. We refer to e.g.~\cite{Zaldarriaga:1996xe} for more details. The spin raising $\up$ and lowering $\down$ operators acting on a generic spin-$s$ function ${}_sf(\theta,\phi) $ defined on a 2D sphere are given by
\begin{align}
\up {}_sf(\theta,\phi) &= -\sin^{s} \theta
\left[\partial_\theta + i\csc \theta
\partial_\phi \right]\sin^{-s} \theta \,\,
{}_s f(\theta,\phi) \, , \nonumber\\
\down {}_sf(\theta,\phi) &= -\sin^{-s} \theta 
\left[ \partial_\theta - i\csc \theta
\partial_\phi \right]\sin^{s} \theta  \,\,{}_s f(\theta,\phi) \, .
\label{eq:edth}
\end{align}
In particular, the new functions $\up {}_sf(\theta,\phi)$ and $\down {}_sf(\theta,\phi)$ have spin $s+1$ and $s-1$, respectively. For example, the spin raising and lowering operators acting twice on a generic
spin-$\pm 2$ function ${}_{\pm 2}f(\theta,\phi)$ which is factorized as ${}_{\pm 2}f(\theta,\phi) = {}_{\pm 2}\tilde{f}(\mu) \, e^{i m \phi}$ (e.g. the CMB linear polarization
fields) can be expressed as
\begin{align}
\down^2 {}_2f(\theta,\phi)&=
\left(-\partial_\mu + \f{m}{1-\mu^2}\right)^2 \left[
(1-\mu^2) \, {}_2 f(\mu,\phi)\right] \, , \nonumber \\ 
\up^2 {}_{-2}f(\theta,\phi) &=
\left(-\partial_\mu - \f{m}{1-\mu^2}\right)^2 \left[(1-\mu^2) \, {}_{-2}
f(\mu,\phi)\right] \, , 
\label{eq:operators1}
\end{align}
where $\mu \equiv \cos \theta$. In this way, just acting with a differential operator, we can easily define spin-0 quantities starting from spin-2 ones. This procedure is used in the case of CMB to switch from the $P^{\pm}$ spin-$\pm2$ linear polarization fields to the $E$ and $B$ modes, which are spin-0 fields.

As an example, using Eqs. \eqref{eq:edth}, we can express the spin-weighted spherical harmonic functions on a 2D sphere, ${}_sY_{\ell m}(\theta,\phi)$, in terms of the common spherical harmonics ${}_{0}Y_{\ell m}(\theta, \phi) = Y_{\ell m}(\theta, \phi)$ by acting with the spin raising/lowering operator as 
\begin{align} \label{eq:spin_harm_sph}
{}_sY_{\ell m}(\theta,\phi) &=
\left[\f{(\ell-s)!}{(\ell+s)!}\right]^{\f{1}{2}}\up^s  Y_{\ell m}(\theta,\phi)
\ \  (0 \leq s \leq \ell) \, , \nonumber  \\ 
{}_sY_{\ell m}(\theta,\phi) &=
\left[\f{(\ell+s)!}{(\ell-s)!}\right]^{\f{1}{2}}(-1)^s 
\down^{-s} Y_{\ell m}(\theta,\phi)
\ \  (-\ell \leq s \leq 0) \, .
\end{align}
It is possible to show the validity of the following relations
\begin{align}
\up {}_sY_{\ell m}(\theta,\phi) &=\left[(\ell-s)(\ell+s+1)\right]^{\f{1}{2}}
\,{}_{s+1}Y_{\ell m}(\theta,\phi) \, , \nonumber \\
\down {}_sY_{\ell m}(\theta,\phi) &=-\left[(\ell+s)(\ell-s+1)\right]^{\f{1}{2}}
\,{}_{s-1}Y_{\ell m}(\theta,\phi) \, , \nonumber \\
\down\up {}_sY_{\ell m}(\theta,\phi) &=-(\ell-s)(\ell+s+1)
\,{}_{s}Y_{\ell m}(\theta,\phi) ~m \, ,
\label{eq:propYs}
\end{align}
which can be used to derive the following explicit expression of the spin-weighted spherical harmonics
\begin{eqnarray}
{}_sY_{\ell m}(\theta, \phi) &=& e^{im\phi}
\left[\f{(\ell+m)!(\ell-m)!}{(\ell+s)!(\ell-s)!}
\f{(2\ell+1)}{4\pi}\right]^{1/2}
\sin^{2\ell}(\theta/2) \nonumber \\
&&\times \sum_r {\ell-s \choose r}{\ell+s \choose r+s-m}
(-1)^{\ell-r-s+m}{\rm cot}^{2r+s-m}(\theta/2) \, .
\label{eq:expl}
\end{eqnarray}

\section{Spin-weighted spherical harmonics: integration and examples} \label{app:Wigner}

In this appendix, we give some useful formulas about spin-weighted spherical harmonics and their integration. We will use $(\theta, \phi)$ or $\hat x$ to denote a given direction on the $2D$ sphere and $d^2 \hat n$ or $d^2 \Omega_x$ to indicate the infinitesimal solid angle on the sphere. We will also review some technical computations of this work. The formulas we provide here are crucial to simplify the expressions for spherical harmonics coefficients when dealing with primordial perturbations from inflation. We refer the reader to e.g. \cite{Okamoto:2002ik,Komatsu:2003iq, Liguori:2005rj,Shiraishi:2012} for more details.

\subsection*{Basic relations}

We start by giving the orthogonality and completeness conditions for the spin-weighted spherical harmonics ${}_sY_{\ell m}(\hat x)$ as 
\begin{align} \label{eq:orto_harm}
\int d^2 \Omega_x \,\, {}_s Y_{\ell m}^*(\hat x)
\, {}_s Y_{\ell' m'}(\hat x) &= \delta_{\ell, \ell'} \, \delta_{m, m'} \, , \\
\sum_{\ell m} \, \, {}_s Y_{\ell m}^*(\hat x) \,
{}_s Y_{\ell m}(\hat x')
&= \delta(\hat x - \hat x') \, ,
\end{align} 
as well as the following properties regarding the transformation under conjugate and parity
\begin{align}
{}_sY^*_{\ell m}(\theta, \phi) &= (-1)^{s + m}{}_{-s}Y_{\ell -m}(\theta, \phi) \, , \nonumber \\
{}_s Y_{\ell m}(\pi - \theta, \phi + \pi) &= (-1)^{\ell} \, {}_{-s} Y_{\ell m}(\theta, \phi) \, . 
\label{eq:parity_Y}\end{align}
We can decompose a weighted spherical harmonics evaluated at an angle between two vectors $\hat k \cdot \hat q$ as (see e.g. \cite{Okamoto:2002ik})
\ba \label{eq:gen_add_rel}
{}_{s} Y_{\ell m}(\hat k \cdot \hat q) = &\sqrt{\frac{4 \pi}{2 \ell +1}} \, (-1)^s \, \sum_{M}  \, {}_{s} Y_{\ell M}(\hat k)  \,  {}_{-m} Y^*_{\ell M}(\hat q)  \nonumber \\
 = &\sqrt{\frac{4 \pi}{2 \ell +1}} \, (-1)^s \, \sum_{M}  \, {}_{s} Y_{\ell M}(\hat q)  \,  {}_{-m} Y^*_{\ell M}(\hat k) \, ,
\ea
which is a variant of the so-called generalized addition relation. 

Another important result is the so-called plane-waves decomposition in terms of spin-0 spherical harmonics 
\ba \label{eq:plane-wave}
e^{i \vec q \cdot \vec x} = & \sum_\ell \sqrt{4 \pi (2 \ell +1)} \, i^\ell  \, j_\ell(q x) \, Y_{\ell 0}(\hat x \cdot \hat q) \\
= & \sum_\ell 4 \pi \, i^\ell  \, j_\ell(q x)  \sum_M \, Y_{\ell M}(\hat q) \, Y^*_{\ell M}(\hat x)\, .
\ea
As a last useful equation, we give the Clebsch-Gordan relation
\begin{align} \label{eq:rel_wigner}
\prod_{i = 1}^2 {}_{s_i}Y_{\ell_i m_i}(\hat x)&= \sum_{\ell_3 m_3 s_3} \,  {}_{s_3}Y^*_{\ell_3 m_3}(\hat x) \, \sqrt{\f{(2 \ell_1 + 1) (2 \ell_2 + 1) (2 \ell_3 + 1)}{4 \pi}}   \nonumber\\
& \times \begin{pmatrix}
	\ell_1 & \ell_2 & \ell_3 \\
	-s_1 & -s_2 & -s_3
	\end{pmatrix}\begin{pmatrix}
	\ell_1 & \ell_2 & \ell_3 \\
	m_1 & m_2 & m_3
	\end{pmatrix} \, ,
\end{align}
which can be used to compose the angular momenta of two separate spherical harmonics evaluated at the same angle. Together with \eqref{eq:plane-wave}, we can employ this result to isolate the radial and angular dependencies of a given expression (see e.g. \cite{Hu:1997hp} for more on this aspect). 

In Eq. \eqref{eq:rel_wigner} we have introduced the Wigner 3-j symbols, that are related to the well-known Clebsch-Gordan coefficients
\be \label{eq:CG}
\mathcal C^{\ell_2 m_3}_{\ell_1 m_1 \ell_2 m_2} = \langle \ell_1  m_1 \ell_2 m_2| \ell_3 m_3  \rangle
\ee
through (see e.g. \cite{Shiraishi:2012})
\be \label{eq:3jsymbols-CG}
\begin{pmatrix}
	\ell_1 & \ell_2 & \ell_3 \\
	m_1 & m_2 & - m_3
	\end{pmatrix} =  \frac{(-1)^{\ell_1 - \ell_2 + m_3}}{\sqrt{2 \ell_3 + 1}} \, \mathcal C^{\ell_2 m_3}_{\ell_1 m_1 \ell_2 m_2} \, .
\ee
Therefore, the 3-j symbols of the form \eqref{eq:3jsymbols-CG} vanish unless the selection rules are satisfied as follows
\ba
&|m_1| \leq \ell_1 \,, \qquad  |m_2|\leq \ell_2 \,, \qquad |m_3|\leq \ell_3 \,, \qquad  m_1 +m_2 =m_3 \, ,\\
&|\ell_1 − \ell_2| \leq \ell_3 \leq \ell_1 + \ell_2 \quad \mbox{(the triangle condition)} \, , \qquad \ell_1 +\ell_2 + \ell_3 \in Z \, .
\ea
Some useful properties of the Wigner 3-j symbols are the transformation rules 
\begin{align} \label{eq:lwigner}
 \begin{pmatrix}
	\ell_1 & \ell_2 & \ell_3 \\
	m_1 & m_2 & m_3
	\end{pmatrix} =& (-1)^{\sum_i \ell_i}\begin{pmatrix}
	\ell_1 & \ell_2 & \ell_3 \\
	 - m_1 & - m_2 & - m_3
	\end{pmatrix}  \nonumber \\
	=& (-1)^{\sum_i \ell_i}\begin{pmatrix}
	\ell_2 & \ell_1 & \ell_3 \\
	  m_2 &  m_1 & m_3
	\end{pmatrix} 
\end{align}
under the $m_i$-sign inversion and odd permutations of columns. On the contrary, these symbols are left invariant by even permutations of columns. 

Another useful property of the Wigner 3-j symbols is the orthogonality condition
\begin{align} \label{eq:sum_m_wigner}
\sum_{m_1, m_2}  \begin{pmatrix}
	\ell_1 & \ell_2 & \ell_3 \\
	m_1 & m_2 & m_3
	\end{pmatrix}\begin{pmatrix}
	\ell_1 & \ell_2 & \ell'_3 \\
	m_1 & m_2 & m'_3
	\end{pmatrix}  = (2 \ell_3 + 1)^{-1} \, \delta_{\ell_3, \ell'_3} \, \delta_{m_3, m'_3} \, .
\end{align}
More properties of the Wigner 3-j symbols can be found in \cite{Shiraishi:2012}.

\subsection*{Integration}

We define the quantity ${}_{s_1 s_2 s_3} \mathcal G_{\ell_1 \ell_2 \ell_3}^{m_1 m_2 m_3}$, which is known as "generalized" Gaunt integral and it represents the angular integral of the product of three (weighted) spherical harmonics. This can be written in terms of Wigner 3-j symbols as (see e.g. \cite{Komatsu:2003iq, Liguori:2005rj})
\begin{align} \label{eq:Gaunt_integral}
{}_{s_1 s_2 s_3}\mathcal G_{\ell_1 \ell_2 \ell_3}^{m_1 m_2 m_3} &= \int d^2 \Omega_x \, {}_{s_1}Y_{\ell_1 m_1}(\hat x) \, {}_{s_2}Y_{\ell_2 m_2}(\hat x)  \,{}_{s_3}Y_{\ell_3 m_3}(\hat x) \nonumber\\
&= \sqrt{\f{(2 \ell_1 + 1) (2 \ell_2 + 1) (2 \ell_3 + 1)}{4 \pi}}   \begin{pmatrix}
	\ell_1 & \ell_2 & \ell_3 \\
	-s_1 & -s_2 & -s_3
	\end{pmatrix}\begin{pmatrix}
	\ell_1 & \ell_2 & \ell_3 \\
	m_1 & m_2 & m_3
	\end{pmatrix} \, .
\end{align}

\subsection*{Examples}

Here we derive some results of the main text. We will take only the GW $I$-mode and the generic CMB $X$-mode. When including all the other polarization modes in the picture, the results follow trivially from those derived here apart for minimal substitutions of CMB transfer functions and sign switches. 

\vspace{0.5cm}
\textbf{\centering Spherical harmonics coefficients of monopolar $\langle \gamma \gamma \zeta \rangle$}
\vspace{0.5cm}

\noindent By taking
\begin{align}
    F_{\rm NL}^{\lambda,\rm tts}(\vk,\vq) = \, \sqrt{4 \pi} \, Y_{0 0}(\hat k \cdot \hat q)  \, f_{\rm NL}^{\lambda, \rm tts}(k,q) = f_{\rm NL}^{\lambda, \rm tts}(k,q) 
\end{align}
and inserting Eq. \eqref{eq_deltaI_tts} into  \eqref{eq:def_sphe_I}, we get
\begin{align}
\delta_{\ell m}^{\rm GW, I, \, tts} = \frac{1}{\mathcal A_t(k)} \, \int\frac{d^3q}{(2\pi)^3}\, \int d^2 \hat n \,  Y^*_{\ell m}(\hat n) \, e^{i d \vq \cdot \hat{n}} \,\left[ \mathcal A^R_t(k) \, f_{\rm NL}^{R, \rm tts}(k,q) + \mathcal A^L_t(k) \, f_{\rm NL}^{L, \rm tts}(k,q) \right]\, \zeta_{\bq} \, .
\end{align}
By expanding the plane-wave using \eqref{eq:plane-wave} and performing the $\hat n$-angular integration with \eqref{eq:orto_harm} we get the result
\begin{align} \label{eq:ttsmono_example}
\delta_{\ell m}^{\rm GW, I, \, tts} = \frac{4 \pi}{\mathcal A_t(k)} \, i^\ell \, \int \frac{d^3q}{(2\pi)^3}\, Y^*_{\ell m}(\hat q) \, j_\ell(q d) \,\left[ \mathcal A^R_t(k) \, f_{\rm NL}^{R, \rm tts}(k,q) + \mathcal A^L_t(k) \, f_{\rm NL}^{L, \rm tts}(k,q) \right]\, \zeta_{\bq} \, .
\end{align}
In an analogous way we get also the spherical harmonics coefficients of the induced contributions, Eq. \eqref{eq:delta_GW_ind}.

\vspace{0.5cm}
\textbf{\centering Spherical harmonics coefficients of quadrupolar $\langle \gamma \gamma \zeta \rangle$}
\vspace{0.5cm}

\noindent By taking
\begin{align}
    F_{\rm NL}^{\lambda,\rm tts}(\vk,\vq) = \sqrt{\frac{4 \pi}{ 5}} \, Y_{2 0}(\hat k \cdot \hat q) \, f_{\rm NL}^{\lambda, \rm tts}(k,q) 
\end{align}
and inserting Eq. \eqref{eq_deltaI_tts} into  \eqref{eq:def_sphe_I}, we get
\begin{align}
\delta_{\ell m}^{\rm GW, I, \, tts} = \frac{1}{\mathcal A_t(k)} \sqrt{\frac{4 \pi}{ 5}} \, \int\frac{d^3q}{(2\pi)^3}\, &\int d^2 \hat n \,  Y^*_{\ell m}(\hat n) \, Y_{2 0}(\hat k \cdot \hat q) \, e^{i d \vq \cdot \hat{n}} \times \nonumber \\
& \times \left[ \mathcal A^R_t(k) \, f_{\rm NL}^{R, \rm tts}(k,q) + \mathcal A^L_t(k) \, f_{\rm NL}^{L, \rm tts}(k,q) \right]\, \zeta_{\bq} \, .
\end{align}
We can expand the plane-wave using \eqref{eq:plane-wave} and decompose the argument dependence of $Y_{2 0}(\hat k \cdot \hat q)$ using \eqref{eq:gen_add_rel}. We get 
\begin{align}
\delta_{\ell m}^{\rm GW, I, \, tts} = &\frac{1}{\mathcal A_t(k)} \frac{16 \pi^2}{ 5} \, \sum_{J,M, m'} \, i^L \left[\int d^2 n \, Y_{\ell m}(\hat n)  \, Y_{J M}(\hat n) \, Y_{2 m'}(\hat n) \right]^* \times \nonumber \\
& \times \int\frac{d^3q}{(2\pi)^3}  \,  Y_{J M}(\hat q) \, Y_{2 m'}(\hat q)  \, j_J(q d) \, \left[ \mathcal A^R_t(k) \, f_{\rm NL}^{R, \rm tts}(k,q) + \mathcal A^L_t(k) \, f_{\rm NL}^{L, \rm tts}(k,q) \right]\, \zeta_{\bq} \, ,
\end{align}
where we have used $Y^*_{2 m'}(\hat k) = Y^*_{2 m'}(-\hat n) = Y^*_{2 m'}(\hat n)$. From here, we can perform the $\hat n$-angular integration by exploiting the Gaunt integral in Eq. \eqref{eq:Gaunt_integral}. We get
\begin{align}  \label{eq:ttsquad_example}
\delta_{\ell m}^{\rm GW, I, \, tts} = \frac{1}{\mathcal A_t(k)} \frac{16 \pi^2}{5} \sqrt{\frac{5}{4 \pi}} \, \sum_{J, M, m'} \, i^J \, \sqrt{(2 J + 1)(2 \ell + 1)} \, \, \begin{pmatrix}
	\ell & J & 2 \\
	0 & 0 & 0
	\end{pmatrix}\begin{pmatrix}
	\ell & J & 2 \\
	m & M & m'
	\end{pmatrix} \,  \times \nonumber \\
\times \int \frac{d^3q}{(2\pi)^3}\, Y_{J M}(\hat q) \, Y_{2 m'}(\hat q) \, j_J(q d) \, \left[ \mathcal A^R_t(k) \, f_{\rm NL}^{R, \rm tts}(k,q) + \mathcal A^L_t(k) \, f_{\rm NL}^{L, \rm tts}(k,q) \right] \, \zeta_{\bq} \, .
\end{align}

\vspace{0.5cm}
\textbf{\centering Spherical harmonics coefficients of quadrupolar $\langle \gamma \gamma \gamma \rangle$}
\vspace{0.5cm}

\noindent By taking
\begin{align}
F_{\rm NL}^{\lambda \lambda',\rm ttt}(\vk,\vq) = - \frac{2}{3} \, \sqrt{\frac{6 \pi}{5}} \, {}_{\pm 2}Y_{20}(\hat k \cdot \hat q)  \, f_{\rm NL}^{\lambda \lambda', \rm ttt}(k,q) = - \epsilon_{ij}^{\lambda'}(q)\,k^i k^j \, f_{\rm NL}^{\lambda \lambda', \rm ttt}(k,q)  
\end{align}
and inserting Eq. \eqref{eq_deltaI_tts} into  \eqref{eq:def_sphe_I}, we get
\begin{align}
\delta_{\ell m}^{\rm GW, I, \, ttt} = -  \frac{1}{\mathcal A_t(k)} \frac{2}{3} \, \sqrt{\frac{6 \pi}{5}} & \, \int\frac{d^3q}{(2\pi)^3} \, \int d^2 \hat n \,  Y^*_{\ell m}(\hat n)  \, e^{i d \vq \cdot \hat{n}} \times \nonumber \\
& \times \left\{ {}_{- 2}Y_{20}(\hat k \cdot \hat q) \,
\left[\mathcal A^R_{t}(k) \, f_{\rm NL}^{RR,\rm ttt}(k,q) + \mathcal A^L_{t}(k) \, f_{\rm NL}^{LR,\rm ttt}(k,q) \right]  \gamma_{\vq}^{R} + \right. \nonumber  \\
&\left. \qquad + {}_{+2}Y_{20}(\hat k \cdot \hat q) \, \left[\mathcal A^L_{t}(k) \, f_{\rm NL}^{LL,\rm ttt}(k,q) + \mathcal A^R_{t}(k) \, f_{\rm NL}^{RL,\rm ttt}(k, q) \right] \gamma_{\vq}^{L} \right\} \, . 
\end{align}
By using the fact that ${}_{\pm2}Y_{20}(\hat k \cdot \hat q) = {}_{\pm2}Y_{20}^*(\hat n \cdot \hat q)$ and expanding the plane-wave using \eqref{eq:plane-wave}, we get
\begin{align} \label{eq:ttt_int}
\delta_{\ell m}^{\rm GW, I, \, ttt} = -  &\sum_{J} \frac{i^J}{\mathcal A_t(k)} \frac{2}{3} \, \sqrt{\frac{6 \pi}{5}}   \, \sqrt{4 \pi (2 J +1)} \,  \int\frac{d^3q}{(2\pi)^3} \, \int d^2 \hat n \,  Y^*_{\ell m}(\hat n)  \, j_J(q d) \, \times \nonumber \\
& \times \left\{ {}_{- 2}Y^*_{20}(\hat n \cdot \hat q) \, Y_{J 0}(\hat n \cdot \hat q)
\left[\mathcal A^R_{t}(k) \, f_{\rm NL}^{RR,\rm ttt}(k,q) + \mathcal A^L_{t}(k) \, f_{\rm NL}^{LR,\rm ttt}(k,q) \right]  \gamma_{\vq}^{R} + \right. \nonumber  \\
&\left. \qquad + {}_{+ 2}Y^*_{20}(\hat n \cdot \hat q) \, Y_{J 0}(\hat n \cdot \hat q) \left[\mathcal A^L_{t}(k) \, f_{\rm NL}^{LL,\rm ttt}(k,q) + \mathcal A^R_{t}(k) \, f_{\rm NL}^{RL,\rm ttt}(k, q) \right] \gamma_{\vq}^{L} \right\} \, . 
\end{align}
Now, we can use \eqref{eq:rel_wigner} to compose the angular momentum of ${}_{\pm 2}Y^*_{20}(\hat n \cdot \hat q)$ with that of $Y_{L 0}(\hat n \cdot \hat q)$ coming from the plane-wave expansion. After making this composition and performing appropriate changes of summation-variable in each new $J$-summation, we can express the final result as a weighted sum of spherical Bessel functions with momenta between $|J − 2|$ and $J + 2$. This can be re-expressed in terms of a single $j_J(q d)$ using the following recursion relation of spherical Bessel functions 
\ba
\frac{j_J(x)}{x} = \frac{1}{2 \ell +1} \left[j_{J-1}(x) + j_{J+1}(x) \right] \, .
\ea
After some straightforward work, we find that the following composition holds
\ba
&\sum_J i^J \, \sqrt{4 \pi (2 J +1)} \, {}_{\pm 2}Y^*_{20}(\hat n \cdot \hat q) \, Y_{J 0}(\hat n \cdot \hat q) \nonumber\\
&= - \sum_J i^J \, \sqrt{4 \pi (2 J +1)} \,  \sqrt{\frac{15(J+2)!}{32 \pi (J-2)!}} \, \frac{j_J(q d)}{(qd)^2} \, {}_{\pm 2}Y^*_{J 0}(\hat n \cdot \hat q) \nonumber \\
&= - \sum_{J, M} i^J \, 4 \pi \,  \sqrt{\frac{3(J+2)!}{8 (J-2)!}} \sqrt{\frac{5}{4\pi}} \, \frac{j_J(q d)}{(qd)^2} \, {}_{\pm 2}Y^*_{J M}(\hat q) \, Y_{J M}(\hat n) \, ,
\ea 
where in the last step we used Eq. \eqref{eq:gen_add_rel} to decompose ${}_{\pm 2}Y^*_{L 0}(\hat n \cdot \hat q)$. By substituting this result in Eq. \eqref{eq:ttt_int} and performing the remaining (now trivial) $\hat n$-integration, we get the final result
\begin{align} \label{eq:tttquad_example}
    \delta^{\rm GW, I, ttt}_{\ell m} = \frac{2 \pi}{\mathcal A_t(k)} \,  i^\ell \, \sqrt{\frac{(\ell+2)!}{(\ell-2)!}} \,& \int \frac{d^3q}{(2\pi)^3} \, \frac{j_\ell(q d)}{(qd)^2} \times \nonumber \\
   & \times \left\{ {}_{- 2}Y_{\ell m}^*(\hat q) \,
\left[\mathcal A^R_{t}(k) \, f_{\rm NL}^{RR,\rm ttt}(k,q) + \mathcal A^L_{t}(k) \, f_{\rm NL}^{LR,\rm ttt}(k,q) \right]  \gamma_{\vq}^{R} + \right. \nonumber  \\
&\left. \qquad + {}_{+2}Y_{\ell m}^*(\hat q) \, \left[\mathcal A^L_{t}(k) \, f_{\rm NL}^{LL,\rm ttt}(k,q) + \mathcal A^R_{t}(k) \, f_{\rm NL}^{RL,\rm ttt}(k, q) \right] \gamma_{\vq}^{L} \right\} \,.
\end{align}

\vspace{0.5cm}
\textbf{\centering GW-GW from monopolar $\langle \gamma \gamma \zeta \rangle$}
\vspace{0.5cm}

\noindent Auto-correlating two quantities \eqref{eq:ttsmono_example}\footnote{Here and in the following we will consider only the diagonal $\ell_1 = \ell_2 = \ell$ correlations. In fact, it is widely known that invariance under rotations implies that only diagonal correlations are non-zero.}, we get
\ba
C_{\ell}^{I I} =& \frac{1}{2 \ell+1} \, \sum_m \langle \delta^{{\rm GW}, \, I}_{\ell m} \delta^{{\rm GW}, \, I *}_{\ell m} \rangle \nonumber \\
= & \frac{1}{2 \ell+1} \, \sum_m \left[\frac{4 \pi}{\mathcal A_t(k)}\right]^2 \, \int \frac{d^3q}{(2\pi)^3}\, \int \frac{d^3p}{(2\pi)^3} \, Y^*_{\ell m}(\hat q) \, Y_{\ell m}(\hat p) \, j_\ell(q d) \, j_\ell(p d) \,\, \langle \zeta(\vec{q}) \, \zeta^*(\vec{p})\rangle \times \nonumber \\
& \times \left[ \mathcal A^R_t(k) \, f_{\rm NL}^{R, \rm tts}(k,q) + \mathcal A^L_t(k) \, f_{\rm NL}^{L, \rm tts}(k,q) \right]\, \left[ \mathcal A^R_t(k) \, f_{\rm NL}^{R, \rm tts}(k,p) + \mathcal A^L_t(k) \, f_{\rm NL}^{L, \rm tts}(k,p) \right] \, .
\ea
Here we can exploit the definition of scalar power spectrum to integrate out one of the two momenta using the Dirac-delta, and get
\ba
C_{\ell}^{I I} =& \frac{1}{2 \ell+1} \, \sum_m \left[\frac{4 \pi}{\mathcal A_t(k)}\right]^2 \, \int dq \, q^2 \, \int \frac{d^2\hat q}{(2\pi)^3} \, Y^*_{\ell m}(\hat q) \, Y_{\ell m}(\hat q)  \, \left[j_\ell(q d)\right]^2 \, P_s(q)  \times \nonumber \\
& \qquad \qquad \times \left[ \mathcal A^R_t(k) \, f_{\rm NL}^{R, \rm tts}(k,q) + \mathcal A^L_t(k) \, f_{\rm NL}^{L, \rm tts}(k,q) \right]^2 \, .
\ea
We can now easily integrate the remnant $\hat q$-angular dependence getting the final result
\ba
C_{\ell}^{I I} = & \frac{2}{\pi} \frac{1}{\mathcal A^2_t(k)} \, \int dq \, q^2 \, \left[j_\ell(q d)\right]^2 \, P_s(q) \, \left[ \mathcal A^R_t(k) \, f_{\rm NL}^{R, \rm tts}(k,q) + \mathcal A^L_t(k) \, f_{\rm NL}^{L, \rm tts}(k,q) \right]^2 \, .
\ea
By employing \eqref{eq:power_inflation} we can switch to dimensionless amplitudes
\ba
C_{\ell}^{I I} = &  \frac{4 \pi}{\mathcal A^2_t(k)} \, \int \frac{dq}{q} \, \left[j_\ell(q d)\right]^2 \, \mathcal A_s(q) \, \left[ \mathcal A^R_t(k) \, f_{\rm NL}^{R, \rm tts}(k,q) + \mathcal A^L_t(k) \, f_{\rm NL}^{L, \rm tts}(k,q) \right]^2 \, .
\ea

\vspace{0.5cm}
\textbf{\centering GW-GW from quadrupolar $\langle \gamma \gamma \zeta \rangle$}
\vspace{0.5cm}

\noindent Auto-correlating two quantities \eqref{eq:ttsquad_example} we get 

\begin{align}
C_{\ell}^{I I} =& \frac{1}{2 \ell+1} \, \sum_m \langle \delta^{{\rm GW}, \, I}_{\ell m} \delta^{{\rm GW}, \, I *}_{\ell m} \rangle \nonumber \\
= & \frac{1}{2 \ell+1} \, \sum_m
\left[\frac{1}{\mathcal A_t(k)}\right]^2 \left(\frac{16 \pi^2}{5}\right)^2 \frac{5}{4 \pi} \, \sum_{J, M, m'}  \sum_{J', M', m'' }\, i^{J-J'} \, (2 \ell + 1) \, \sqrt{(2 J + 1)(2 J' + 1)} \times \nonumber \\
& \times \begin{pmatrix}
	\ell & J & 2 \\
	0 & 0 & 0
	\end{pmatrix}\begin{pmatrix}
	\ell & J & 2 \\
	m & M & m'
	\end{pmatrix} \,  \times \begin{pmatrix}
	\ell & J' & 2 \\
	0 & 0 & 0
	\end{pmatrix}\begin{pmatrix}
	\ell & J' & 2 \\
	m & M' & m''
	\end{pmatrix} \,  \nonumber \\
&\times \int \frac{d^3q}{(2\pi)^3}\, \int \frac{d^3p}{(2\pi)^3} \, Y_{J M}(\hat q) \, Y_{2 m'}(\hat q) \, j_J(q d) \, Y^*_{J' M'}(\hat p) \, Y^*_{2 m''}(\hat p) \, j_{J'}(p d) \times \nonumber \\
&\times \left[ \mathcal A^R_t(k) \, f_{\rm NL}^{R, \rm tts}(k,q) + \mathcal A^L_t(k) \, f_{\rm NL}^{L, \rm tts}(k,q) \right] \, \left[ \mathcal A^R_t(k) \, f_{\rm NL}^{R, \rm tts}(k,p) + \mathcal A^L_t(k) \, f_{\rm NL}^{L, \rm tts}(k,p) \right] \langle \zeta_{\bq} \, \zeta_{\mathbf p}^*\rangle \, .
\end{align}
By exploiting the definition of scalar power spectrum we can integrate out one of the two momenta using the Dirac-delta, and get
\begin{align} \label{eq:quad_app}
C_{\ell}^{I I} =& \frac{1}{2 \ell+1} \, \sum_m
\left[\frac{1}{\mathcal A_t(k)}\right]^2 \left(\frac{16 \pi^2}{5}\right)^2 \frac{5}{4 \pi} \, \sum_{J, M, m'}  \sum_{J', M', m'' }\, i^{J-J'} \, (2 \ell + 1) \, \sqrt{(2 J + 1)(2 J' + 1)} \times \nonumber \\
& \times \begin{pmatrix}
	\ell & J & 2 \\
	0 & 0 & 0
	\end{pmatrix}\begin{pmatrix}
	\ell & J & 2 \\
	m & M & m'
	\end{pmatrix} \,  \times \begin{pmatrix}
	\ell & J' & 2 \\
	0 & 0 & 0
	\end{pmatrix}\begin{pmatrix}
	\ell & J' & 2 \\
	m & M' & m''
	\end{pmatrix} \, \times \nonumber \\
&\times \int \frac{d^3q}{(2\pi)^3}\, Y_{J M}(\hat q) \, Y_{2 m'}(\hat q) \, Y^*_{J' M'}(\hat q) \, Y^*_{2 m''}(\hat q) \, \, j_{J}(q d) \, j_{J'}(q d) \, \times \nonumber \\
&\times \, P_s(q) \, \left[ \mathcal A^R_t(k) \, f_{\rm NL}^{R, \rm tts}(k,q) + \mathcal A^L_t(k) \, f_{\rm NL}^{L, \rm tts}(k,q) \right]^2  \, .
\end{align}
By exploiting Eq. \eqref{eq:rel_wigner} we can rewrite the following product 
\ba
Y^*_{J' M'}(\hat q) \, Y^*_{2 m''}(\hat q) = \sum_{J'', M''} Y_{J'' M''}(\hat q) \, \sqrt{\frac{5}{4 \pi}} \, \sqrt{(2 J' + 1) (2 J'' + 1)} \times \begin{pmatrix}
	J' & 2 & J'' \\
	0 & 0 & 0
	\end{pmatrix}\begin{pmatrix}
	J' & 2 & J'' \\
	M' & m'' & M''
	\end{pmatrix}
\ea
Inserting this last equation in \eqref{eq:quad_app}, we get
\begin{align} 
C_{\ell}^{I I} =&  \frac{1}{2 \ell+1} \, \sum_m
\left[\frac{1}{\mathcal A_t(k)}\right]^2 \left(\frac{16 \pi^2}{5}\right)^2 \left(\frac{5}{4 \pi}\right)^{3/2} \times  \nonumber \\
& \times \sum_{J, M, m'}  \sum_{J', M', m'' } \sum_{J'', M''} \, i^{J-J'} \, (2 \ell + 1) \, (2 J' + 1) \, \sqrt{(2 J + 1)(2 J'' + 1)} \times \nonumber \\
& \times \, \begin{pmatrix}
	\ell & J & 2 \\
	0 & 0 & 0
	\end{pmatrix}\begin{pmatrix}
	\ell & J & 2 \\
	m & M & m'
	\end{pmatrix} \,  \times \begin{pmatrix}
	\ell & J' & 2 \\
	0 & 0 & 0
	\end{pmatrix}\begin{pmatrix}
	\ell & J' & 2 \\
	m & M' & m''
	\end{pmatrix} \, \times \begin{pmatrix}
	J' & 2 & J'' \\
	0 & 0 & 0
	\end{pmatrix}\begin{pmatrix}
	J' & 2 & J'' \\
	M' & m'' & M''
	\end{pmatrix} \times \nonumber \\
&\times \, \int dq \, q^2 \, \int \frac{d^2\hat q}{(2\pi)^3} \, Y_{J M}(\hat q) \, Y_{2 m'}(\hat q) \, Y_{J'' M''}(\hat q) \, \, j_{J}(q d) \, j_{J'}(q d) \, \times \nonumber \\
&\times \, P_s(q) \, \left[ \mathcal A^R_t(k) \, f_{\rm NL}^{R, \rm tts}(k,q) + \mathcal A^L_t(k) \, f_{\rm NL}^{L, \rm tts}(k,q) \right]^2  \, .
\end{align}
We can perform the remnant $\hat q$-integration by exploiting again Eq. \eqref{eq:Gaunt_integral}. We get
\begin{align} 
C_{\ell}^{I I} =&  \frac{1}{2 \ell+1} \, \sum_m
\left[\frac{1}{\mathcal A_t(k)}\right]^2 \left(\frac{16 \pi^2}{5}\right)^2 \left(\frac{5}{4 \pi}\right)^{2} \times  \nonumber \\
& \times \sum_{J, M, m'}  \sum_{J', M', m'' } \sum_{J'', M''} \, i^{J-J'} \, (2 \ell + 1) \, (2 J' + 1) \, (2 J + 1) \, (2 J'' + 1) \times \nonumber \\
& \times \, \begin{pmatrix}
	\ell & J & 2 \\
	0 & 0 & 0
	\end{pmatrix}\begin{pmatrix}
	\ell & J & 2 \\
	m & M & m'
	\end{pmatrix} \,  \times \begin{pmatrix}
	\ell & J' & 2 \\
	0 & 0 & 0
	\end{pmatrix}\begin{pmatrix}
	\ell & J' & 2 \\
	m & M' & m''
	\end{pmatrix} \, \times 
	 \begin{pmatrix}
	J' & 2 & J'' \\
	0 & 0 & 0
	\end{pmatrix}  \begin{pmatrix}
	J & 2 & J'' \\
	0 & 0 & 0
	\end{pmatrix} \times \nonumber \\
&\times \, \sum_{m'', M''} \begin{pmatrix}
	J' & 2 & J'' \\
	M' & m'' & M''
	\end{pmatrix} 
	\begin{pmatrix}
	J & 2 & J'' \\
	M & m' & M''
	\end{pmatrix} \times \, \int \frac{d q \, q^2 }{(2\pi)^3} \, j_{J}(q d) \, j_{J'}(q d) \, P_s(q) \, \times \nonumber \\
	& \, \times  \left[ \mathcal A^R_t(k) \, f_{\rm NL}^{R, \rm tts}(k,q) + \mathcal A^L_t(k) \, f_{\rm NL}^{L, \rm tts}(k,q) \right]^2  \, .
\end{align}
Now, using the properties \eqref{eq:lwigner}, we can re-order the Wigner 3-j symbols as
\begin{align} 
C_{\ell}^{I I} =&  \frac{1}{2 \ell+1} \, \sum_m
\left[\frac{1}{\mathcal A_t(k)}\right]^2 \left(\frac{16 \pi^2}{5}\right)^2 \left(\frac{5}{4 \pi}\right)^{2} \times  \nonumber \\
& \times \sum_{J}  \sum_{J'} \sum_{J'',M''} \, i^{J-J'} \, (2 \ell + 1) \, (2 J' + 1) \, (2 J + 1) \, (2 J'' + 1) \times \nonumber \\
& \times \, \begin{pmatrix}
	\ell & J & 2 \\
	0 & 0 & 0
	\end{pmatrix} \,  \begin{pmatrix}
	\ell & J' & 2 \\
	0 & 0 & 0
	\end{pmatrix} \,
	\begin{pmatrix}
	J' & 2 & J'' \\
	0 & 0 & 0
	\end{pmatrix} \, 
	\begin{pmatrix}
	J & 2 & J'' \\
	0 & 0 & 0
	\end{pmatrix}
	\left[\sum_{M, m'}
	\begin{pmatrix}
	J &  2 &  \ell  \\
	M & m' &   m
	\end{pmatrix} \,
	\begin{pmatrix}
	J & 2 & J'' \\
	M & m' & M''
	\end{pmatrix} \right]  \times \nonumber \\
&\times \, \left[\sum_{m'', M'} \begin{pmatrix}
	2 & J' &  \ell \\
	m'' & M' &  m
	\end{pmatrix}
	\begin{pmatrix}
	 2 & J' & J'' \\
	 m'' & M' & M''
	\end{pmatrix} \right]
	 \times \, \int \frac{d q \, q^2 }{(2\pi)^3} \, j_{J}(q d) \, j_{J'}(q d) \, P_s(q) \, \times \nonumber \\
	& \, \times  \left[ \mathcal A^R_t(k) \, f_{\rm NL}^{R, \rm tts}(k,q) + \mathcal A^L_t(k) \, f_{\rm NL}^{L, \rm tts}(k,q) \right]^2  \, .
\end{align}
We can perform the summations inside square-parenthesis using Eq. \eqref{eq:sum_m_wigner}. We obtain
\begin{align} 
C_{\ell}^{I I} =&  \frac{1}{2 \ell+1} \, \sum_m
\left[\frac{1}{\mathcal A_t(k)}\right]^2 \left(\frac{16 \pi^2}{5}\right)^2 \left(\frac{5}{4 \pi}\right)^{2} \times  \nonumber \\
& \times \sum_{J}  \sum_{J'} \, i^{J-J'}  \, (2 J' + 1) \, (2 J + 1) \, \times \, \begin{pmatrix}
	\ell & J & 2 \\
	0 & 0 & 0
	\end{pmatrix} \,  \begin{pmatrix}
	\ell & J' & 2 \\
	0 & 0 & 0
	\end{pmatrix} \,
    \begin{pmatrix}
	J' & 2 & \ell \\
	0 & 0 & 0
	\end{pmatrix} \, 
	\begin{pmatrix}
	J & 2 & \ell \\
	0 & 0 & 0
	\end{pmatrix} \, 
	\times \nonumber \\
&\times \, \int \frac{d q \, q^2 }{(2\pi)^3} \, j_{J}(q d) \, j_{J'}(q d) \, P_s(q) \, \left[ \mathcal A^R_t(k) \, f_{\rm NL}^{R, \rm tts}(k,q) + \mathcal A^L_t(k) \, f_{\rm NL}^{L, \rm tts}(k,q) \right]^2  \, .
\end{align}
Simplifying and re-ordering various objects we get the final result
\begin{align}
C_{\ell}^{I I} =  \frac{2}{\pi} \frac{1}{\mathcal A^2_t(k)} & \sum_{J, J'} \, i^{J - J'} \, (2 J +1) \, (2 J' +1) \, \left[\begin{pmatrix}
	\ell & J & 2 \\
	0 & 0 & 0
	\end{pmatrix}\right]^2 \, \left[\begin{pmatrix}
	\ell & J' & 2 \\
	0 & 0 & 0
	\end{pmatrix}\right]^2 \times  \nonumber \\
	& \times \int \, dq \, q^2 \, j_{J}(q d) \, j_{J'}(q d) \, P_s(q) \, \left[ \mathcal A^R_t(k) \, f_{\rm NL}^{R, \rm tts}(k,q) + \mathcal A^L_t(k) \, f_{\rm NL}^{L, \rm tts}(k,q)\right]^2 \, .
\end{align}
Switching to dimensionless amplitudes we get
\begin{align}
C_{\ell}^{I I} =   \frac{4 \pi}{\mathcal A^2_t(k)} & \sum_{J, J'} \, i^{J - J'} \, (2 J +1) \, (2 J' +1) \, \left[\begin{pmatrix}
	\ell & J & 2 \\
	0 & 0 & 0
	\end{pmatrix}\right]^2 \, \left[\begin{pmatrix}
	\ell & J' & 2 \\
	0 & 0 & 0
	\end{pmatrix}\right]^2 \times  \nonumber \\
	& \times \int \, \frac{dq}{q} \, j_{J}(q d) \, j_{J'}(q d) \, \mathcal A_s(q) \, \left[ \mathcal A^R_t(k) \, f_{\rm NL}^{R, \rm tts}(k,q) + \mathcal A^L_t(k) \, f_{\rm NL}^{L, \rm tts}(k,q)\right]^2 \, .
\end{align}

\vspace{0.5cm}
\textbf{\centering GW-GW from quadrupolar $\langle \gamma \gamma \gamma \rangle$}
\vspace{0.5cm}

\noindent Auto-correlating two quantities \eqref{eq:tttquad_example} and taking only the parts related to non-mixed tensor power spectra\footnote{The mixed tensor power spectra are vanishing under the assumption of rotational symmetry.}, we get
\ba
C_{\ell}^{I I} =& \frac{1}{2 \ell+1} \, \sum_m \langle \delta^{{\rm GW}, \, I}_{\ell m} \delta^{{\rm GW}, \, I *}_{\ell m} \rangle \\
= & \frac{1}{2 \ell+1} \, \sum_m \left[\frac{2 \pi}{\mathcal A_t(k)}\right]^2 \,  \,\frac{(\ell+2)!}{(\ell-2)!} \, \int \frac{d^3q}{(2\pi)^3} \,\int \frac{d^3p}{(2\pi)^3}  \, \frac{j_\ell(q d)}{(qd)^2} \, \frac{j_\ell(p d)}{(pd)^2} \times \nonumber \\
& \times \Big\{ {}_{- 2}Y_{\ell m}^*(\hat q) \, {}_{- 2}Y_{\ell m}(\hat p) 
\left[\mathcal A^R_{t}(k) \, f_{\rm NL}^{RR,\rm ttt}(k,q) + \mathcal A^L_{t}(k) \, f_{\rm NL}^{LR,\rm ttt}(k,q) \right] \times \nonumber \\
& \qquad \times \left[\mathcal A^R_{t}(k) \, f_{\rm NL}^{RR,\rm ttt}(k,p) + \mathcal A^L_{t}(k) \, f_{\rm NL}^{LR,\rm ttt}(k,p) \right] \langle\gamma_{\vq}^{R} \, \gamma_{\mathbf{p}}^{R *}\rangle +  \nonumber  \\
& \qquad + {}_{+2}Y_{\ell m}^*(\hat q) \, {}_{+ 2}Y_{\ell m}(\hat p)  \left[\mathcal A^L_{t}(k) \, f_{\rm NL}^{LL,\rm ttt}(k,q) + \mathcal A^R_{t}(k) \, f_{\rm NL}^{RL,\rm ttt}(k, q) \right]  \times \nonumber  \\
&  \qquad \qquad \times \left[\mathcal A^L_{t}(k) \, f_{\rm NL}^{LL,\rm ttt}(k,p) + \mathcal A^R_{t}(k) \, f_{\rm NL}^{RL,\rm ttt}(k,p) \right]  \langle\gamma_{\vq}^{L} \, \gamma_{\mathbf{p}}^{L *}\rangle \Big\} \, .
\ea
By exploiting the definition of tensor power spectra we can integrate out one of the two momenta using the Dirac-delta, and get
\ba
C_{\ell}^{I I} =&  \frac{1}{2 \ell+1} \, \sum_m \left[\frac{2 \pi}{\mathcal A_t(k)}\right]^2 \,  \,\frac{(\ell+2)!}{(\ell-2)!} \, \int dq \, q^2 \, \int \frac{d^2\hat q}{(2\pi)^3} \,  \left[\frac{j_\ell(q d)}{(qd)^2}\right]^2 \, \times \nonumber \\
& \times \Big\{ {}_{- 2}Y_{\ell m}^*(\hat q) \, {}_{- 2}Y_{\ell m}(\hat q) 
\, P^R_t(q) \left[\mathcal A^R_{t}(k) \, f_{\rm NL}^{RR,\rm ttt}(k,q) + \mathcal A^L_{t}(k) \, f_{\rm NL}^{LR,\rm ttt}(k,q) \right]^2 +  \nonumber  \\
& \qquad + {}_{+2}Y_{\ell m}^*(\hat q) \, {}_{+ 2}Y_{\ell m}(\hat q)  \, P^L_t(q) \left[\mathcal A^L_{t}(k) \, f_{\rm NL}^{LL,\rm ttt}(k,q) + \mathcal A^R_{t}(k) \, f_{\rm NL}^{RL,\rm ttt}(k, q) \right]^2 \Big\} \, .
\ea
We can now easily integrate the remnant $\hat q$-angular dependence getting the final result
\begin{align}
C_{\ell}^{I I} =  \frac{1}{2 \pi} \frac{1}{\mathcal A^2_t(k)} \, \frac{(\ell + 2)!}{(\ell - 2)!} \int \, dq \, q^2 \, & \left[\frac{j_\ell(q d)}{(q d)^2}\right]^2 \, \Big\{ P^R_t(q) \left[ \mathcal A^R_t(k) \, f_{\rm NL}^{RR, \rm ttt}(k,q) + \mathcal A^L_t(k) \, f_{\rm NL}^{LR, \rm ttt}(k,q)\right]^2 + \nonumber \\
&+ P^L_t(q) \left[ \mathcal A^L_t(k) \, f_{\rm NL}^{LL, \rm ttt}(k,q) + \mathcal A^R_t(k) \, f_{\rm NL}^{RL, \rm ttt}(k,q)\right]^2 \Big\} \, .
\end{align}
Switching to dimensionless amplitudes we get
\begin{align}
C_{\ell}^{I I} =   \frac{\pi}{\mathcal A^2_t(k)} \, \frac{(\ell + 2)!}{(\ell - 2)!} \int \, \frac{dq}{q}  \, & \left[\frac{j_\ell(q d)}{(q d)^2}\right]^2 \, \Big\{ \mathcal A^R_t(q) \left[ \mathcal A^R_t(k) \, f_{\rm NL}^{RR, \rm ttt}(k,q) + \mathcal A^L_t(k) \, f_{\rm NL}^{LR, \rm ttt}(k,q)\right]^2 + \nonumber \\
&+ \mathcal A^L_t(q) \left[ \mathcal A^L_t(k) \, f_{\rm NL}^{LL, \rm ttt}(k,q) + \mathcal A^R_t(k) \, f_{\rm NL}^{RL, \rm ttt}(k,q)\right]^2 \Big\} \, .
\end{align}

\vspace{0.5cm}
\textbf{\centering GW-CMB from monopolar $\langle \gamma \gamma \zeta \rangle$}
\vspace{0.5cm}

\noindent The logical steps are similar to the GW-GW from monopolar $\langle \gamma \gamma \zeta \rangle$ case, therefore we skip the derivation.  We just point out that when GW anisotropies from a squeezed $\langle \gamma \gamma \zeta \rangle$ are cross-correlated with CMB $B$ modes, then the resultant cross-correlation is vanishing as scalar perturbations do not source $B$ modes at linear level.

\vspace{0.5cm}
\textbf{\centering GW-CMB from quadrupolar $\langle \gamma \gamma \zeta \rangle$}
\vspace{0.5cm}

\noindent Cross-correlating one quantity \eqref{eq:ttsquad_example} with one quantity \eqref{eq:a_CMB_scalar}, we get
\begin{align}
C_{\ell}^{I X} =& \frac{1}{2 \ell+1} \, \sum_m \langle \delta^{{\rm GW}, \, I}_{\ell m} a^{{\rm X}*}_{\ell m} \rangle \nonumber \\
= & \frac{1}{2 \ell+1} \, \sum_m
\left[\frac{4 \pi}{\mathcal A_t(k)}\right] \left(\frac{16 \pi^2}{5}\right) \sqrt{\frac{5}{4 \pi}} \, \sum_{J, M, m'}  \, i^{J-\ell}  \, \sqrt{(2 J + 1) (2 \ell + 1)} \times \nonumber \\
& \times \begin{pmatrix}
	\ell & J & 2 \\
	0 & 0 & 0
	\end{pmatrix}\begin{pmatrix}
	\ell & J & 2 \\
	m & M & m'
	\end{pmatrix} \,  \times  \int \frac{d^3q}{(2\pi)^3}\, \int \frac{d^3p}{(2\pi)^3} \, Y_{J M}(\hat q) \, Y_{2 m'}(\hat q) \, Y_{\ell m}(\hat p) \, j_J(q d) \, {\cal T}_{\ell(s)}^{X}(p) \,  \times \nonumber \\
&\times \left[ \mathcal A^R_t(k) \, f_{\rm NL}^{R, \rm tts}(k,q) + \mathcal A^L_t(k) \, f_{\rm NL}^{L, \rm tts}(k,q) \right] \, \langle \zeta_{\bq} \, \zeta_{\mathbf{p}}^*\rangle \, .
\end{align}
We can exploit the definition of scalar power spectrum to integrate out one of the two momenta using the Dirac-delta. We get
\begin{align}
C_{\ell}^{I X} =&  \frac{1}{2 \ell+1} \, \sum_m
\left[\frac{4 \pi}{\mathcal A_t(k)}\right] \left(\frac{16 \pi^2}{5}\right) \sqrt{\frac{5}{4 \pi}} \, \sum_{J, M, m'}  \, i^{J-\ell}  \, \sqrt{(2 J + 1) (2 \ell + 1)} \times \nonumber \\
& \times \begin{pmatrix}
	\ell & J & 2 \\
	0 & 0 & 0
	\end{pmatrix}\begin{pmatrix}
	\ell & J & 2 \\
	m & M & m'
	\end{pmatrix} \,  \times  \int dq \, q^2 \, \int \frac{d^2\hat q}{(2\pi)^3}  \, Y_{J M}(\hat q) \, Y_{2 m'}(\hat q) \, Y_{\ell m}(\hat q) \, j_J(q d) \, {\cal T}_{\ell(s)}^{X}(q) \,  \times \nonumber \\
&\times \, P_s(q)  \left[ \mathcal A^R_t(k) \, f_{\rm NL}^{R, \rm tts}(k,q) + \mathcal A^L_t(k) \, f_{\rm NL}^{L, \rm tts}(k,q) \right] \, .
\end{align}
We can perform the $\hat q$-integration by exploiting Eq. \eqref{eq:Gaunt_integral}. We get 
\begin{align}
C_{\ell}^{I X} =&  \frac{1}{2 \ell+1} \, \sum_m
\left[\frac{4 \pi}{\mathcal A_t(k)}\right] \left(\frac{16 \pi^2}{5}\right) \frac{5}{4 \pi} \, \sum_{J}  \, i^{J-\ell}  \, (2 J + 1) (2 \ell + 1) \times \nonumber \\
& \times \left[\begin{pmatrix}
	\ell & J & 2 \\
	0 & 0 & 0
	\end{pmatrix}\right]^2
	\sum_{M, m'} \left[\begin{pmatrix}
	\ell & J & 2 \\
	m & M & m'
	\end{pmatrix}\right]^2 \,  \times  \int \frac{dq \, q^2}{(2\pi)^3} \, j_J(q d) \, {\cal T}_{\ell(s)}^{X}(q) \,  \times \nonumber \\
&\times \, P_s(q)  \left[ \mathcal A^R_t(k) \, f_{\rm NL}^{R, \rm tts}(k,q) + \mathcal A^L_t(k) \, f_{\rm NL}^{L, \rm tts}(k,q) \right] \, .
\end{align}
By performing the $M, m'$-summation using \eqref{eq:sum_m_wigner} and making some simplifications, we get the final result
\begin{align}
C_{\ell}^{I X} =  \frac{2}{\pi} \frac{1}{\mathcal A_t(k)} & \sum_J \, i^{J - \ell} \, (2 J +1) \, \left[\begin{pmatrix}
	\ell & J & 2 \\
	0 & 0 & 0
	\end{pmatrix}\right]^2 \times \nonumber \\
	& \times \int \, dq \, q^2 \, j_J(q d) \, {\cal T}_{\ell(s)}^{X}(q) \, P_s(q) \, \left[ \mathcal A^R_t(k) \, f_{\rm NL}^{R, \rm tts}(k,q) + \mathcal A^L_t(k) \, f_{\rm NL}^{L, \rm tts}(k,q)\right] \, .
\end{align}
Switching to dimensionless amplitudes we get
\begin{align}
C_{\ell}^{I X} =  \frac{4 \pi}{\mathcal A_t(k)} & \sum_J \, i^{J - \ell} \, (2 J +1) \, \left[\begin{pmatrix}
	\ell & J & 2 \\
	0 & 0 & 0
	\end{pmatrix}\right]^2 \times \nonumber \\
	& \times \int \, \frac{dq}{q} \, j_J(q d) \, {\cal T}_{\ell(s)}^{X}(q) \,  \mathcal A_s(q) \, \left[ \mathcal A^R_t(k) \, f_{\rm NL}^{R, \rm tts}(k,q) + \mathcal A^L_t(k) \, f_{\rm NL}^{L, \rm tts}(k,q)\right] \, .
\end{align}
Again, when $X = B$ the resultant cross-correlation is vanishing as scalar perturbations do not source $B$ modes at linear level.

\vspace{0.5cm}
\textbf{\centering GW-CMB from quadrupolar $\langle \gamma \gamma \gamma \rangle$}
\vspace{0.5cm}

\noindent The logical steps are similar to the GW-GW from monopolar $\langle \gamma \gamma \gamma \rangle$ case, therefore we omit the derivation.

\bibliography{references.bib}
\end{document}